\documentclass{aa}  

\usepackage{graphicx}
\usepackage{natbib}
\bibpunct{(}{)}{;}{a}{}{,} % to follow the A&A style

\usepackage{txfonts}
\begin{document} 

\title{The {\it Herschel}\thanks{{\it Herschel} is an ESA space observatory with
science instruments provided by European-led Principal Investigator consortia
and with important participation from NASA.} Virgo Cluster Survey} 
\subtitle{XX. Dust and gas in the foreground Galactic cirrus}

\author{
S. Bianchi\inst{1},
C. Giovanardi\inst{1},
M. W. L.  Smith\inst{2},
J. Fritz\inst{3},
J. I. Davies\inst{2},
M. P. Haynes\inst{4},
R. Giovanelli\inst{4},
M. Baes\inst{5},
M. Bocchio\inst{6},
S. Boissier\inst{7},
M. Boquien\inst{8},
A. Boselli\inst{7},
V. Casasola\inst{1},
C. J. R. Clark\inst{2},
I. De Looze\inst{9},
S. di Serego Alighieri\inst{1},
M. Grossi\inst{10},
A. P. Jones\inst{6},
T. M. Hughes\inst{11},
L. K. Hunt\inst{1},
S. Madden\inst{12},
L. Magrini\inst{1},
C. Pappalardo\inst{13,14},
N. Ysard\inst{6},
S. Zibetti\inst{1}
}

\institute{
INAF--Osservatorio Astrofisico di Arcetri, Largo E. Fermi, 5, I-50125 Firenze, Italy. 
\email{sbianchi@arcetri.astro.it}
\and
School of Physics and Astronomy, Cardiff University, Queens Building, The Parade, Cardiff, CF24 3AA, UK
\and
Instituto de Radioastronom\'ia y Astrof\'isica, UNAM, Antigua Carretera a P\'atzcuaro \#8701, Morelia, Michoac\'an, Mexico 
\and
Cornell Center for Astrophysics and Planetary Science, Space Sciences Bldg., Cornell University, Ithaca, NY 14853
\and
Sterrenkundig Observatorium, Universiteit Gent, Krijgslaan 281-S9, B-9000 Gent, Belgium
\and
Institut d’Astrophysique Spatiale (IAS), UMR 8617, CNRS/Université Paris-Sud, 91405 Orsay, France
\and
Aix Marseille Universit\'e, CNRS, LAM (Laboratoire d’Astrophysique de Marseille), UMR 7326, 13388 Marseille, France
\and
Unidad de Astronomia, Facultad de Ciencias Basicas, Universidad de Antofagasta, Avenida Angamos 601, Antofagasta 1270300, Chile 
\and
Dept. of Physics \& Astronomy, University College London, Gower Street, London WC1E 6BT, UK
\and
Observat\'orio do Valongo, Universidade Federal do Rio de Janeiro, Ladeira Pedro Ant\^onio 43, Rio de Janeiro, Brazil
\and
Instituto de F\'{i}sica y Astronom\'{i}a, Universidad de Valpara\'{i}so, Avda. Gran Breta\~{n}a 1111, Valpara\'{i}so, Chile
\and
Laboratoire AIM, CEA/DSM - CNRS - Universit\'e Paris Diderot, IRFU/Service d'Astrophysique, CEA Saclay, 91191 Gif-sur-Yvette, France
\and
Instituto de Astrof\'isica e Ci\^encias do Espa\c{c}o, Universidade de Lisboa, Tapada da Ajuda, 1349-018 Lisboa, Portugal
\and
Observat\'orio Astron\'omico de Lisboa, Tapada da Ajuda, 1349-018 Lisboa, Portugal
 \\
}

\date{Received ; accepted }

% \abstract{}{}{}{}{} 
% 5 {} token are mandatory
 
\abstract{
We study the correlation between far-infared/submm dust emission and atomic gas 
column density in order to derive the properties of the high Galactic latitude,
low density, Milky Way cirrus in the foreground of the Virgo cluster of galaxies.
Dust emission maps from 60 to 850~$\mu$m are obtained from SPIRE observations 
carried out within the {\it Herschel} Virgo Cluster Survey, complemented 
by IRAS-IRIS and {\em Planck}-HFI maps. Data from the Arecibo legacy Fast ALFA 
Survey is used to derive atomic gas column densities for two broad velocity 
components, low and intermediate velocity clouds.
Dust emissivities are derived
for each gas component and each far-infared/submm band. For the low velocity
clouds, we measure
an average emissivity $\epsilon^\mathrm{LVC}_\nu = (0.79\pm0.08) \times 10^{-20}$ 
MJy sr$^{-1}$ cm$^2$ at 250~$\mu$m.
After fitting a modified blackbody to the available bands, we estimated a dust 
absorption cross-section
$\tau^\mathrm{LVC}_\nu/{N_\ion{H}{i}} = (0.49\pm0.13) \times 10^{-25}$ cm$^2$ H$^{-1}$ 
at 250~$\mu$m (with dust temperature $T=20.4\pm1.5$K and spectral index 
$\beta=1.53\pm0.17$). The results are in excellent agreement with those
obtained by {\em Planck} over a much larger 
coverage of the high Galactic latitude cirrus (50\% of the sky vs 0.2\% in our
work).
For dust associated with intermediate velocity gas, we confirm earlier {\em Planck} results and 
find a higher temperature and lower emissivity and cross-section.
After subtracting the modelled components, we find 
{
regions at scales smaller than 20$\arcmin$ where the residuals deviate significantly 
from the average, cosmic-infrared-background dominated, scatter.
These large residuals} are most likely due to local variations in the cirrus dust properties (and/or 
the dust/atomic-gas correlation) or to high-latitude
molecular clouds with average $N_{\mathrm{H}_2} \lesssim 10^{20}$ cm$^{-2}$.
{
We find no conclusive evidence for intracluster dust emission in Virgo.
}
}

\keywords{
dust, extinction - radiation mechanisms: thermal - infrared, submillimeter, radio lines: ISM - local interstellar matter
}

\titlerunning{Dust and gas in the Virgo cirrus}
\authorrunning{S.\ Bianchi et al.}

\maketitle
%
%________________________________________________________________

\section{Introduction}

Diffuse far-infrared (FIR) emission from high-Galactic latitude dust (the
Milky Way {\em cirrus}) is an important test bed for models of interstellar 
grains. The thinness of the Galactic disk, our peripheric position in it, 
and a viewing direction away from the Galactic plane offer clear advantages
to modelling: the main source of dust heating is provided by the ambient
radiation \citep[the Local Interstellar Radiation Field - LISRF;][]{MathisA&A1983};
the diffuse radiation gradients perpendicular to the disk being small
\citep[as shown by radiative transfer models of galactic disks; ][]{BianchiA&A2000b,BocchioA&A2013,PopescuMNRAS2013};
thus the mixing of dust at different temperatures (and/or with different
properties) along a single line of sight is limited.

Starting from the first detection of the cirrus emission by the IRAS satellite
\citep{LowApJL1984}, a tight correlation of the FIR surface brightness with 
the atomic gas column density was observed \citep{BoulangerApJ1988}.
The cirrus surface brightness per \ion{H}{i} column density (a quantity known as 
{\em emissivity}) measured in the 60 and 100~$\mu$m IRAS bands was then used to verify 
the predictions or constrain the properties of early interstellar grain models 
\citep{DraineApJ1984,DesertA&A1990}. 

With the advent of the instruments aboard the COBE satellite, and in particular
the FIRAS spectrophotometer, it was possible to study the Spectral Energy
Distribution (SED) of the cirrus up to the peak of thermal emission and beyond.
\citet{BoulangerA&A1996} found a tight correlation with \ion{H}{i} column density
and measured the emissivity up to 1~mm. After fitting a modified blackbody (MBB) to the
data, the dust absorption cross-section per unit \ion{H}{i} column density was retrieved, which
was found to be in agreement with the available dust models. Since then, the FIRAS
emissivity has been one of the major constraints { to dust models, which
were formulated} to reproduce its SED levels and shape, resulting in an average
dust cross-section $\propto \nu^\beta$, with $\beta\approx 1.8-2$
\citep[see, e.g., ][]{DraineARA&A2003,ZubkoApJS2004,CompiegneA&A2011,JonesA&A2013}.

The recent analysis of data from the {\em Planck} satellite, however, revealed 
an inconsistency between {\em Planck} and FIRAS, which prompted a recalibration
on planet fluxes of the 350 and 550~$\mu$m data from the High Frequency Instrument 
\citep[HFI; ][]{Planck2013VIII}. As a result, the newly determined
emissivities have been found to be lower than the FIRAS values, and with
a reduced dust cross-section spectral index \citep[$\beta\approx 1.6$;][]{Planck2013XI,PlanckIntermediateXVII}.
{ These results suggest that a re-evaluation of the dust model parameters may 
be required,}
and in particular of the grain size distributions and relative contribution
of grains of different chemical composition to the SED at 100~$\mu$m$ \la \lambda \la$ 1 mm.
Furthermore, the higher resolution of {\em Planck} data with respect to FIRAS
highlighted local dust emissivity variations 
with the environment \citep{Planck2013XI,PlanckIntermediateXVII}.

A tenuous cirrus is seen in the foreground of the Virgo cluster
of galaxies ($l=283.8^\circ, b=74.4^\circ$) through scattered starlight
in deep optical \citep{RudickApJ2010,MihosProc2015} and UV images 
\citep{CorteseMNRAS2010,BoissierA&A2015}; it corresponds to low
extinction, with $0.02 < E(B-V) < 0.1$ \citep{BoissierA&A2015}.
Its emission is clearly detected at 250, 350 and
500~$\mu$m in images obtained by the SPIRE instrument  
\citep{GriffinA&A2010} aboard the {\em Herschel} Space Observatory
\citep{PilbrattA&A2010}, as part of the {\it Herschel} Virgo Cluster
Survey \citep[HeViCS; ][]{DaviesA&A2010,DaviesMNRAS2012,AuldMNRAS2013}.
In this work, we will derive the dust emissivity of the HeViCS cirrus in 
the SPIRE bands.
This will allow to verify the latest {\em Planck} results, since the
SPIRE calibration also is based on planet models. Besides, the 250~$\mu$m
band will fill the gap between the IRAS and {\em Planck} data,
and complement the lower sensitivity and resolution COBE-DIRBE
emissivity at 240~$\mu$m \citep{PlanckIntermediateXVII}. When compared
to the {\it Planck} large area estimates, the relatively small sky coverage 
of the HeViCS field will also highlight local emissivity 
variations.

The emissivities will be derived using $\ion{H}{i}$ observations from the Arecibo
Legacy Fast ALFA survey \citep[hereafter ALFALFA][]{GiovanelliAJ2005}. The high resolution
of the ALFALFA data (FWHM=3$\farcm$5) is comparable to that of the longest wavelength
data dominated by dust thermal emission (the 850~$\mu$m {\em Planck}-HFI band,
with 4$\farcm$8);
 it is also a factor two better than other $\ion{H}{i}$ 
data used to derive dust emissivities: previous large-area studies
\citep{BoulangerA&A1996,Planck2013XI} used observations at FWHM=0.6$^\circ$ 
from the Leiden/Argentine/Bonn (LAB) Survey of Galactic $\ion{H}{i}$ \citep{KalberlaA&A2005}.
The FWHM=14$\farcm$5 GASS $\ion{H}{i}$ survey of the southern sky has been used in
\citet{PlanckIntermediateXVII}, while \citet{PlanckEarlyXXIV}
exploited Green Bank Telescope maps at FWHM=9$\farcm$1. Only in the recent work
of \citet{ReachApJ2015} Arecibo telescope data from the Galactic ALFA 
\citep[GALFA; ][]{PeekApJS2011} survey have been 
used to study the correlation between $\ion{H}{i}$ column density of
isolated high-latitude clouds and {\em Planck}-based dust column densities.

{
The resolution of the ALFALFA survey thus allows a characterization of the
residuals to the dust-$\ion{H}{i}$ correlation at smaller scales. 
}
In fact,
one of the aims of HeViCS is the detection of dust emission from the intracluster 
medium (ICM), once the structure of the foreground cirrus is subtracted using Galactic
$\ion{H}{i}$ as a template. 
{ 
The presence of the cirrus is indeed the limiting 
factor in these studies: so-far the only claim of the detection of FIR
emission from the ICM is that on the Coma cluster by \citet{StickelA&A1998}, using 
data from the ISOPHOT instrument aboard the ISO satellite. However, the analysis
of five other clusters with analogous data and techniques
yielded no detection \citep{StickelA&A2002}; a following analysis of
Coma using data from the {\em Spitzer} satellite dismissed the putative 
emission as cirrus contamination \citep{KitayamaApJ2009}.
}

This paper is organised as follows: in Sect.~\ref{sec:data} we present the
ALFALFA and HeViCS data, together with other ancillary maps used in this
work. In Sect.~\ref{sec:ana} we describe the method used to derive the 
correlation between dust and $\ion{H}{i}$. The fitted emissivities
are presented in Sect.~\ref{sec:res}, while Sect.~\ref{sec:sed} is dedicated 
to the derivation of the mean opacity cross section from  the
emissivity SEDs. The possible Galactic origin of the largest residuals is
discussed in Sect.~\ref{sec:resi}, while in Sect.~\ref{sec:icd} we put
constrains on the dust emission from the Virgo ICM.
The work is summarised in Sect.~\ref{sec:sum}.

\section{Data sets}
\label{sec:data}

{
The HeViCS project obtained FIR/submm images of about 84 deg$^2$ over the 
Virgo cluster, using the photometers aboard {\em Herschel}.
The survey area was covered with four overlapping
fields. The survey strategy and observations are described in detail in several 
previous publication of the HeViCS series
\citep[e.g. ][]{DaviesA&A2010,DaviesMNRAS2012,AuldMNRAS2013}.
}

In this section we describe the \ion{H}{i} and FIR/submm data sets we 
have used to derive the emissivity of MW dust in the HeViCS footprint.

\subsection{\ion{H}{i} ALFALFA Data}
\label{sec:alfalfa}

{ 
ALFALFA \citep{GiovanelliAJ2005} is a
spectro-photometric, wide-field survey {at the Arecibo radiotelescope}  aimed to delivering the properties of galaxies in the
local Universe, as revealed by the 21~cm line of interstellar atomic Hydrogen. It covers a
spectral range of 100 MHz with a resolution of 25 KHz ($\sim$5.1 km/s), including the spectral
domain of interstellar and perigalactic emission. It covers uniformly the sky footprint of
the HeViCS survey, with an angular resolution of 3.5$\arcmin$ FWHM. Data have been taken in drift mode,
i.e. with the telescope scanning tracks of constant declination at the sidereal rate. Two
independent polarization components of the data are recorded. Basic data units are three-dimensional.
Each data cube is complemented by a two-dimensional map of the continuum background and a 3-d
array carrying the data quality of each pixel.

The use of the ALFALFA data for this work requires two important differences in the generation
of the data cubes, with respect to the pipeline processing adopted for extragalactic applications:
(i) the spectral stretch of the data cubes is restricted to the range between -1000 and +1050
km/s, which included 400 spectral channels spaced by 25 KHz (5.1 km/s) and (ii) the flux scale
is calibrated in Kelvin degrees of antenna temperature T$_\mathrm{A}$. The sky footprint of each cube
retains the same size as the  standard ALFALFA data cubes, i.e. $2.4^\circ \times 2.4^\circ$, sampled
at 1$\arcmin$ pixel separation. The array dimensions of each data cube are then $400\times 2\times 144
\times 144$, where the
'2' refers to the two polarization channels. Each of the four HeViCS fields is fully covered by
10 ALFALFA data cubes, with significant coverage overlap: each point in the HeViCS sky footprint
is contained in at least 2 different ALFALFA data cubes.
}

We then proceeded to the baselining. For each grid and for each polarization,
we assembled all the spectra at a given declination in an image,
thus obtaining the equivalent of a longslit image.
Upon inspection of the central 1000~km s$^{-1}$,
and excluding the range occupied by the Galactic line 
at levels over the noise, we fitted to each spectrum 
and then removed a polynomial baseline of order up to 6.
During this phase we also removed from the ``longslit" images
any striping due to residual RFI or autocorrelator faults.
{ 
The processing pipeline was overall satisfactory, except for lines of sight within 15'
of the the center of galaxy M87, where the strong continuum emission of Virgo A drastically
increases noise and associated standing  waves makes spectral baselining impossible.
}

The next phase was the flatfielding of the channel maps.
Again separately for the two polarizations, we assembled
the grid data for each velocity channel into a different map. 
These maps were singly inspected and, if required, corrected for
residual striping or stepping at the junction between different
drifts. The flatfielding procedure was repeated twice: first to eliminate
the most obvious defects, and then again for refinement and a double check
of the results.

Finally we used an IDL procedure to produce, for each grid, FITS 2D images of the
single fully reduced channel maps, or of a velocity range of choice;
such a procedure also averages the two polarizations.
The {\tt swarp} software \citep{swarp}
was used to mosaic the FITS output for the grids into
maps covering each single HeViCS field and the whole HeViCS footprint.
The final maps were converted into units of main beam temperature.

\begin{figure}
\includegraphics[width=\hsize,trim = 0 0 30bp 20bp]{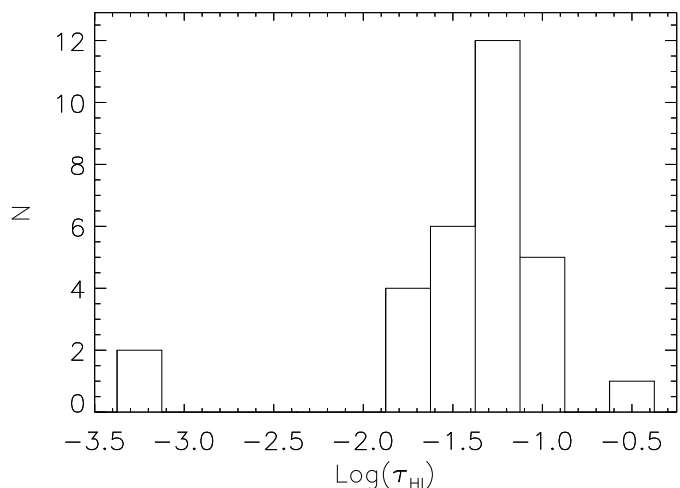}
\caption{Distribution histogram of the 21~cm line opacity computed for the 
30 absorption features detected on the line of sight to NVSS continuum 
radiosources.}
\label{fig:tauhist}
\end{figure}

In order to check the \ion{H}{i} 21~cm line optical depths in the area,
we selected from the NVSS catalog \citep{CondonAJ1998} all 1400~MHz
radiosources with a flux density in excess of 500~mJy
located within the HeViCS footprint.
This flux limit was computed, given the noise figures of the
ALFALFA spectra, to ensure a clear detection (S/N>5) of absorption features
from column densities $N_{\ion{H}{i}} > 5 \times 10^{18}$~cm$^{-2}$
with average conditions for spin temperature (100~K)
and velocity dispersion (5~km~s$^{-1}$).
We rejected those sources located within 20$\arcmin$ of M87,
and those not isolated enough or extended,
ending up with a list of 22 objects.
For each source, the central spectrum was obtained by summing
the spectra within a radius of 3$\arcmin$.
The emission spectrum was estimated as the median of six similar apertures
surrounding in an hexagonal pattern the central source at distance 
of 10$\arcmin$.  The absorption spectrum was derived by subtracting 
the emission spectrum from the central one, and analysed using
standard techniques \citep[see, e.g., ][]{SpitzerBook1978}.
Absorption was detected in all 22 spectra, sometimes
with multiple features. 
{
The optical depths, which we derive with uncertainty of $\sim$0.003, span
from 0.01 to 0.4 but with an average of only $0.058 \pm 0.058$;
their distribution is illustrated in Fig.~\ref{fig:tauhist}.
The spin temperatures, measured with uncertainty of $\sim$10~K, are low, 
between 50 and 100~K, with only one case exceeding 150~K; their
mean value is $55\pm28$~K. 
The $\ion{H}{i}$ column densities 
range between $2 \times 10^{18}$ and $2 \times 10^{20}$~cm$^{-2}$,
with a log mean of $19.60 \pm 0.54$.
Given that the peak brightness temperature of the line is
around 3~K, and the noise in the spectra is 0.003K, even taking into
account the error in the optical depth corrections, the uncertainty in the 
total column density is always better than 4\%. The average correction factor for
optical thickness is 1.021$\pm$0.029; consequently, throughout the paper,
we derived the \ion{H}{i} column
densities at the optically thin limit. 
}

\begin{figure}
\includegraphics[width=\hsize]{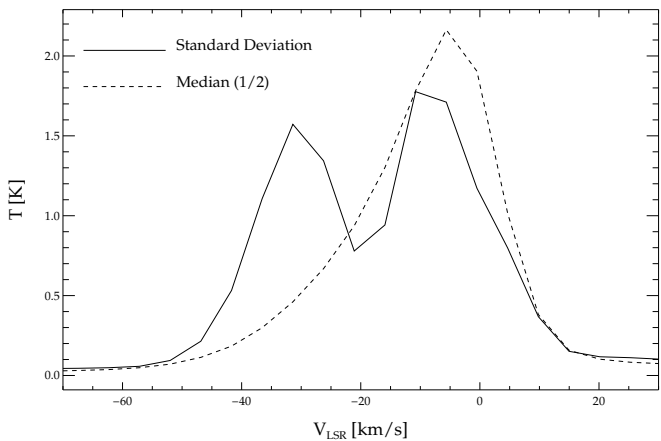}
\caption{Standard deviation (solid line) and median (dashed line) ALFALFA spectra for the HeViCS 
field.  The median spectrum has been divided by two for ease of presentation}
\label{medstd}
\end{figure}

\citet{PlanckEarlyXXIV} has shown that the \ion{H}{i} emission can be decomposed into 
various velocity components by inspecting the { channel-to-channel variations of the
median and of the standard deviation; the second in particular is more sensitive 
to structure variation across the field.}
In Fig.~\ref{medstd} we plot these median and standard deviation spectra for the full HeViCS 
field. The standard deviation spectrum shows that there are two components, partially 
overlapping in velocity at $v_\mathrm{LSR}\approx -20$~km s$^{-1}$. 
{ 
Following the approach of \citet{PlanckEarlyXXIV}, we thus produced two separate 
maps of column density using
\[
N_\ion{H}{i} = 1.823\times 10^{18} \sum T_b(v) \Delta v \,\,\mathrm{cm}^{-2}
\]
with $T_b$ the main beam temperature spectrum for each pixel. The
summation was extended over the velocity range $-20 \la v_\mathrm{LSR}$~(km s$^{-1}$) $\la 100$
for Low Velocity Clouds (LVC); over $-100 \la v_\mathrm{LSR}$~(km s$^{-1}$) $\la -20$, for 
Intermediate Velocity Clouds (IVC). 
No other significant velocity components are found in the HeViCS field:
in particular, there are no conspicuous High Velocity Clouds at the galactic
latitude of Virgo \citep[see, e.g., Fig.1 in ][]{WakkerARA&A1997} and indeed
the velocity channels in the range 
$-200 \la v_\mathrm{LSR}$~(km s$^{-1}$) $\la -100$ are devoid of large-scale
emission features.
}

We have verified the consistency of the column density values measured on our ALFALFA 
map with those from the LAB survey \citep{KalberlaA&A2005}.  In particular, the Arecibo telescope 
is known to have significant stray-radiation contamination, which can 
affect precise determination of the total column densities of large-scale features
\citep[see, e.g.\ ][]{PeekApJS2011}. Instead, the LAB data have been corrected for stray 
radiation. We checked the contamination of the ALFALFA map by smoothing it to the
LAB resolution and performing a linear correlation between the column densities of the 
two dataset, for both the LVC and IVC channels. We found that the ALFALFA map of the
LVC channel has a positive offset of $3.4\times 10^{19}$ cm$^{-2}$ with respect to 
LAB, and gain close to unity, 1.06. For the IVC channel, the gain is higher
(1.16) and the offset smaller ($1.3\times 10^{19}$ cm$^{-2}$). Since in this work we
compare our determination of dust emissivities with those obtained using the LAB
\citep{Planck2013VIII,Planck2013XI} we corrected our maps for those offsets and gains.
We will comment on these corrections in Sect.~\ref{sec:res}.

\subsection{HeViCS SPIRE images}
\label{sec:spire}

Images at 250, 350 and 500~$\mu$m  were obtained using the SPIRE instrument
\citep{GriffinA&A2010} aboard {\em Herschel}.
The SPIRE data were reduced using the dedicated software HIPE,
version 11 \citep{OttProc2010} and following the recommended
procedures for maps of extended emission\footnote{ For information on
the adopted calibration, beam sizes and color corrections, see
the SPIRE Handbook v.2.5 (2014), available at the {\em Herschel} 
Astronomer' website.}. In previous HeViCS papers we used a 
custom method for the temperature drift correction and residual baseline 
subtraction \citep[Bright Galaxy Adaptive Element - BriGAdE;][]{SmithThesis2012}, 
which is optimised for galaxies. 
For the larger scale of the cirrus emission, however, BriGAdE tends to 
flatten the surface brightness gradients. Thus, for the work of this paper 
we preferred to use the standard {\em destriper} available within HIPE,
adopting a constant offset for each bolometer.

Because of memory limitations and of the relatively small overlap of the
four tiles of which the survey is composed, it was not possible to run
the {\em destriper} on the full HeViCS area. 
In any case, we have preferred
to conduct our analysis on the individual fields, to study possible variations
of the dust properties across the survey area. This also minimizes the impact of
larger scale gradients resulting from diffuse Galactic emission and zodiacal light, 
which are poorly sampled by our limited-size tiles.
As SPIRE is only sensitive to 
relative field variations, each field is offset from the others 
because of their different median surface brightness levels. While it 
is possible to set an absolute flux offset by correlating {\em Herschel} 
and {\em Planck} observations within HIPE, { we retained 
the original levels and relied on our own determination of the zero levels 
with respect to \ion{H}{i} column density (see Sect.~\ref{sec:ana}).
}

Images were created in Jy/beam with pixel sizes of 6, 8 and 12" at 250, 
350 and 500~$\mu$m.  These values are about 1/3 of the main beam FWHM, 
which, for the adopted pixel sizes, is about 18, 24.5 and 36". 
Units were finally converted to MJy sr$^{-1}$.
The calibration uncertainty is 7\% at all SPIRE wavelengths, including
uncertainties in the model, in the measurements of the calibrator flux, and
in the estimate of the beam areas.

We did not use the PACS \citep{PoglitschA&A2010} data taken in parallel with SPIRE.
The cirrus is indeed detected at very faint levels at 160~$\mu$m, but the images 
contain large scale artifacts of the data reduction process that prevented their
usage.

\begin{figure*}
\includegraphics[height=9cm]{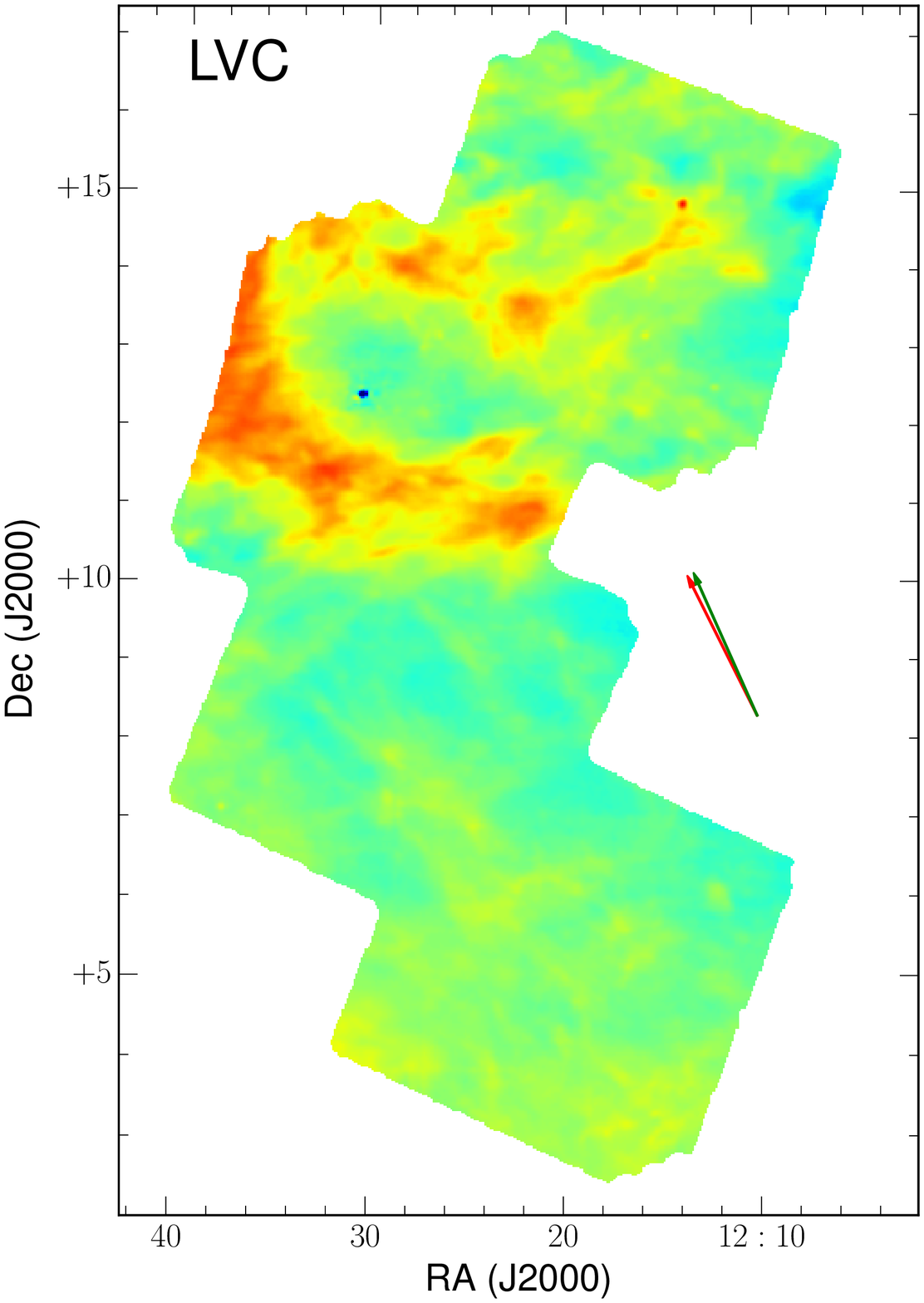}
\includegraphics[height=9cm]{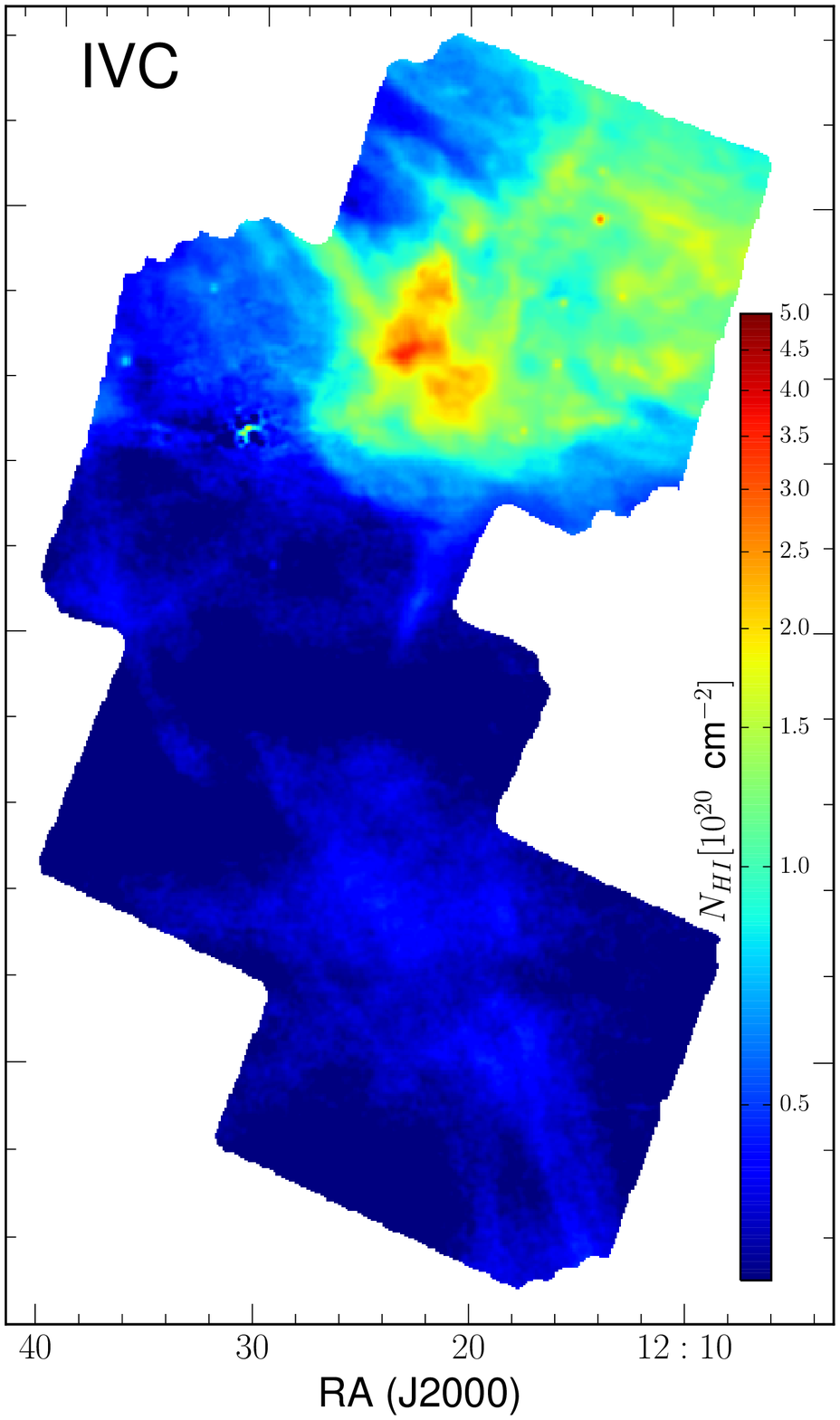}
\includegraphics[height=9cm]{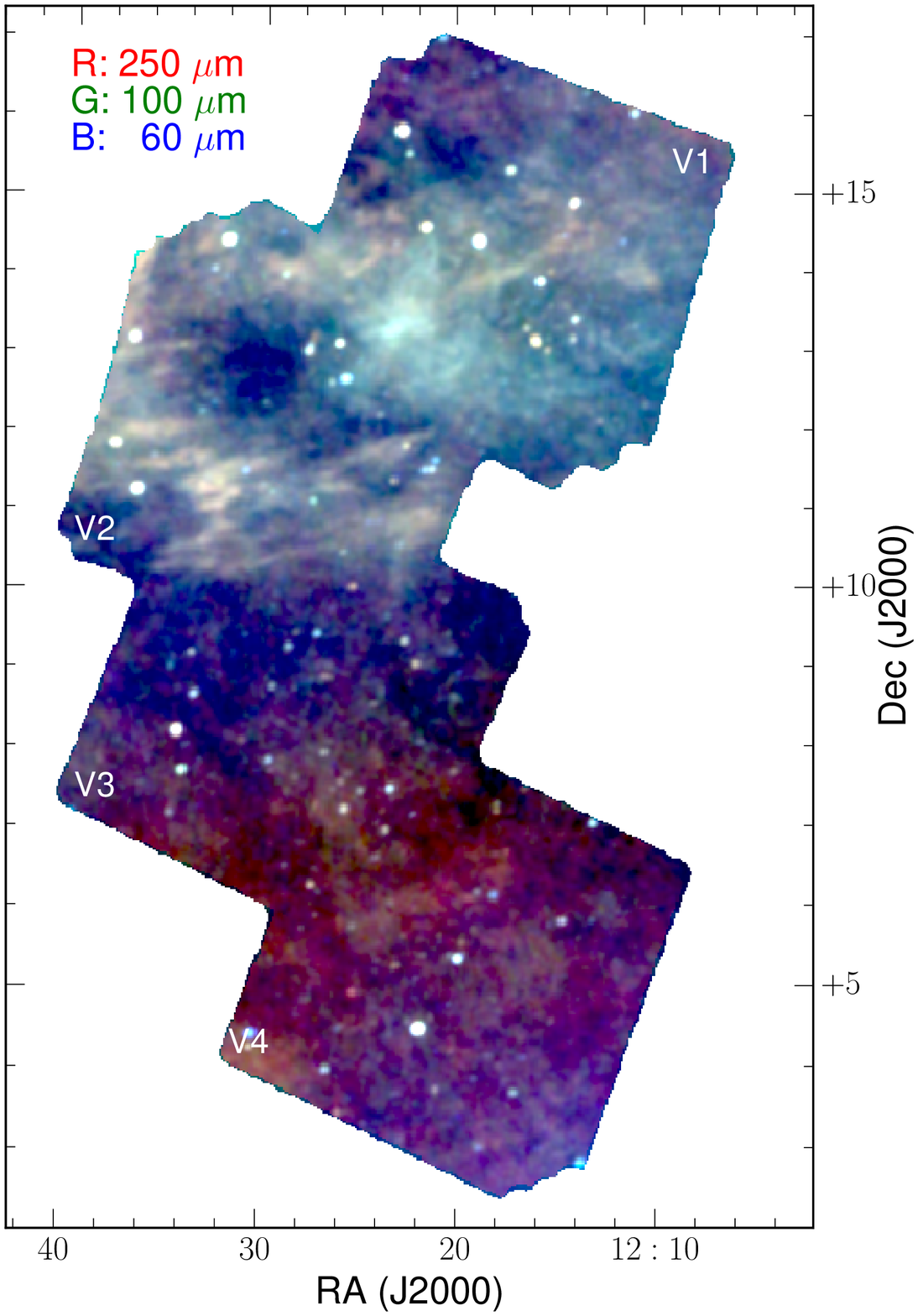}
\caption{The HeViCS field: ALFALFA \ion{H}{i} column density of the LVC component 
with $-20 < v_\mathrm{LSR}$/ km s$^{-1}  < 100$ (left panel); and of the IVC 
component with $-100 < v_\mathrm{LSR}$/ km s$^{-1} < -20$ (central panel).
{ Both components are shown with the same color scale.}
The RGB composite in the right panel shows dust emission at 60~$\mu$m (blue),
100~$\mu$m (green) and 250~$\mu$m (red). All images are convolved to a resolution 
of FWHM=4$\farcm$8. For this figure, the full $250\mu$m map was 
obtained by removing the mutual offsets between the HeViCS fields;
the quantitative analysis is instead done separately on 
each of the four fields, as motivated in Sect.~\ref{sec:spire}. In the
left panel, the red and green arrows indicate the direction of the Galactic 
and ecliptic North, respectively.
}
\label{fig:maps}
\end{figure*} 

\subsection{IRAS  \& {\em Planck}}

We used 60 and 100~$\mu$m maps from the IRAS satellite \citep{NeugebauerApJL1984}.
Images in MJy sr$^{-1}$ with pixels size 1$\farcm$5 were obtained using the dataset and procedures from 
the {\em Improved Reprocessing of the IRAS Survey} \citep[IRIS;][]{MivilleDeschenesApJS2005}.
The PSF of the IRIS maps is estimated to be  Gaussian with FWHM= 4$\farcm$0 and 4$\farcm$3 
at 60 and 100~$\mu$m, respectively. At both wavelengths, calibration uncertainties 
amount to about 15\% and results from the uncertainties in the gain and color correction 
with respect to data from the poorer-resolution, better-photometric accuracy DIRBE instrument 
on the COBE satellite, which was used to recalibrate the IRAS data.

Maps at 350, 550 and 850~$\mu$m (857, 545 and 353 GHz, respectively) produced
by the High Frequency Instrument (HFI) aboard the {\em Planck} satellite were taken
from the 2013 data release \citep{Planck2013I}. These channels are dominated by foreground dust 
emission; the 850~$\mu$m map allows us to extend the analysis to longer 
wavelengths than those covered by SPIRE, while the maps at 350 and 550~$\mu$m 
provide a consistency check between SPIRE and HFI (and in general, a sanity check with the analysis in \citealt{PlanckEarlyXXIV} and \citealt{Planck2013XI}). We have used the data 
corrected for Zodiacal emission, and regridded them from the original HEALpix format
to a RA/Dec frame centered on the HeViCS field.

While maps at 350 and 550~$\mu$m are already provided in MJy sr$^{-1}$, the 
map at 850~$\mu$m was converted into those units by multiplying it by the 
band-average unit conversion coefficient as described in \citet{Planck2013IX}.
The map at 850~$\mu$m was found to be significantly contaminated by the 
Cosmic Microwave Background (CMB). A few {\em Planck}-based models of the CMB are 
available, obtained with different methods \citep{Planck2013XII}. We
first used the CMB model produced with the leading method, SMICA, which
was found successful in subtracting the CMB contamination from maps of 
M31 \citep{PlanckA&A2015}. However, in the SMICA map the CMB fluctuations 
are not defined in an aperture of 1$^\circ$ around M87; large residuals
are left in the area after the subtraction, which affect the derivation 
of the 850~$\mu$m emissivity in field V2. Thus, we preferred the
NILC model, which shows CMB features also close to M87. For the CMB
subtraction (and emissivity analysis) in the other HeViCS fields, the 
use of the NILC map produces equivalent results to SMICA.

The PSF of {\em Planck} images can be described by Gaussians with
FWHM very similar to those of IRAS: 4$\farcm$3, 4$\farcm$7 and 4$\farcm$8 at 350, 550 and 850~$\mu$m,
respectively \citep{Planck2013I}. Calibration uncertainties are of 1\% for the CMB
calibrated 850~$\mu$m maps, and 10\% for the two shorter wavelengths maps, whose 
calibration is based on planets: in the latter case the calibration uncertainty is 
not completely uncorrelated to that of SPIRE, both sharing a 5\% uncertainty in the
model flux of one of the HFI's calibrators, the planet Neptune \citep{Planck2013VIII}.

\subsection{Convolved and regridded maps}

For each of the HeViCS fields (named V1, V2, V3 and V4 from North to South) we have regridded all
data sets to a common J2000 RA \& Dec frame { in gnomonic projection}, with pixels size 1$\farcm$5. All 
maps were convolved
to the poorer resolution of {\em Planck} 850~$\mu$m (4$\farcm$8) using simple Gaussian kernels. 
Excluding border regions with incomplete PSF sampling, the total area analysed amounts
to 76~deg$^2$ { (64~deg$^2$ when excluding the masked sources)}. In Fig.~\ref{fig:maps} we show the gas column density of the LVC component (left panel), 
the IVC component (central panel) and an RGB representation of dust emission (right 
panel) of the full HeViCS area.

The left panel of Fig.~\ref{fig:maps} shows that LVC gas is present in all the HeViCS tiles. 
It is more diffuse in the southernmost fields V3 and V4, with an average column density 
$N_\ion{H}{i}=1.2\times 10^{20}$ cm$^{-2}$. Most of the gas is in the ring-like structure 
of field V2, coincidentally centered on the center of the Virgo cluster, where the peak 
column density is $3.3\times 10^{20}$ cm$^{-2}$. A long filament is the main structure in 
the center of field V1\footnote{The bright point source present in both LVC and IVC maps 
for field V1 is the Virgo spiral NGC4192 (M98), with Heliocentric radial velocity -139 km s$^{-1}$
and velocity width $W_{50}=461$ km s$^{-1}$ \citep{GiovanelliAJ2007}. The artifacts due to M87 are
also visible as a lower (higher) than average column density feature in the LVC (IVC) map 
(see Sect.~\ref{sec:alfalfa}).}.

Most of the IVC component (central panel) is instead concentrated in field V1 and V2, with 
a peak column density $N_\ion{H}{i}=3.5\times 10^{20}$ cm$^{-2}$ in the overlap of 
the two tiles. This region corresponds to a column density enhancement at the
border of a loose cloud known as {\em clump S1}, part of a larger Galactic structure 
named {\em Intermediate-Velocity (IV) Spur} \citep{KuntzApJ1996}. The analysis of interstellar 
absorption lines
towards the direction of the IV-Spur has shown that this gas resides at large distance
from the Galactic plane, with a distance bracket between 0.3 and 2.1 kpc 
\citep{KuntzApJ1996,WakkerApJS2001}. Because of this structure, the average column density
of the IVC component in field V1 is comparable to that of LVC ($N_\ion{H}{i}=1.1\times 10^{20}$ 
vs $1.4\times 10^{20}$cm$^{-2}$, respectively).  Fields V3 and V4 have instead a tenuous
IVC component, with an average column density $N_\ion{H}{i}=3.2\times 10^{19}$ cm$^{-2}$.
Also, in these fields IVC and LVC show a larger overlap in velocity space, and
our simple $v_\mathrm{LSR}$ cut cannot provide a clear separation of the two components.

The RGB map in the right panel of Fig.~\ref{fig:maps} shows the overall correlation between
dust and atomic gas: in particular, 
{ 
the ring-like feature of field V2 is also visible in dust 
emission (as it is in UV scattered light; \citealt{BoissierA&A2015}). Also,
}
dust emission associated to IVC gas appears to be bluer
than that coming from LVC gas. The correlation is evident also 
in fields V3 and V4, where dust emission is tenuous.

\section{Analysis}
\label{sec:ana}

Our analysis follows the approach of \citet{PlanckEarlyXXIV}. The surface brightness map of 
dust emission at frequency $\nu$, $I_\nu$, can be written as:
\begin{equation}
I_\nu = \epsilon^\mathrm{LVC}_\nu \times N^\mathrm{LVC}_{\ion{H}{i}} + \epsilon^\mathrm{IVC}_\nu \times N^\mathrm{IVC}_{\ion{H}{i}} +
        O_\nu + E_\nu \times (b-b_0)+R_\nu,
\label{eq:fitted}
\end{equation}
where $N^\mathrm{LVC}_{\ion{H}{i}}$ and $N^\mathrm{IVC}_{\ion{H}{i}}$ are the $\ion{H}{i}$ 
column density maps for the low and
intermediate velocity cloud components, respectively, $\epsilon^\mathrm{LVC}_\nu$ and
$\epsilon^\mathrm{IVC}_\nu$ are the emissivities of dust associated to them. As (most of)
the dust emission maps have a relative calibration, 
{ 
the absolute surface brightness levels are unknown: thus, 
}
the term $O_\nu$ is used
to describe an offset (or {\em zero-level}), which will also include
any large-scale contribution not directly correlated to the atomic gas. Since SPIRE
images are not corrected for zodiacal light (and other data might have residual
contribution from this foreground), we also included a term $E_\nu$ to describe
a gradient in ecliptic latitude
$b$ ($b_0$ being the central latitude in each map). 
{
The analysis consists in fitting the first four terms in right-hand side of Eq.~\ref{eq:fitted}
to the observed surface brightness $I_\nu$.  The last term in Eq.~\ref{eq:fitted}, $R_\nu$, 
is the map of the residuals, i.e.\ the difference between the observed and modelled surface brightness.
}

{

We neglected in Eq.~\ref{eq:fitted} the contribution of ionized gas,
for which there is no available data at a comparable angular scale.
Available, larger-scale, H$\alpha$ surveys \citep{HaffnerApJS2003,FinkbeinerApJS2003}
only show diffuse low-level features in the HeViCS footprint,
barely correlated to our $\ion{H}{i}$ and dust emission maps.
Converting the H$\alpha$ intensities to ionised gas column densities 
\citep[as done in][]{LagacheA&A2000}, we estimated that, within each field, 
the local column density can differ from the field average by up to 
$N_{\ion{H}{ii}}\approx 3\times 10^{19}$ cm$^{-2}$. While dust emission
associated with the average values would be hidden into $O_\nu$, we predicted
localised emission to be of the order of the standard deviation of the 
residuals to our fit \citep[we assumed similar dust properties in the neutral 
and ionised gas;][]{LagacheA&A1999,LagacheA&A2000}; it is probably smaller, 
since a significant fraction of high-latitude H$\alpha$ emission could be due
to scattered light rather than to ionized gas \citep{WittApJ2010}. Thus,
dust associated to ionised gas could contribute to the scatter in the
$I_\nu$ / $N_{\mathrm{H}}$ correlation, but it is not likely to produce
any large deviation from the assumed model.

}

The model parameters were estimated using the IDL function {\tt regress} which
performs a multiple linear regression fit. Fits were done separately for each of the HeViCS 
fields. We thus obtain, for each band, four values for each of the
parameters, $\epsilon^\mathrm{LVC}$, $\epsilon^\mathrm{IVC}$, $O_\nu$, $E_\nu$.
We excluded from the fits the regions with significant FIR emission from
extragalactic sources, by masking circular apertures of diameter 
4$\times$FWHM around the brightest (F > 1.1 Jy) IRAS point sources \citep{IrasProc1988}.
The flux limit was selected in order to keep the masked areas to a reasonable size; 
we will discuss later in Sect.~\ref{sec:resi} the effect of other unmasked sources 
on the residuals.
{
Another mask was devised to exclude large deviations from the assumed model: after a first 
fit was performed to the high signal-to-noise 250$\mu$m data, we inspected the probability 
distribution function (PDF) of $R_\nu$ and, by fitting a Gaussian to it, estimated the 
standard deviation of the residuals $\sigma_R$.  This second mask thus excludes all pixels
where residuals at 250$\mu$m are $> 3\sigma_R$. 
}

{
Since the fluctuations in residual maps do not behave as white noise}, mainly because of the contribution of
the Cosmic Infrared Background (CIB), the parameter uncertainties 
derived from the fit are underestimated (see \citealt{PlanckEarlyXXIV} for a discussion). 
Thus, we estimated the uncertainties using a Monte Carlo technique.
Simulated images were created starting from a mock map of the residuals, with the
same power spectrum of $R_\nu$, but with random phase, to which the other terms
in Eq.~\ref{eq:fitted} were added, using the observed hydrogen column density maps and
the parameters estimated by the fit. The linear regression was repeated on the 
simulated images, after adding random representations of the instrumental noise 
to $I_\nu$ and the two column density maps. The procedure was repeated a thousand
times, and a standard deviation retrieved from the resulting PDFs of each of the
four parameters. The instrumental noise used in this process was estimated on the 
convolved and regridded FIR/submm images by measuring the standard deviation of 
difference maps: for IRAS, we used the difference between observations
of the first and second surveys (HCON-1 and HCON-2, the area was not observed
during HCON-3), which were retrieved using the IRIS software; for SPIRE, we used 
the difference between two maps created combining four scans each; for {\em Planck}, we 
took the two {\em ringhalf} maps provided within the 2013 data release. The
error is rather uniform across the HeViCS footprint and amounts to 0.03, 0.05, 
0.05, 0.025, 0.015, 0.02, 0.03 and 0.035 MJy sr$^{-1}$ for the IRAS 60 and 100~$\mu$m,
SPIRE 250, 350 and 500~$\mu$m, {\em Planck} 350, 550 and 850~$\mu$m bands, respectively.

The instrumental uncertainty on the $\ion{H}{i}$ column density was directly estimated from the
convolved ALFALFA maps, measuring the standard deviation of the featureless 
regions (fields V3 and V4) in the IVC map range. It is 1.0$\times 10^{18}$ cm$^{-2}$
for both the IVC and LVC channels. For the typical emissivities found here, the 
contribution of instrumental noise in \ion{H}{i} column density maps is 
smaller than that from FIR/submm images (reaching a maximum of 0.005
MJy sr$^{-1}$ in the 350~$\mu$m bands).

In all the dust emission maps we used, monochromatic flux densities were derived
from wide band measurements assuming a flat spectrum, $F_\nu \propto 1/\nu$.
We estimated color corrections assuming the average SED reconstructed using 
Eq.~\ref{eq:fitted} as the {\em true} spectrum. They were found to be small, within less than 2\% for 
most bands, with the exception of the {\em Planck} 550~$\mu$m and 850~$\mu$m bands. In the following,
we only correct the results obtained on these bands, by multiplying them by a factor
0.9. A similar result can be obtained using the {\em Planck} color correction software 
\citep{Planck2013IX} and assuming the {\em Planck} MBB fit to high
latitude Galactic emission \citep[$T$=20.3 K, $\beta$=1.59;][]{Planck2013XI}.

\section{Results}
\label{sec:res}

The parameters derived by fitting Eq.~\ref{eq:fitted} to each image of dust emission,
for each field, are given in Table~\ref{table:fit}, together with the errors estimated
from { the Monte Carlo analysis}. The dust emission associated with LVC gas is estimated at a significant level
in all fields and bands ($\epsilon^\mathrm{LVC}_\nu> 3\sigma$). In contrast,
dust associated with IVC gas is significantly detected only in fields V1 and V2, while
in fields V3 and V4 $\epsilon^\mathrm{IVC}_\nu < 3\sigma$ in most cases (the only exceptions
are the marginal detections in the two IRAS bands for V4). This was not unexpected,
given the low column density of IVC gas in the two southernmost fields already shown in
Fig.~\ref{fig:maps}. 

\begin{sidewaystable*}
\caption{Dust emissivities $\epsilon^\mathrm{LVC}_\nu$ and $\epsilon^\mathrm{IVC}_\nu$ for the two \ion{H}{i}
components, offsets $O_\nu$ and ecliptic latitude gradients $E_\nu$, derived for each field and
band considered in this work. Errors are derived with the
Monte Carlo technique described in Sect.~\ref{sec:ana}. Estimates below $3\sigma$ are in italics.
$\sigma_R$ is the standard deviation of the residuals $R_\nu$. $\sigma_C$ is derived from $\sigma_R$ 
after removing the contribution to the uncertainties due to instrumental noise in dust emission
and \ion{H}{i} maps. For $\epsilon^\mathrm{LVC}_\nu$ and $\epsilon^\mathrm{IVC}_\nu$, the average over
the fields is also reported (excluding low S/N values), together with its standard deviation.
 }
\label{table:fit}
\centering 
\small
\begin{tabular}{l c | c  c | c c c | c c c}
\hline                                   % inserts single horizontal line
 &       & \multicolumn{2}{c |}{IRAS} & \multicolumn{3}{c |}{{\it Herschel} - SPIRE} &
\multicolumn{3}{c}{{\em Planck} - HFI} \\
 & field & 60~$\mu$m & 100~$\mu$m & 250~$\mu$m & 350~$\mu$m & 500~$\mu$m & 350~$\mu$m & 550~$\mu$m & 850~$\mu$m \\
\hline                                   % inserts single horizontal line
       &V1 & {             0.163$\pm$     0.012} & {             0.774$\pm$     0.041} & {             0.747$\pm$     0.045} & {             0.341$\pm$     0.034} & {             0.134$\pm$     0.024} & {             0.399$\pm$     0.028} & {             0.101$\pm$     0.012} & {            0.0284$\pm$    0.0016} \\
$\epsilon^{LVC}_\nu$ &V2 & {            0.2051$\pm$    0.0089} & {             0.773$\pm$     0.029} & {             0.844$\pm$     0.037} & {             0.414$\pm$     0.021} & {             0.163$\pm$     0.011} & {             0.462$\pm$     0.023} & {            0.1288$\pm$    0.0066} & {           0.03434$\pm$   0.00089} \\
{\scriptsize $10^{-20}$ MJy sr$^{-1}$ cm$^2$} &V3 & {             0.135$\pm$     0.014} & {             0.680$\pm$     0.034} & {             0.859$\pm$     0.039} & {             0.415$\pm$     0.025} & {             0.175$\pm$     0.011} & {             0.484$\pm$     0.026} & {            0.1495$\pm$    0.0098} & {            0.0409$\pm$    0.0015} \\
       &V4 & {             0.123$\pm$     0.025} & {             0.508$\pm$     0.047} & {             0.698$\pm$     0.061} & {             0.412$\pm$     0.037} & {             0.212$\pm$     0.022} & {             0.486$\pm$     0.048} & {             0.154$\pm$     0.019} & {            0.0464$\pm$    0.0026} \\
 \cline{2-          10}
       &avg. & {             0.157$\pm$     0.036} & {              0.68$\pm$      0.12} & {             0.787$\pm$     0.077} & {             0.395$\pm$     0.036} & {             0.171$\pm$     0.032} & {             0.458$\pm$     0.041} & {             0.133$\pm$     0.024} & {            0.0375$\pm$    0.0078} \\
\hline
       &V1 & {             0.245$\pm$     0.011} & {             0.706$\pm$     0.029} & {             0.447$\pm$     0.027} & {             0.204$\pm$     0.022} & {             0.059$\pm$     0.014} & {             0.209$\pm$     0.020} & {            0.0500$\pm$    0.0079} & {            0.0139$\pm$    0.0012} \\
$\epsilon^{IVC}_\nu$ &V2 & {             0.228$\pm$     0.010} & {             0.696$\pm$     0.027} & {             0.370$\pm$     0.031} & {             0.168$\pm$     0.020} & {            0.0586$\pm$    0.0097} & {             0.189$\pm$     0.022} & {            0.0479$\pm$    0.0074} & {            0.0123$\pm$    0.0012} \\
{\scriptsize $10^{-20}$ MJy sr$^{-1}$ cm$^2$} &V3 & {\em          0.038$\pm$     0.071} & {\em           0.37$\pm$      0.19} & {\em           0.14$\pm$      0.19} & {\em           0.06$\pm$      0.13} & {\em          0.038$\pm$     0.057} & {\em          -0.03$\pm$      0.12} & {\em         -0.023$\pm$     0.049} & {\em        -0.0002$\pm$    0.0073} \\
       &V4 & {             0.317$\pm$     0.066} & {              0.51$\pm$      0.13} & {\em           0.24$\pm$      0.23} & {\em           0.26$\pm$      0.13} & {\em          0.013$\pm$     0.073} & {\em           0.13$\pm$      0.17} & {\em          0.049$\pm$     0.059} & {\em         0.0145$\pm$    0.0093} \\
 \cline{2-          10}
       &avg. & {             0.263$\pm$     0.047} & {              0.64$\pm$      0.11} & {             0.408$\pm$     0.054} & {             0.186$\pm$     0.026} & {           0.05880$\pm$   0.00023} & {             0.199$\pm$     0.014} & {            0.0489$\pm$    0.0015} & {            0.0131$\pm$    0.0011} \\
\hline
       &V1 & {             0.201$\pm$     0.021} & {             0.989$\pm$     0.063} & {            -1.529$\pm$     0.064} & {            -0.695$\pm$     0.047} & {            -0.248$\pm$     0.031} & {             0.238$\pm$     0.038} & {             0.181$\pm$     0.016} & {            0.0955$\pm$    0.0020} \\
$O_\nu$ &V2 & {             0.185$\pm$     0.018} & {             1.051$\pm$     0.056} & {            -1.677$\pm$     0.072} & {            -0.811$\pm$     0.040} & {            -0.314$\pm$     0.021} & {\em          0.105$\pm$     0.044} & {             0.113$\pm$     0.013} & {            0.0811$\pm$    0.0019} \\
MJy sr$^{-1}$ &V3 & {             0.231$\pm$     0.026} & {             1.220$\pm$     0.054} & {            -1.101$\pm$     0.068} & {            -0.531$\pm$     0.043} & {            -0.226$\pm$     0.017} & {             0.188$\pm$     0.046} & {             0.124$\pm$     0.016} & {            0.0784$\pm$    0.0024} \\
       &V4 & {             0.120$\pm$     0.033} & {             1.303$\pm$     0.071} & {             -0.97$\pm$      0.11} & {            -0.619$\pm$     0.065} & {            -0.286$\pm$     0.035} & {\em          0.146$\pm$     0.079} & {             0.101$\pm$     0.029} & {            0.0667$\pm$    0.0043} \\
\hline
       &V1 & {           -0.0206$\pm$    0.0036} & {            -0.062$\pm$     0.011} & {\em          0.029$\pm$     0.011} & {            0.0303$\pm$    0.0091} & {\em         0.0135$\pm$    0.0060} & {            0.0472$\pm$    0.0068} & {            0.0228$\pm$    0.0029} & {           0.00909$\pm$   0.00033} \\
$E_\nu$ &V2 & {\em        -0.0021$\pm$    0.0039} & {\em        -0.0006$\pm$    0.0096} & {\em         -0.012$\pm$     0.012} & {\em        -0.0116$\pm$    0.0079} & {\em        -0.0042$\pm$    0.0044} & {\em         0.0104$\pm$    0.0091} & {\em         0.0009$\pm$    0.0033} & {\em        0.00051$\pm$   0.00045} \\
MJy sr$^{-1}$ deg$^{-1}$ &V3 & {            0.0769$\pm$    0.0034} & {            0.0475$\pm$    0.0080} & {           -0.0729$\pm$    0.0089} & {           -0.0370$\pm$    0.0055} & {           -0.0158$\pm$    0.0022} & {           -0.0231$\pm$    0.0056} & {           -0.0077$\pm$    0.0019} & {\em       -0.00075$\pm$   0.00027} \\
       &V4 & {           -0.0802$\pm$    0.0042} & {\em        -0.0139$\pm$    0.0056} & {\em         0.0154$\pm$    0.0077} & {\em         0.0113$\pm$    0.0046} & {\em         0.0081$\pm$    0.0031} & {            0.0230$\pm$    0.0062} & {\em         0.0066$\pm$    0.0025} & {           0.00172$\pm$   0.00035} \\
\hline
       &V1 &    0.048 (   0.037 ) &     0.11 (    0.10 ) &     0.14 (    0.13 ) &     0.10 (    0.10 ) &    0.061 (   0.059 ) &     0.10 (   0.099 ) &    0.057 (   0.048 ) &    0.037 (   0.012 ) \\
$\sigma_R (\sigma_C)$ &V2 &    0.049 (   0.039 ) &     0.13 (    0.12 ) &     0.16 (    0.15 ) &    0.096 (   0.093 ) &    0.054 (   0.052 ) &     0.11 (    0.10 ) &    0.057 (   0.048 ) &    0.037 (   0.012 ) \\
MJy sr$^{-1}$ &V3 &    0.047 (   0.036 ) &    0.097 (   0.083 ) &     0.12 (    0.11 ) &    0.087 (   0.083 ) &    0.050 (   0.048 ) &    0.093 (   0.091 ) &    0.057 (   0.048 ) &    0.037 (   0.012 ) \\
       &V4 &    0.045 (   0.033 ) &    0.088 (   0.072 ) &     0.11 (    0.10 ) &    0.083 (   0.079 ) &    0.050 (   0.048 ) &    0.091 (   0.089 ) &    0.054 (   0.045 ) &    0.037 (   0.012 ) \\
\hline

\end{tabular}
\end{sidewaystable*}

\newcommand\dirres{./}
\newcommand\dirresh{./}
\newcommand\sizep{8.8cm}

\begin{figure*}
\center
\includegraphics[width=\sizep]{\dirres/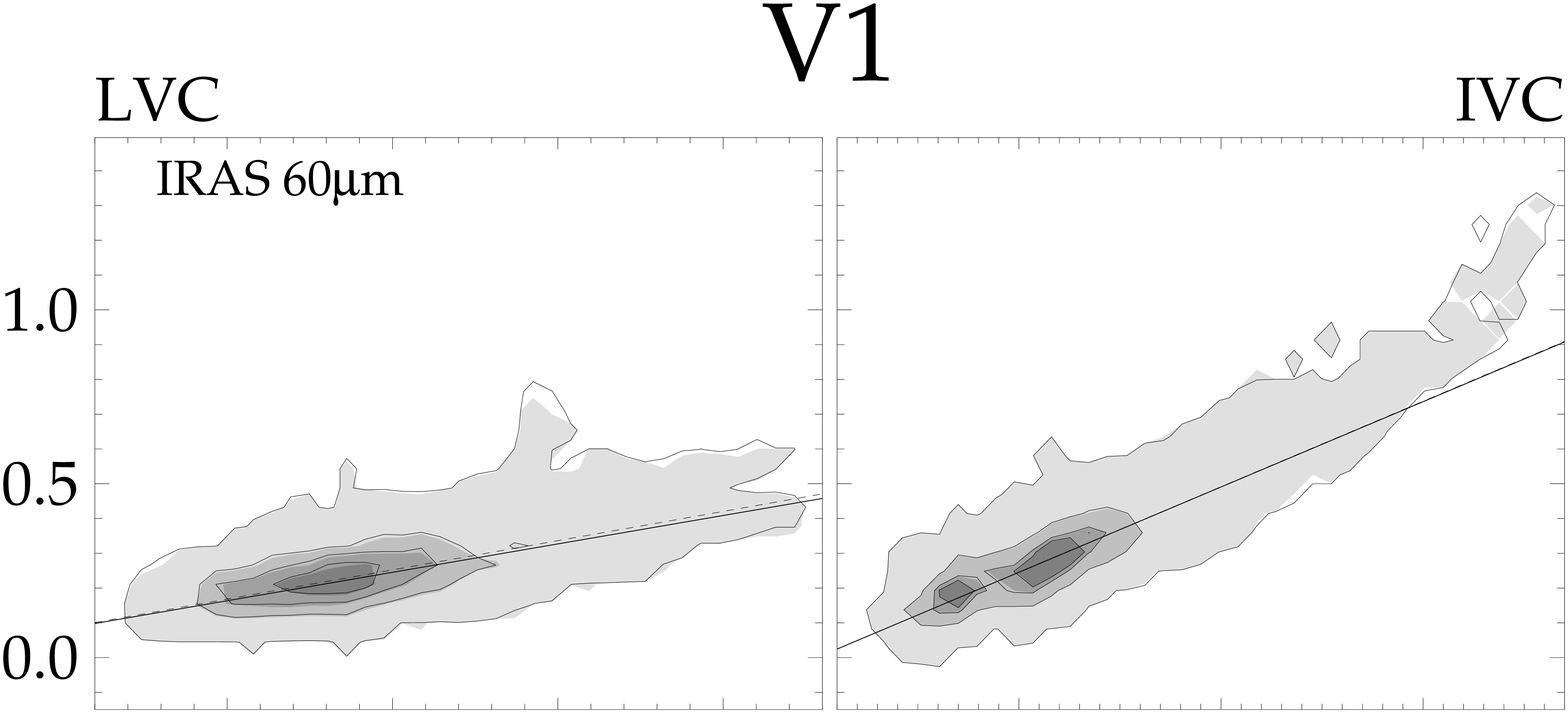}
\includegraphics[width=\sizep]{\dirres/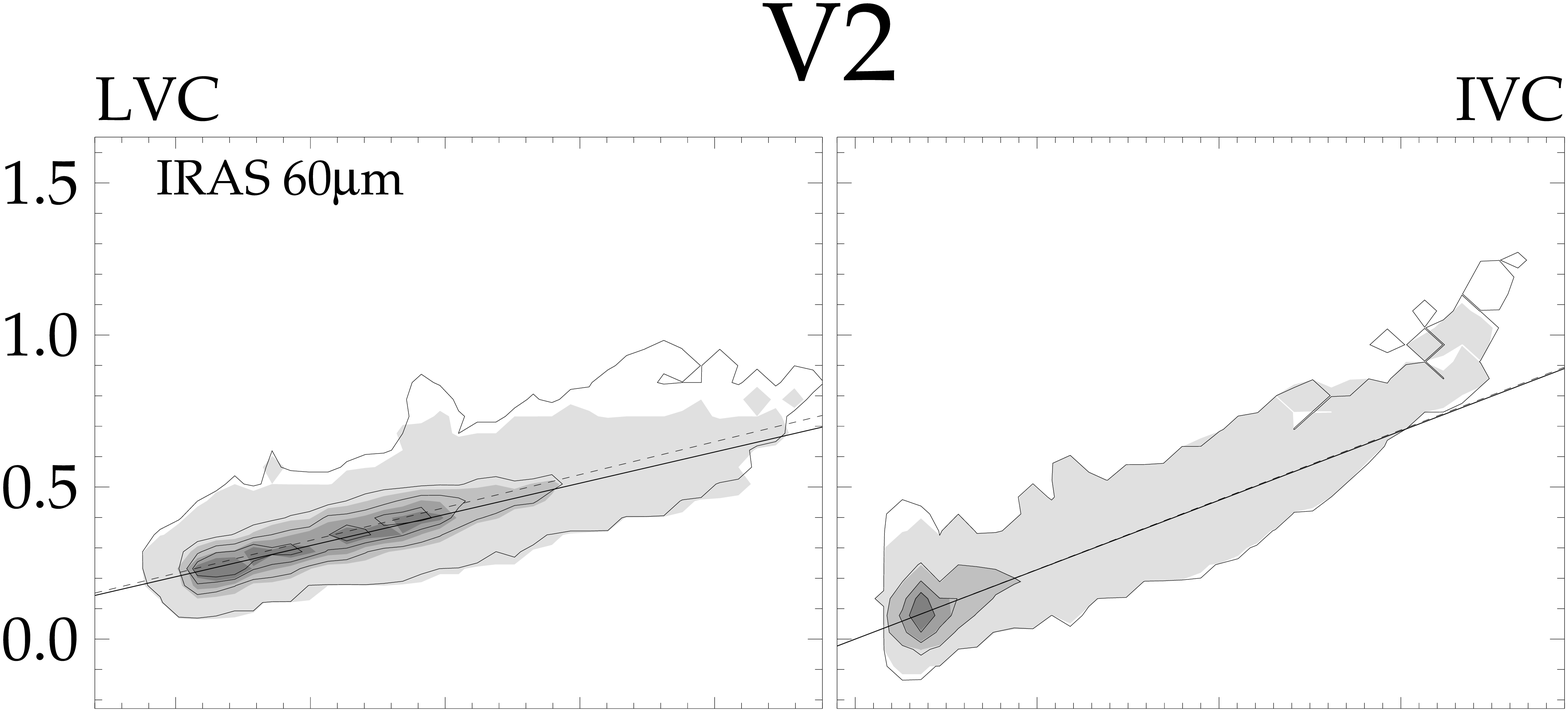}

\includegraphics[width=\sizep]{\dirres/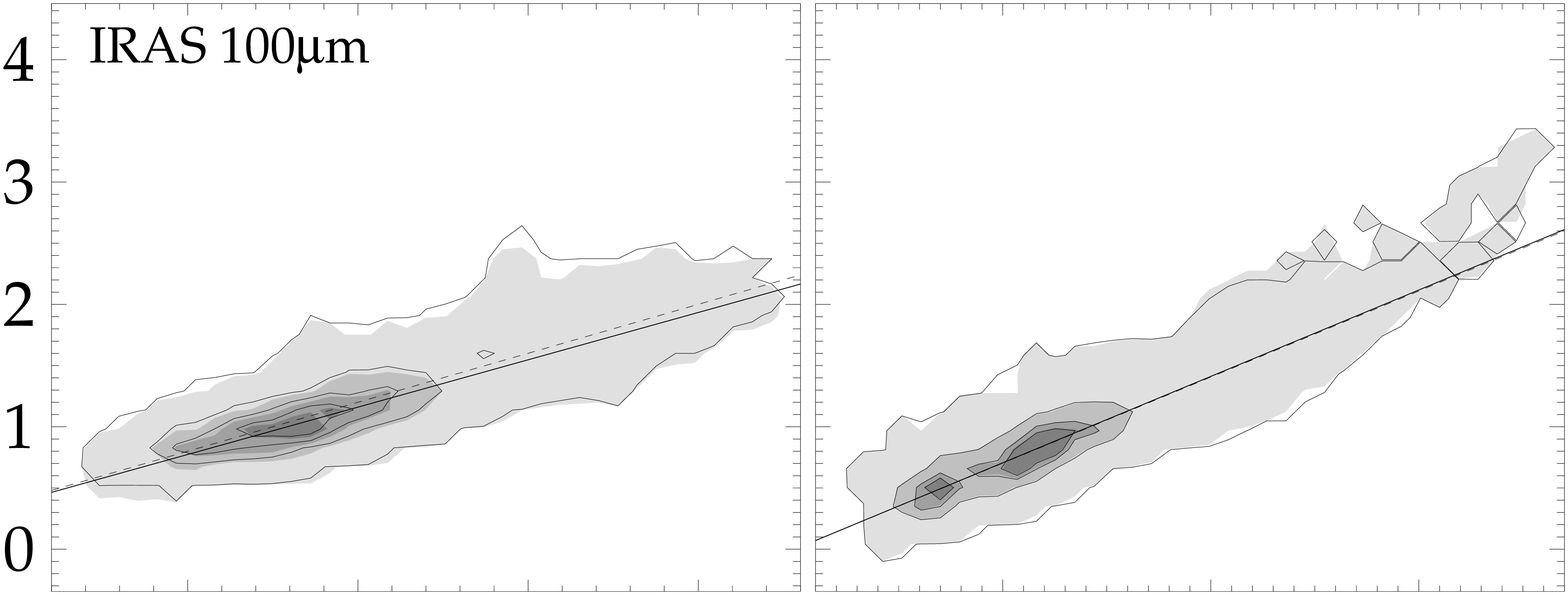}
\includegraphics[width=\sizep]{\dirres/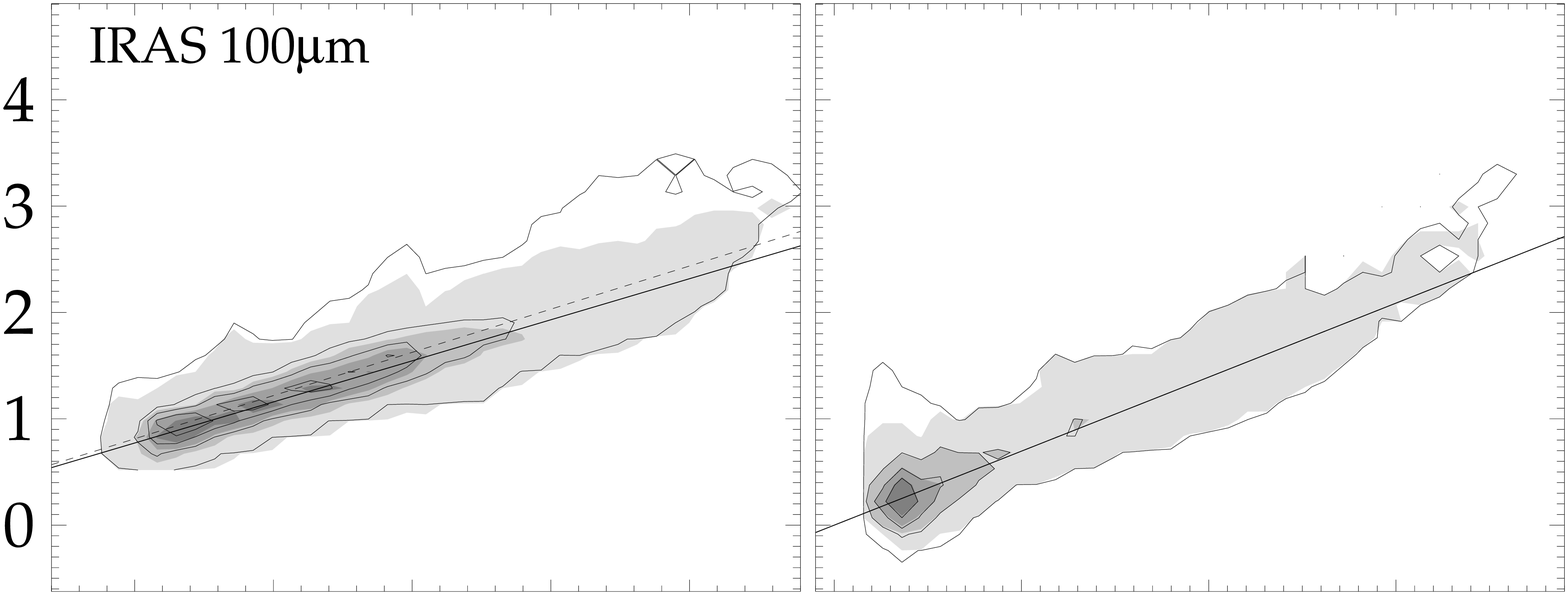}

\includegraphics[width=\sizep]{\dirresh/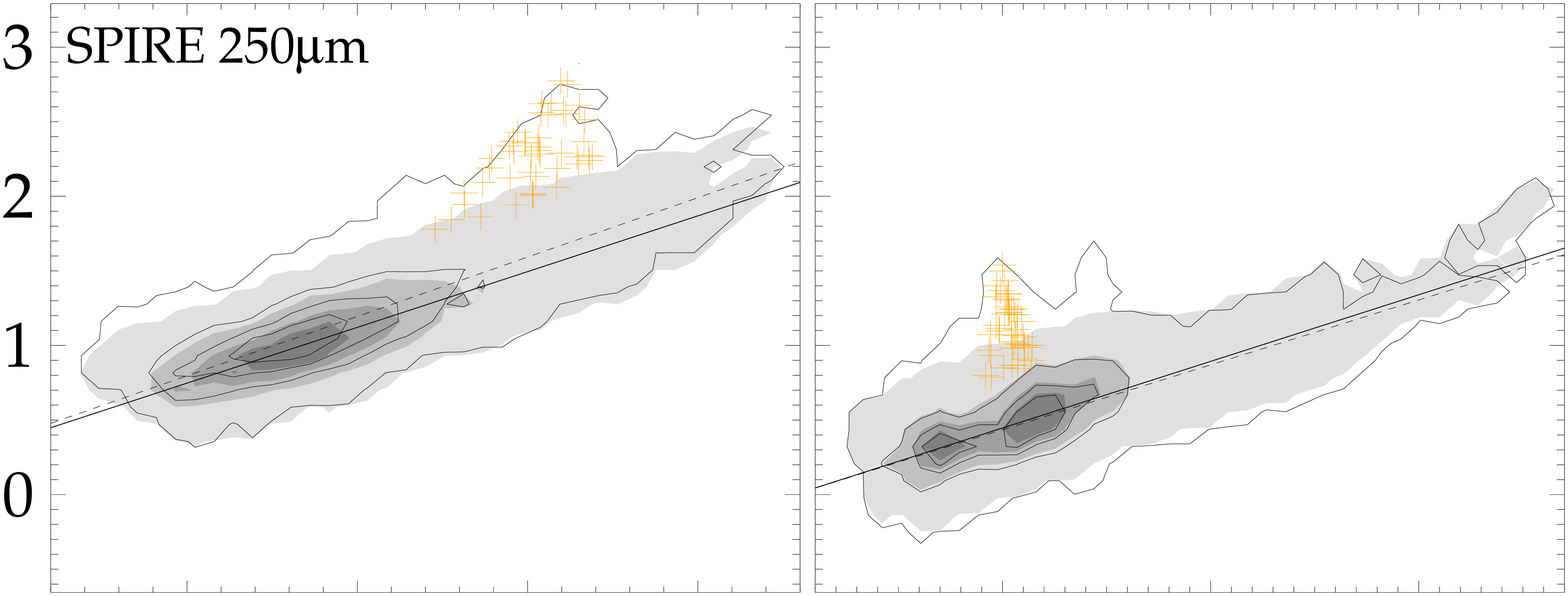}
\includegraphics[width=\sizep]{\dirresh/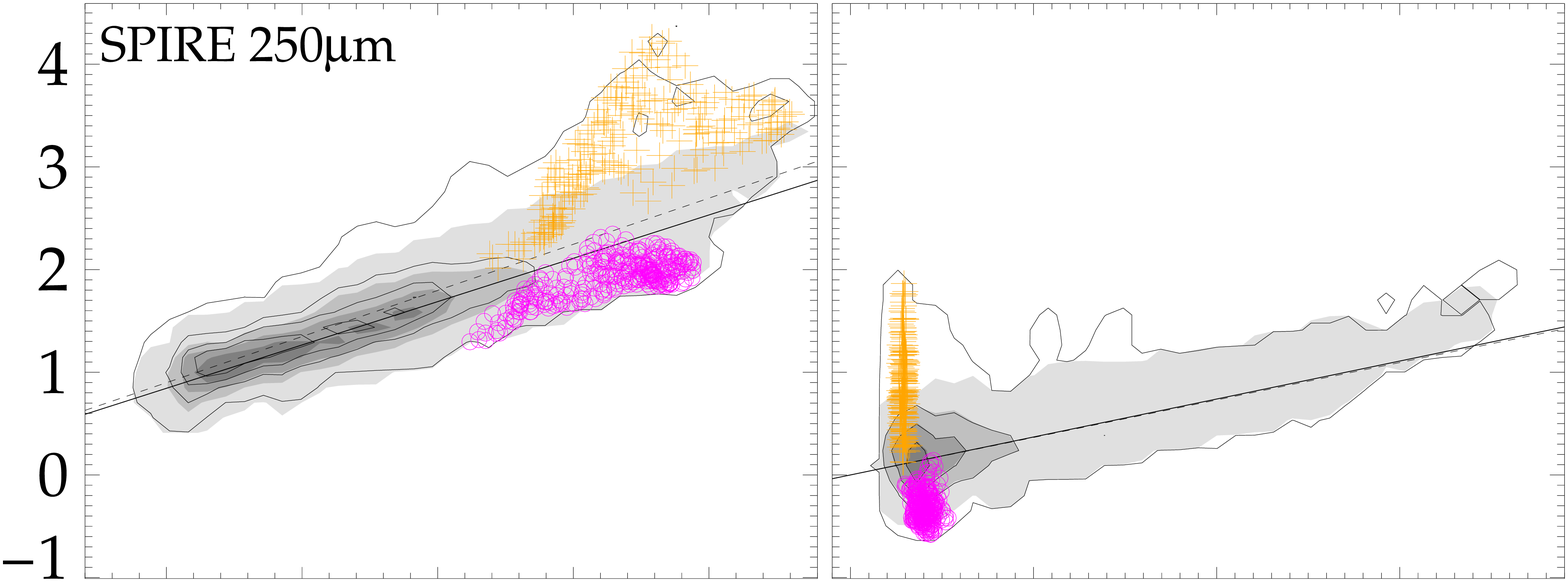}

\includegraphics[width=\sizep]{\dirres/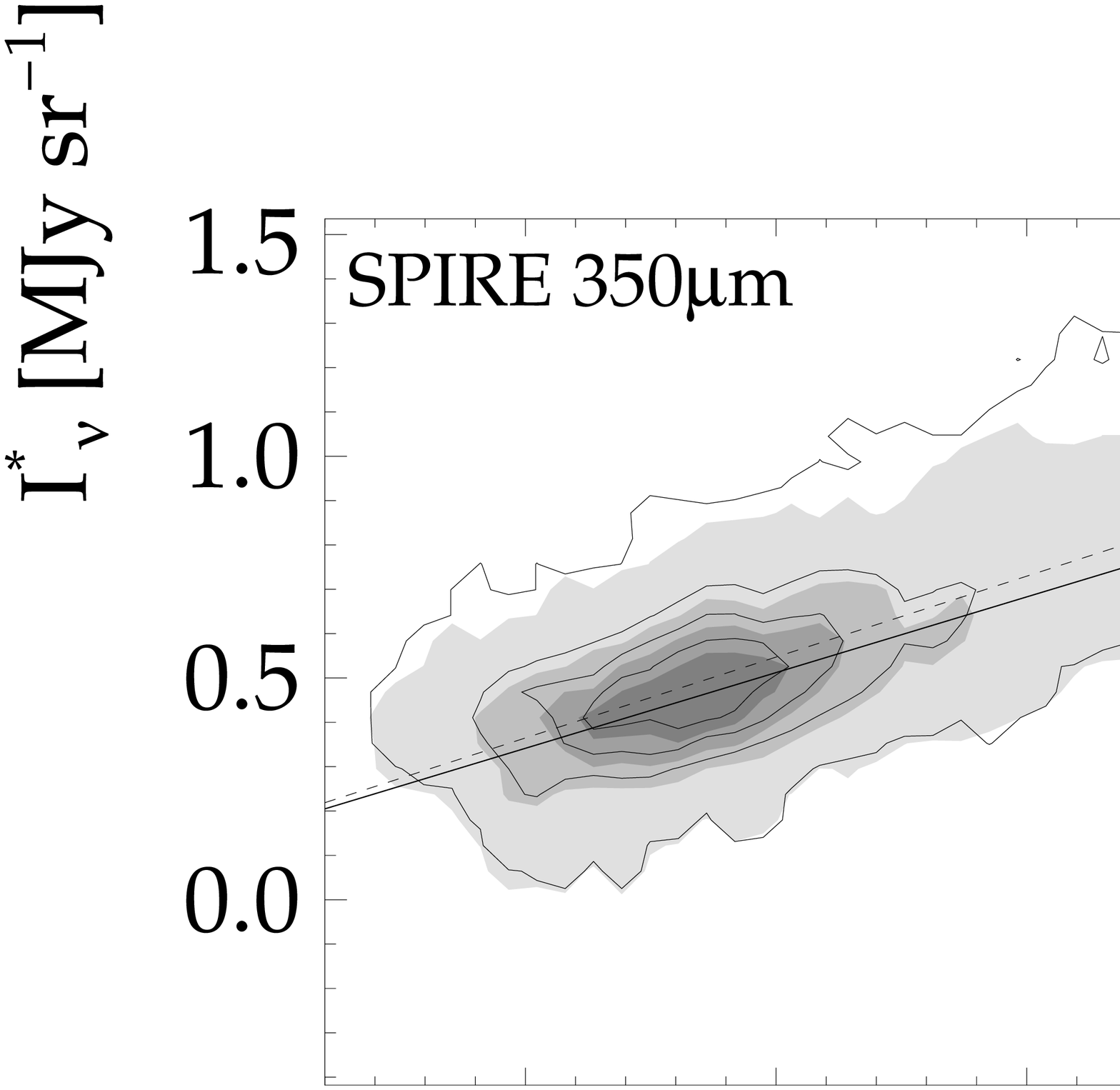}
\includegraphics[width=\sizep]{\dirres/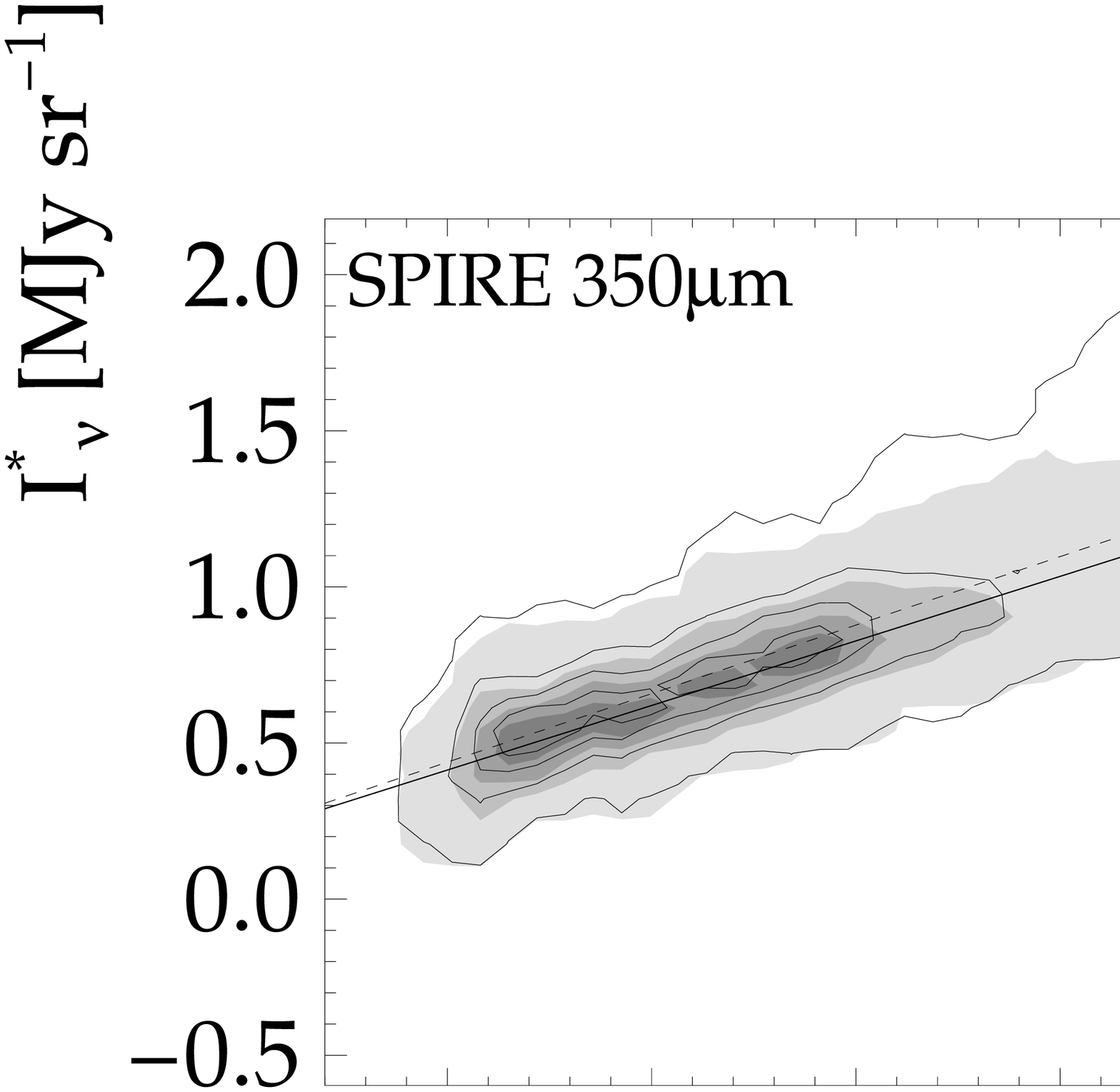}

\includegraphics[width=\sizep]{\dirres/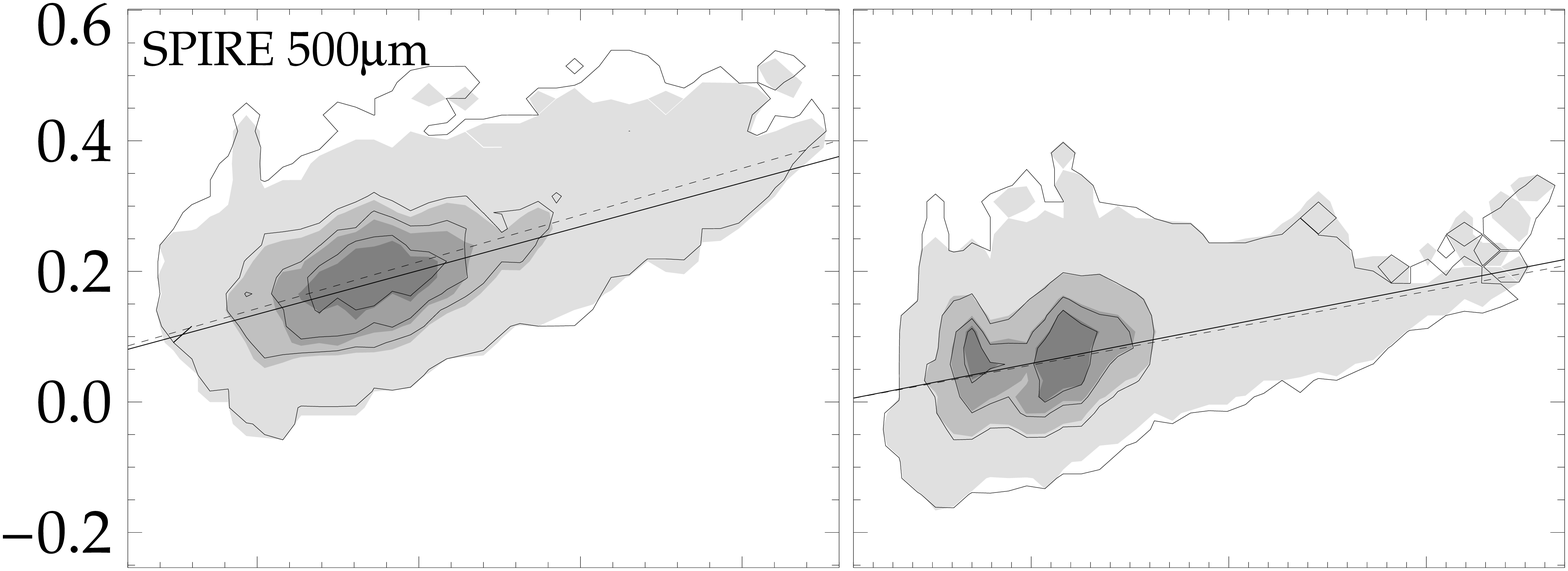}
\includegraphics[width=\sizep]{\dirres/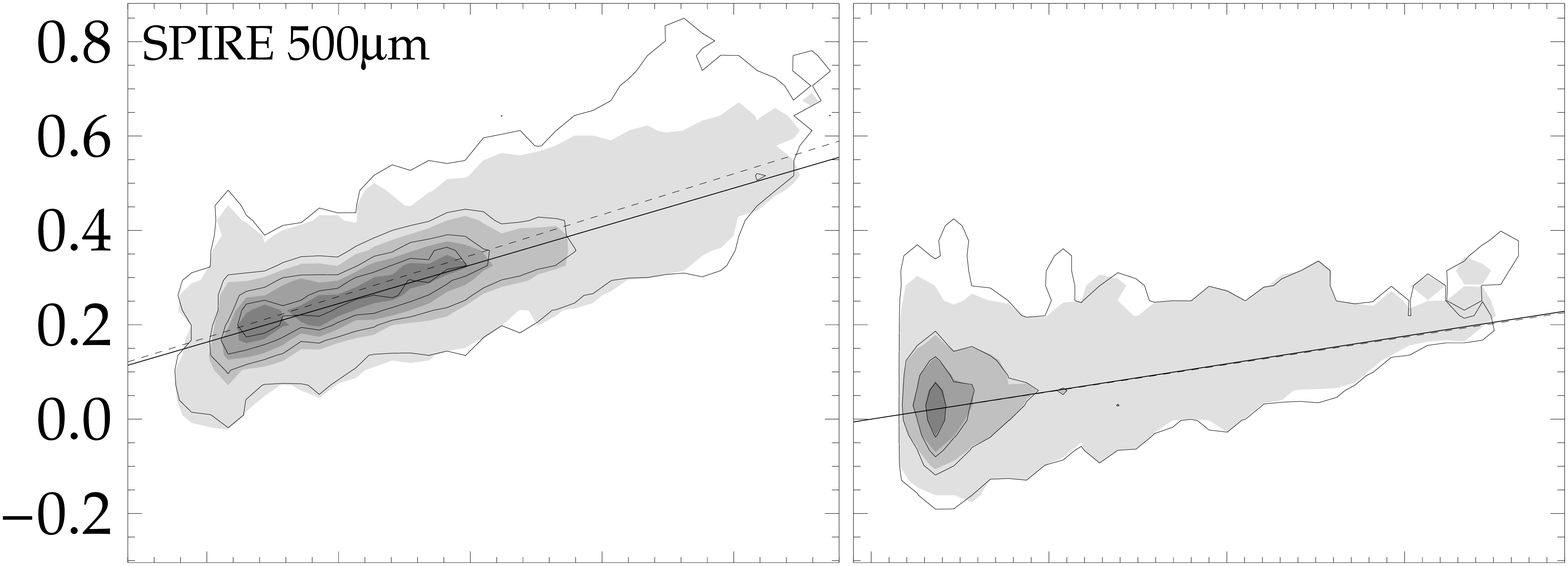}

\includegraphics[width=\sizep]{\dirres/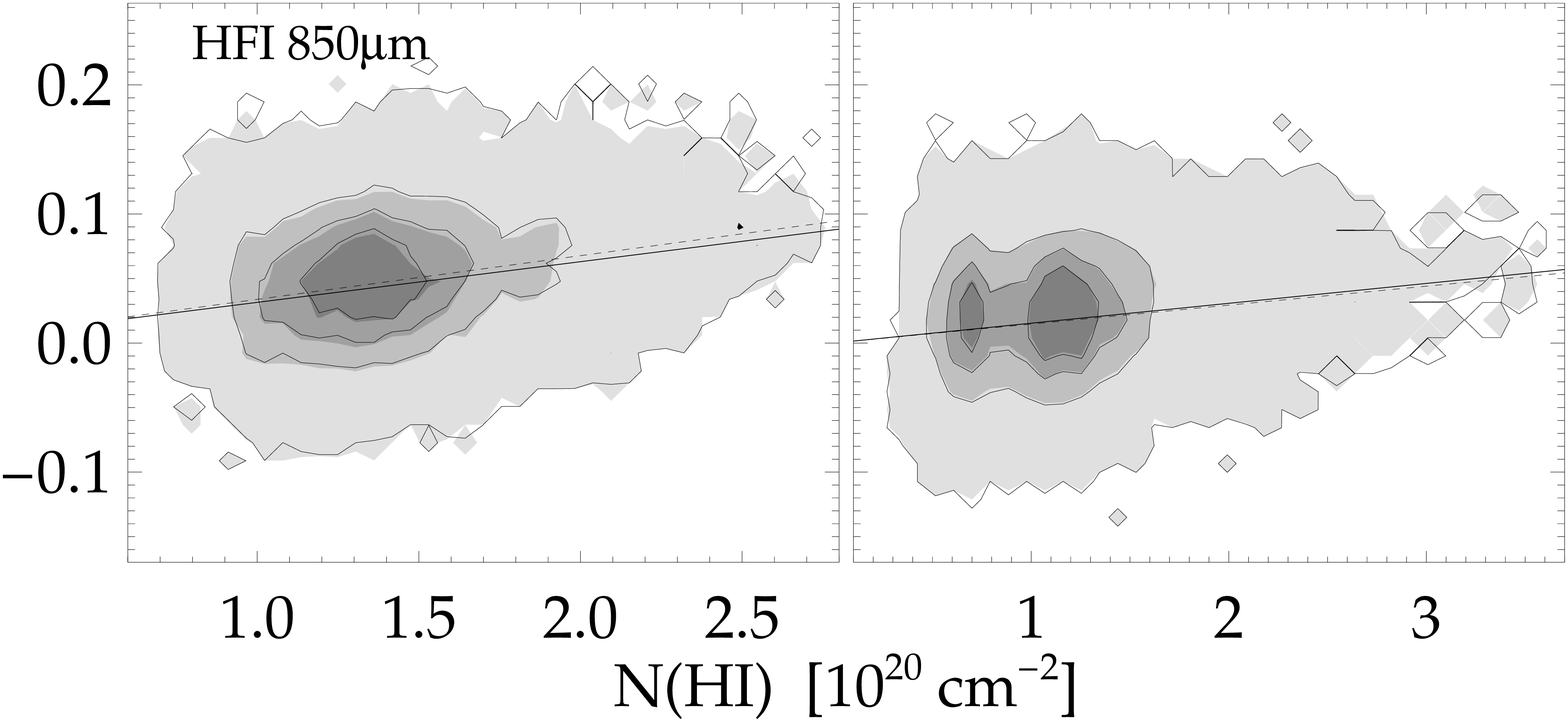}
\includegraphics[width=\sizep]{\dirres/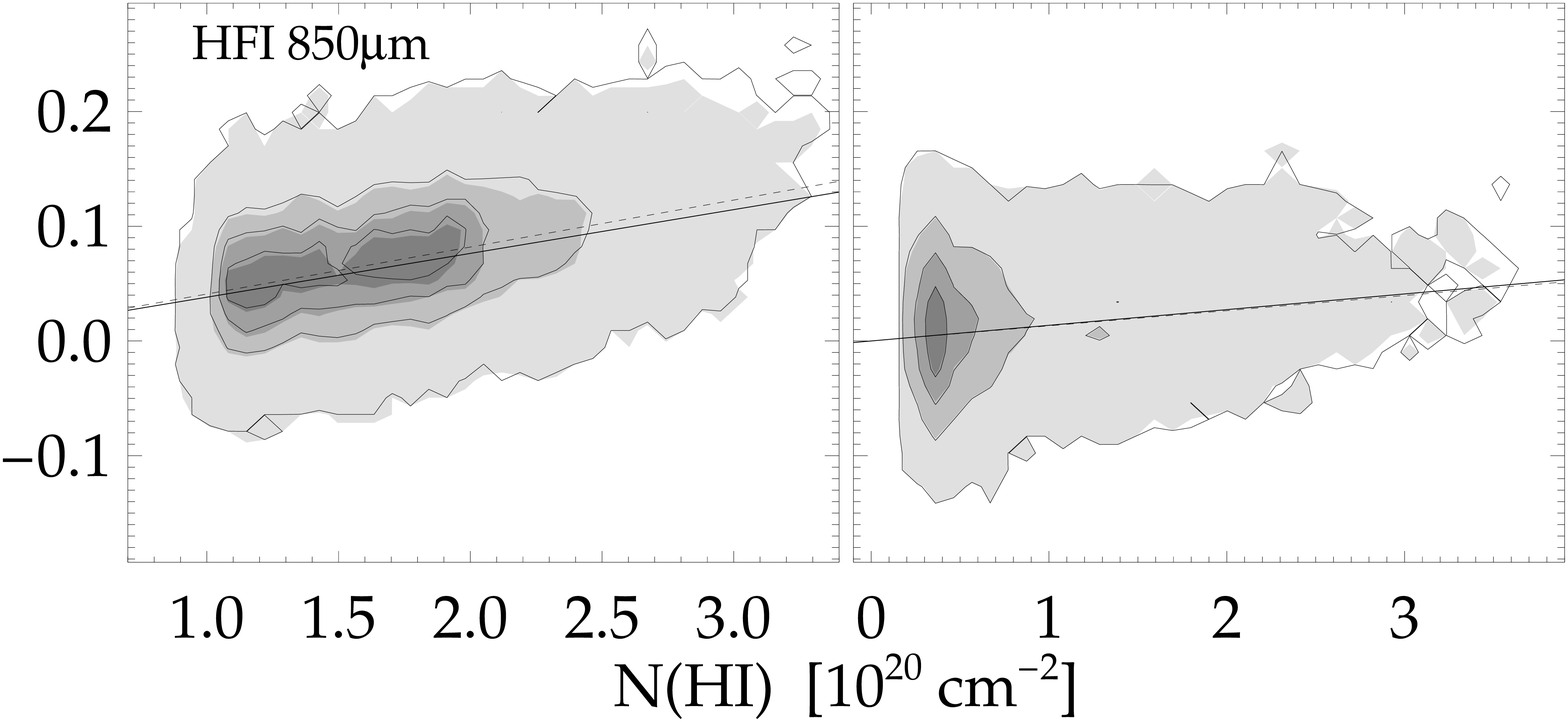}

\caption{Pixel-by-pixel correlation between dust emission and atomic gas column density, for field V1 and V2 
in the IRAS, SPIRE and {\em Planck}-HFI $850\mu$m bands. For the LVC (IVC) gas component, the dust surface brightness
$I_\nu^{*}$ is corrected to exclude the fitted contribution of the IVC (LVC) component; also, the offset and 
ecliptic latitude component is removed. Contours of constant 
pixel density enclosing 99.9, 75, 50 and 25\% of the pixels are plotted. The gray-scale filled contours show 
the same after masking regions deviant from the model by more than $3\sigma_R$ at 250 $\mu$m. Solid and dashed lines
show the fitted correlation before and after the masking, respectively. Small changes in the fitted
$O_\nu$ before and after the masking result in slight displacements of the line and filled contours.
Orange crosses in the $250\mu$m panels refer to pixels within the isophotal ellipses of object 6 (in V1) and
14 (in V2). Magenta circles in the V2 $250\mu$m panels show region 'N' of negative residuals (see Sect.~\ref{sec:resi}
for details).
}
\label{fig:corre1}
\end{figure*} 

\begin{figure*}
\center
\includegraphics[width=\sizep]{\dirres/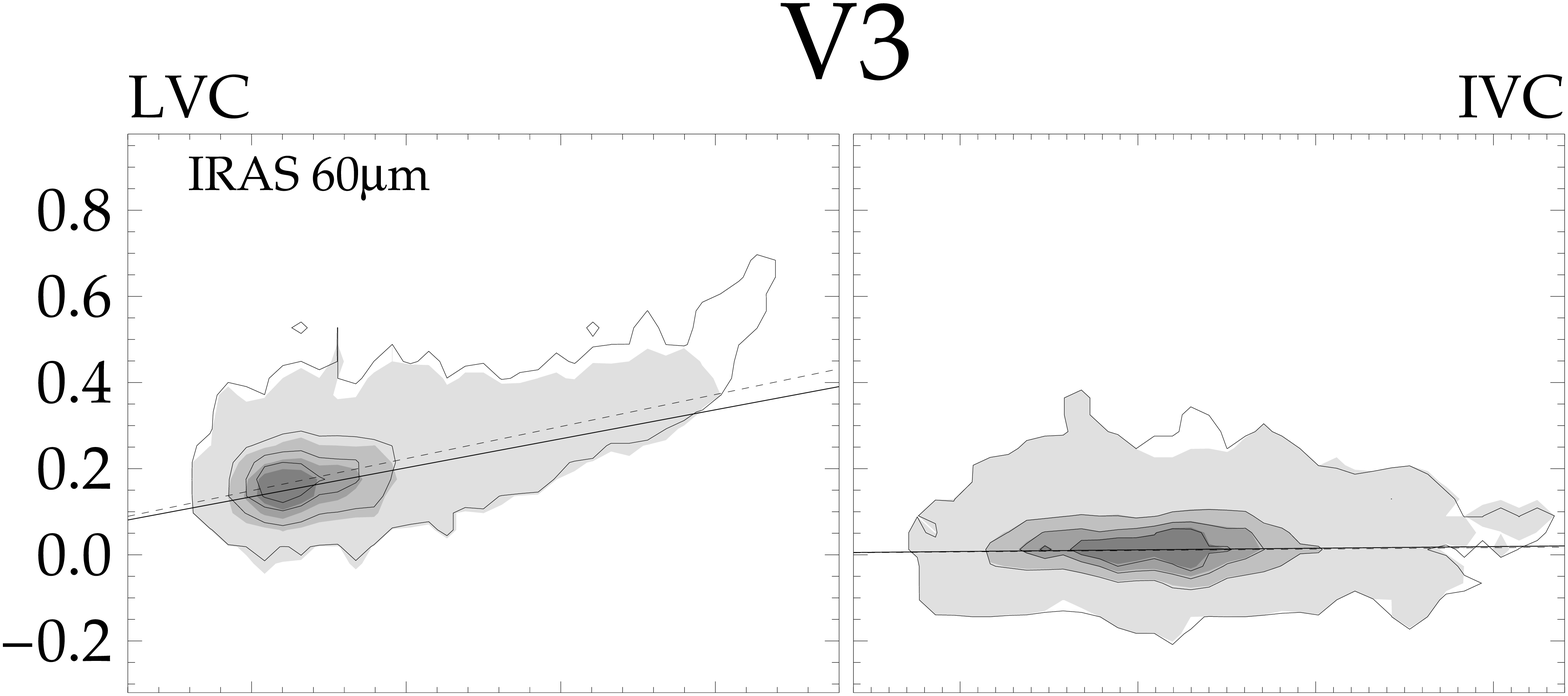}
\includegraphics[width=\sizep]{\dirres/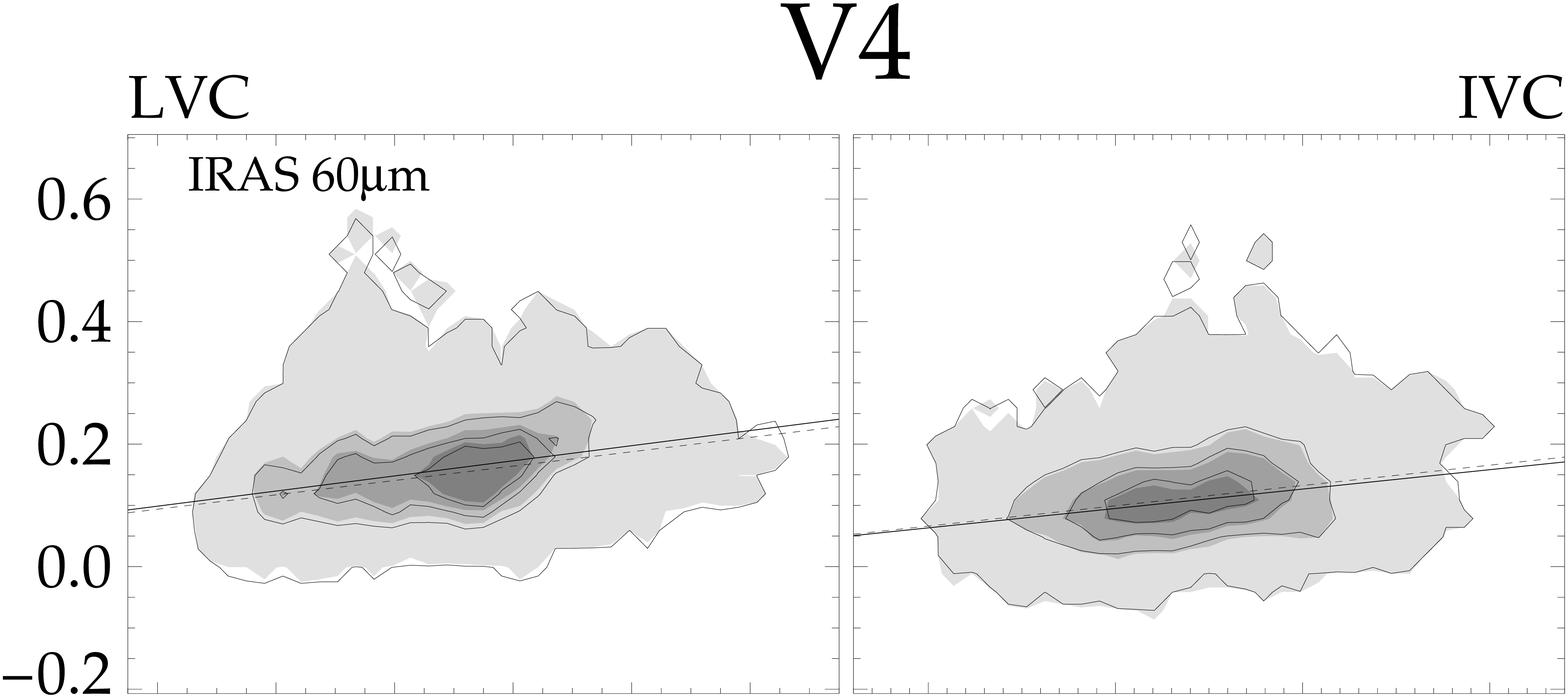}

\includegraphics[width=\sizep]{\dirres/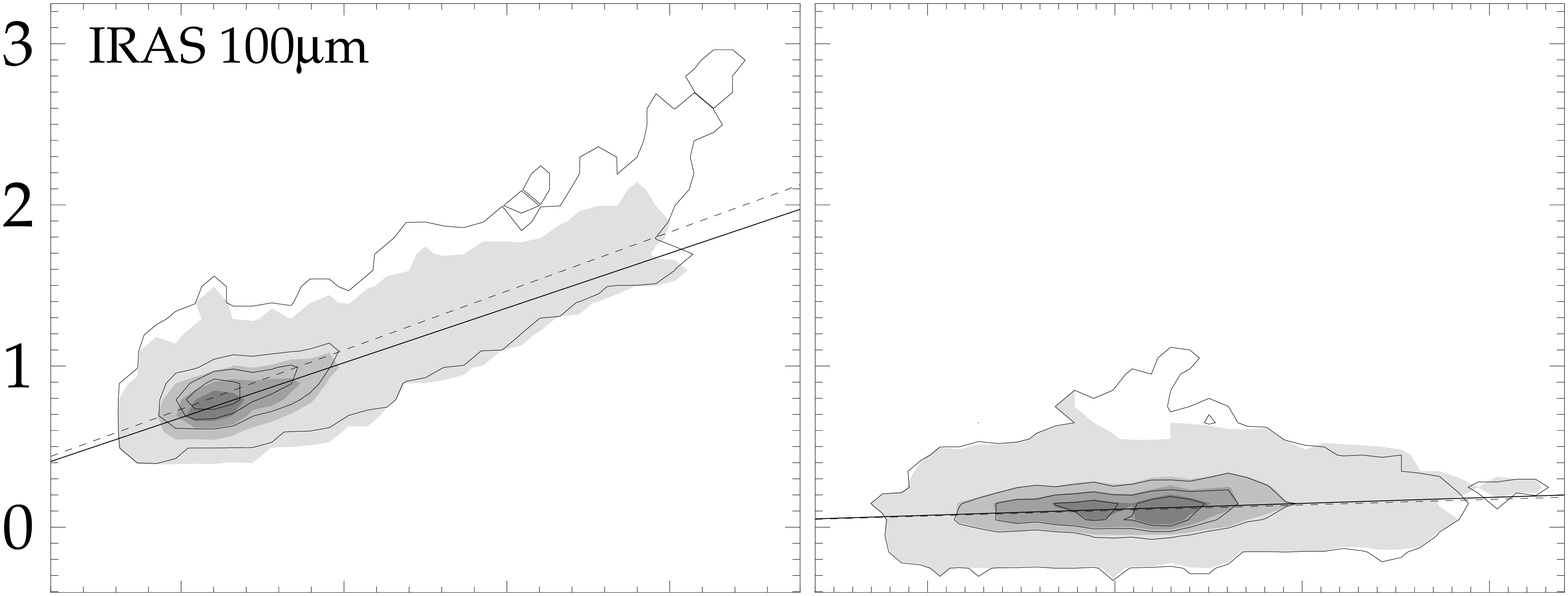}
\includegraphics[width=\sizep]{\dirres/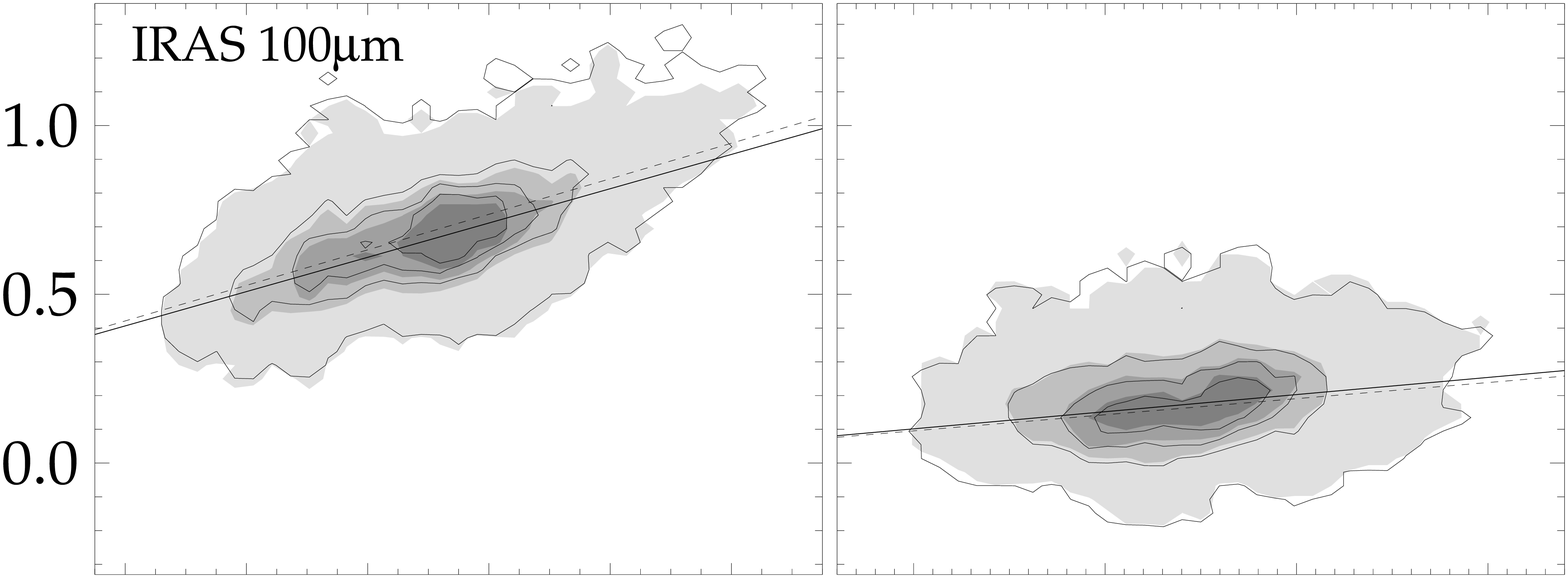}

\includegraphics[width=\sizep]{\dirresh/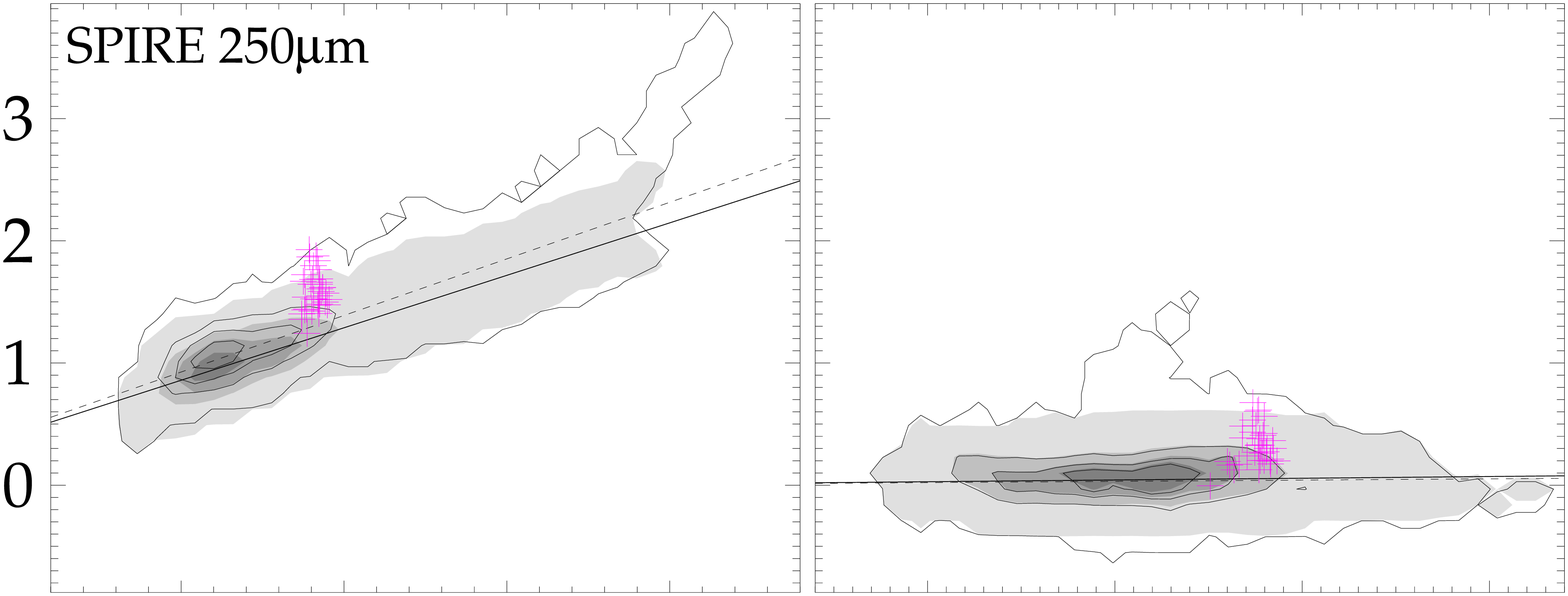}
\includegraphics[width=\sizep]{\dirres/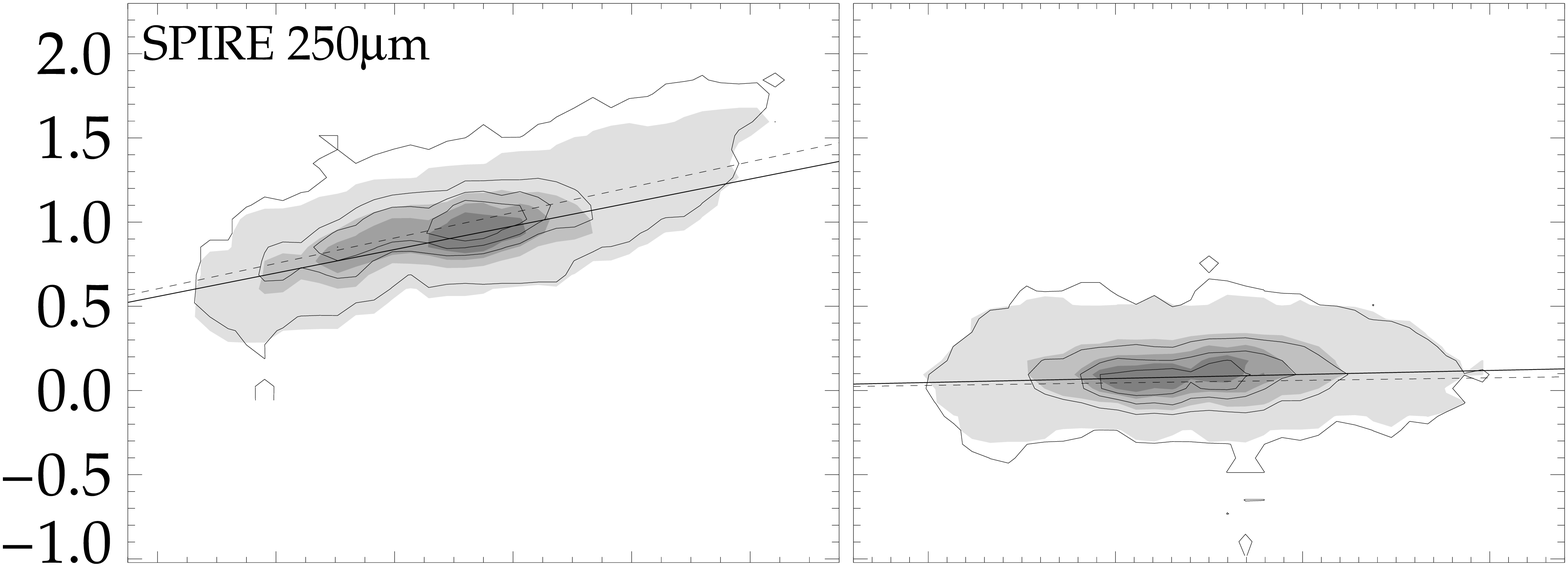}

\includegraphics[width=\sizep]{\dirres/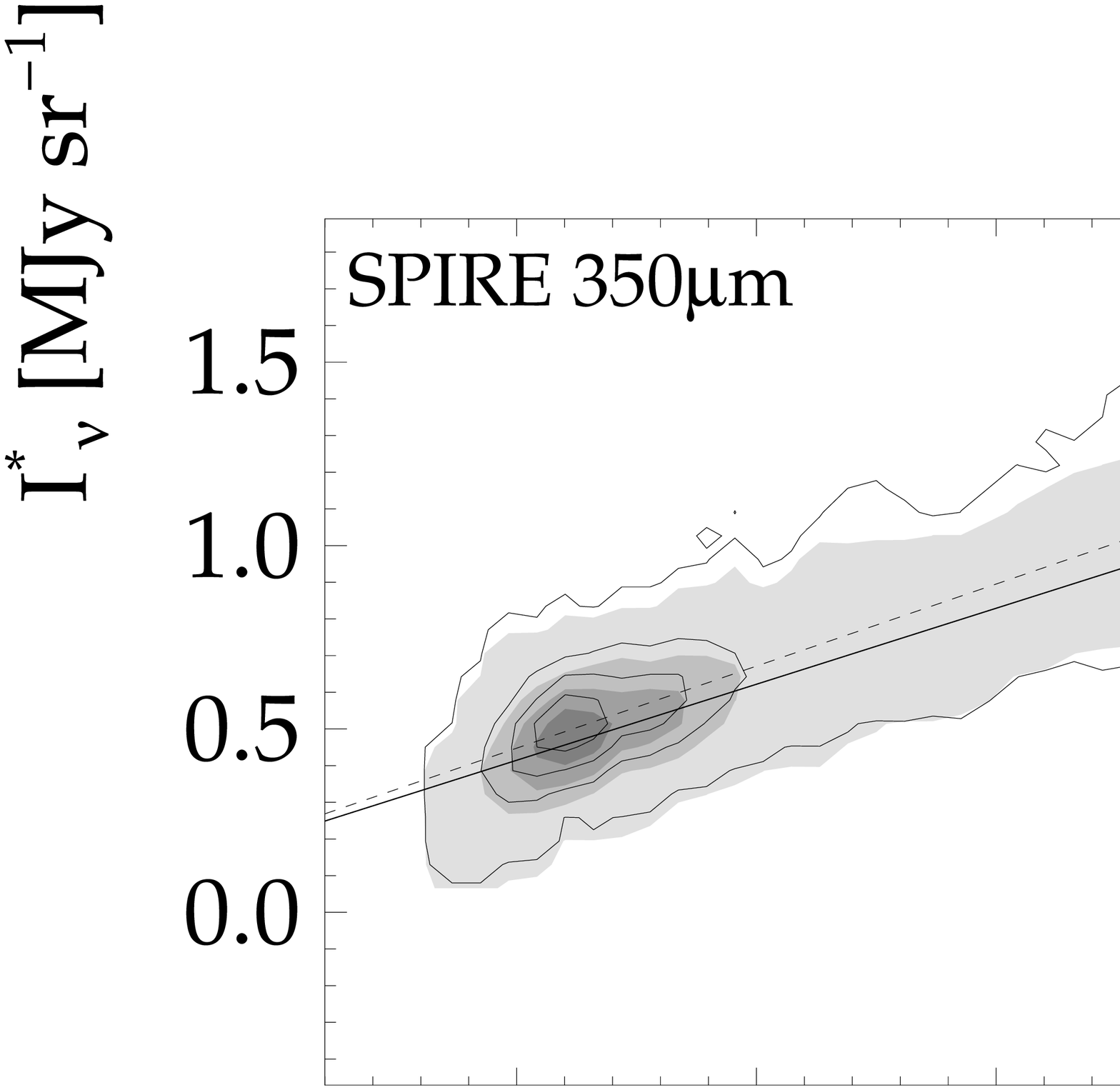}
\includegraphics[width=\sizep]{\dirres/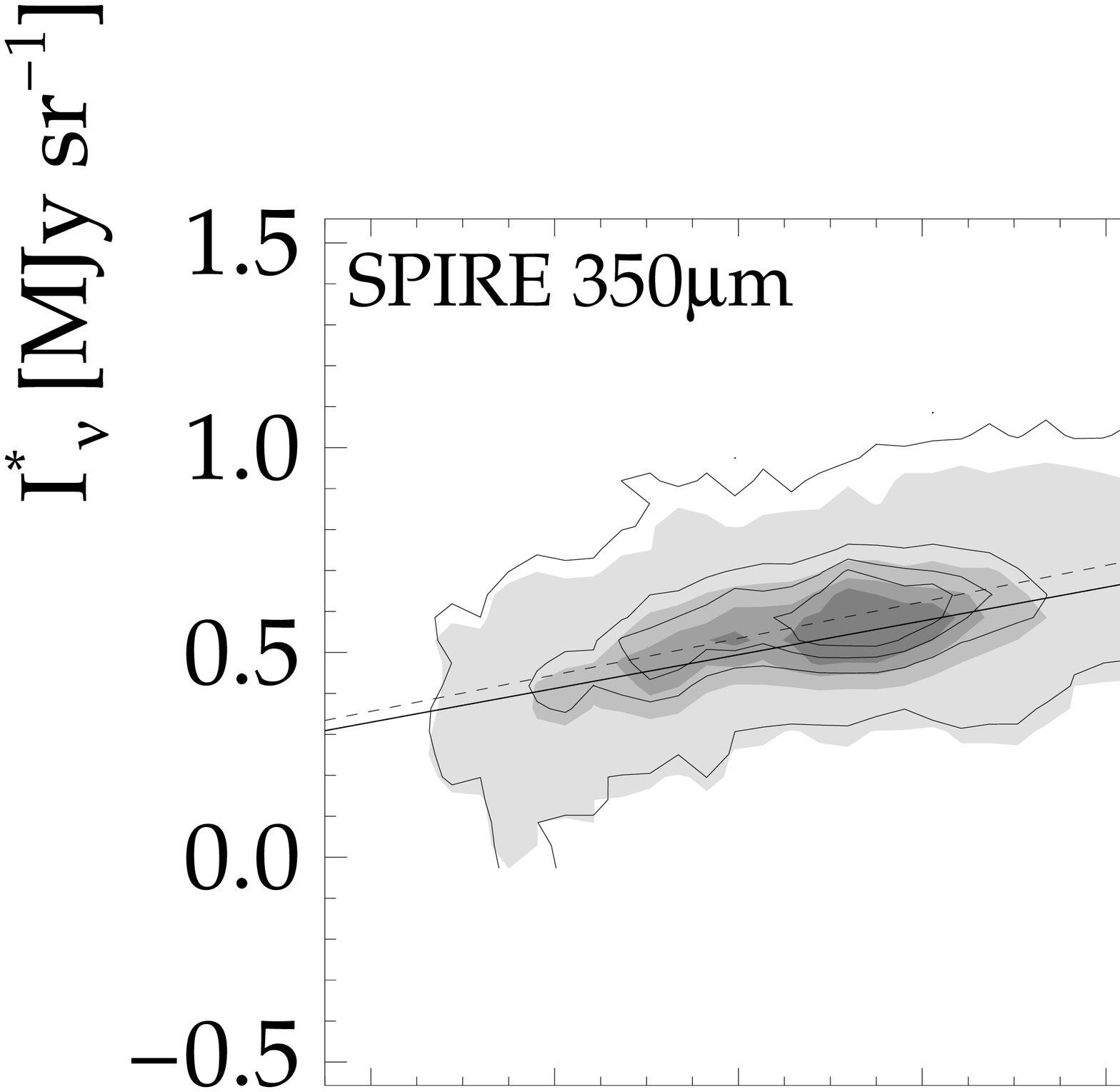}

\includegraphics[width=\sizep]{\dirres/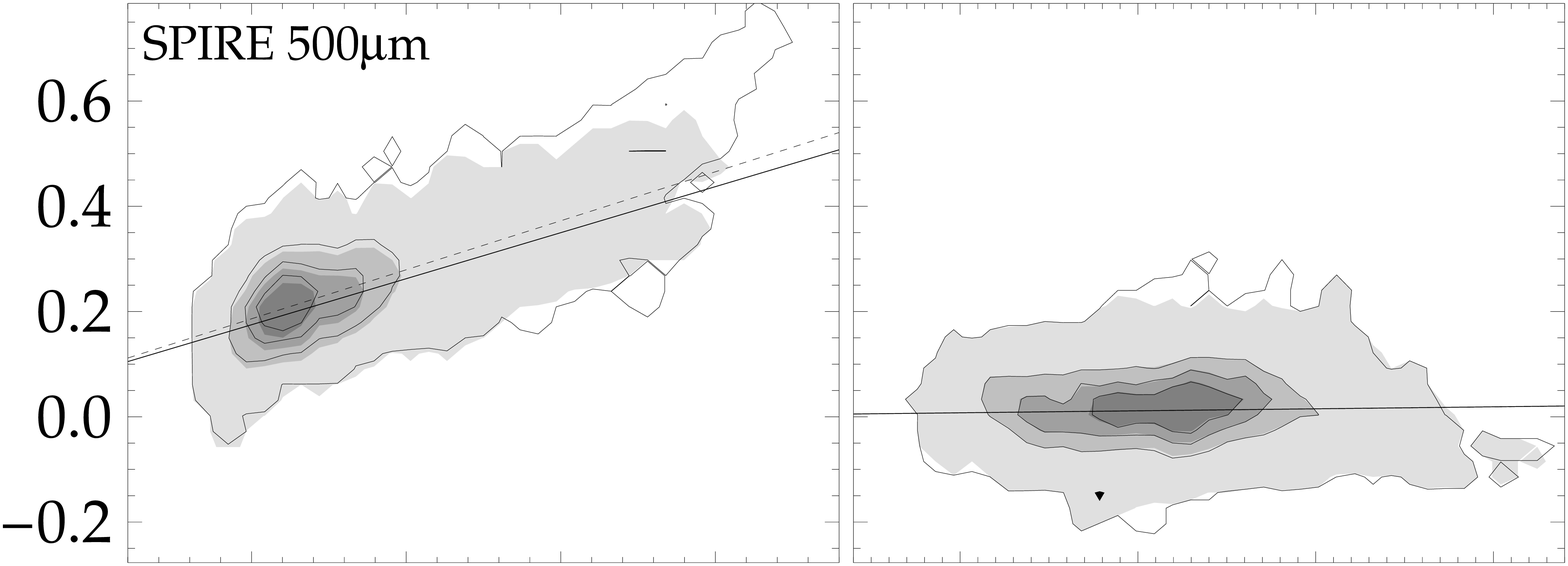}
\includegraphics[width=\sizep]{\dirres/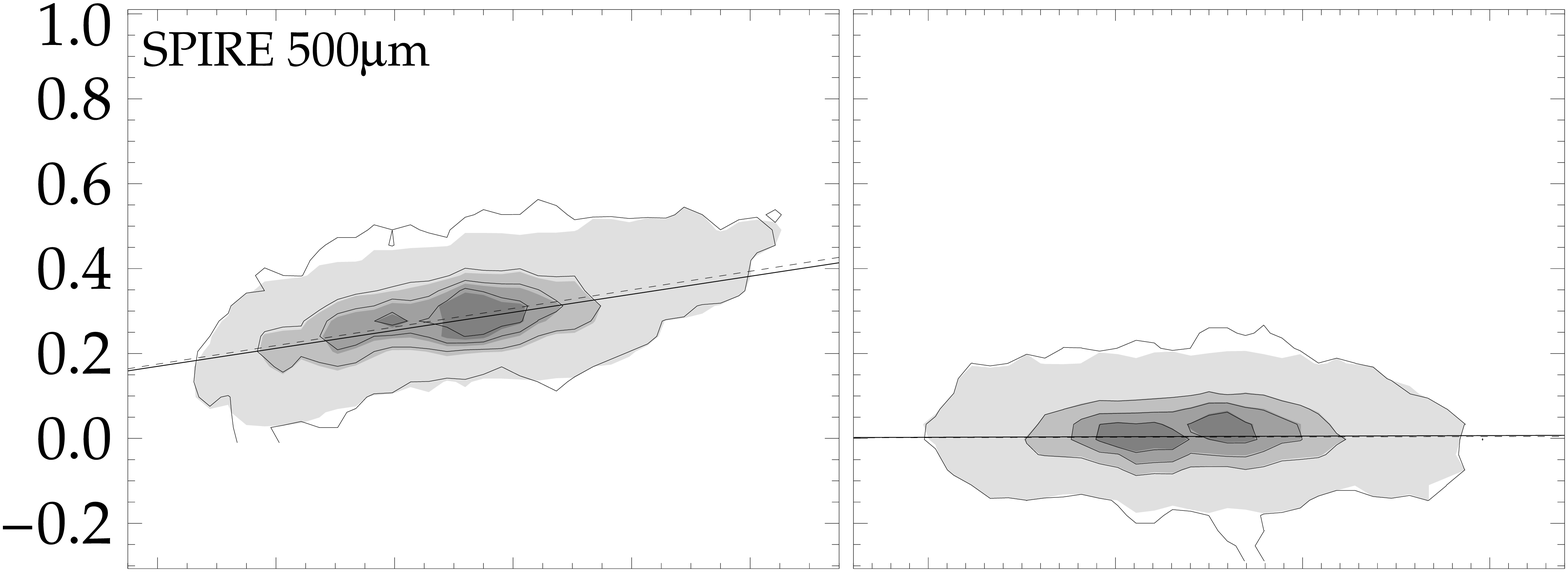}

\includegraphics[width=\sizep]{\dirres/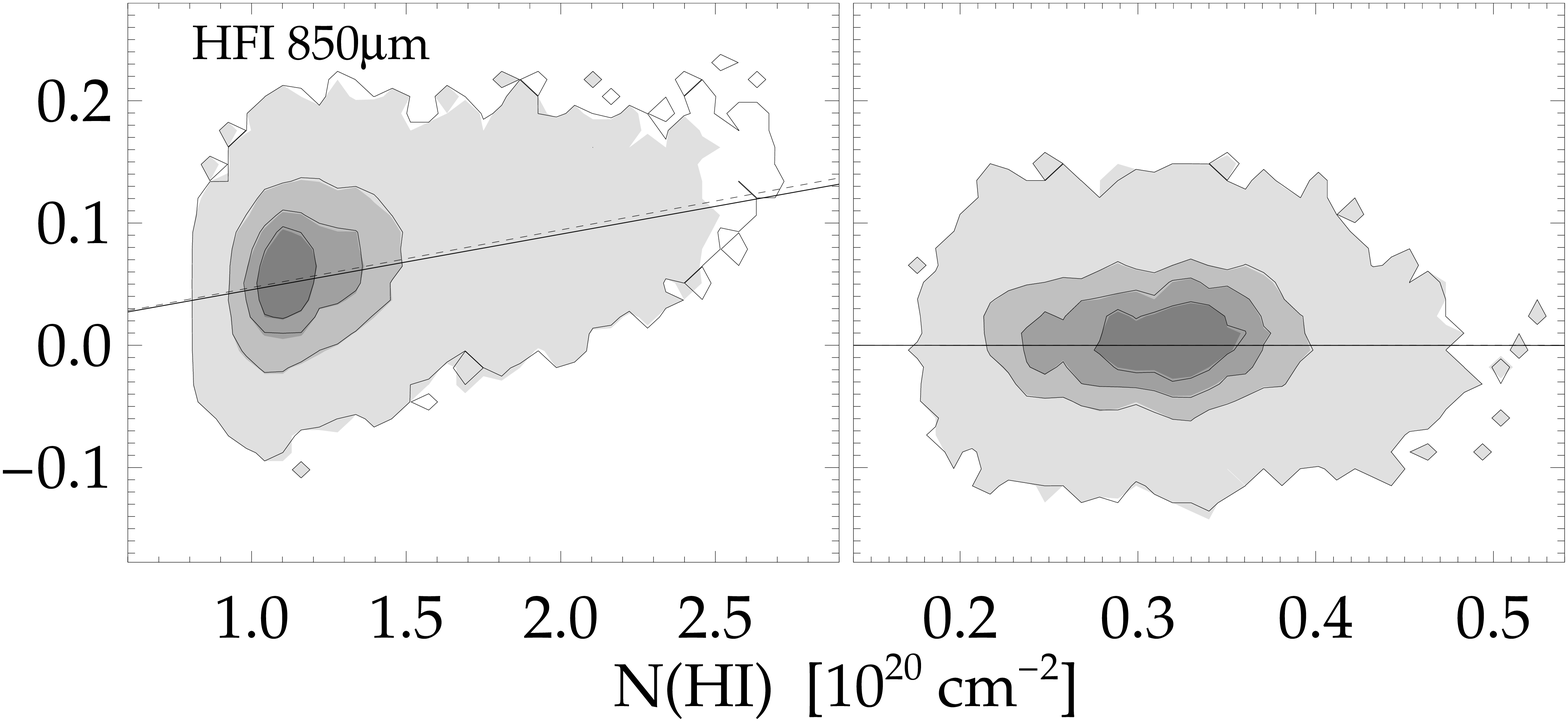}
\includegraphics[width=\sizep]{\dirres/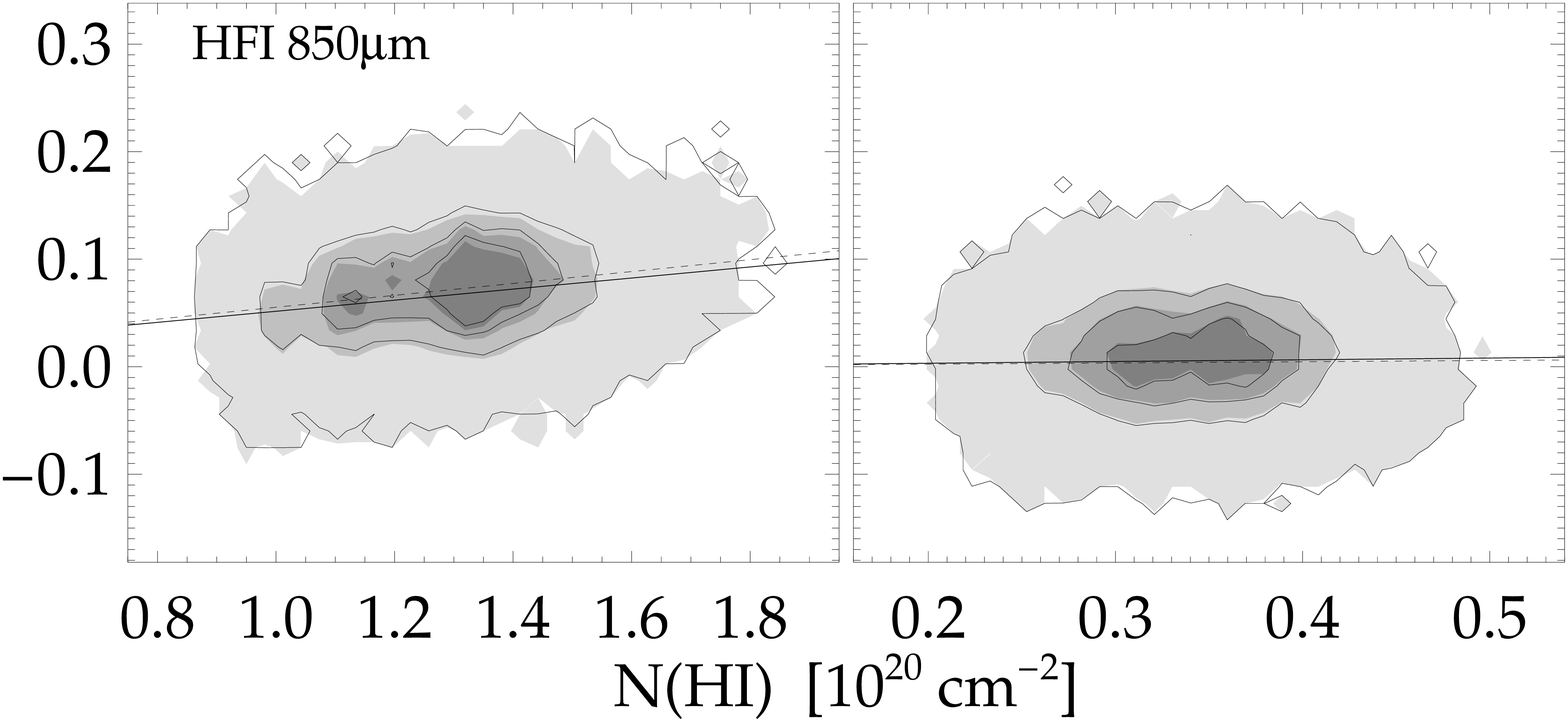}

\caption{Same as Fig.~\ref{fig:corre1}, but for field V3 and V4. Magenta crosses in the V3 
$250\mu$m panels show the pixels of region 'G' (the galaxy VCC975; see Sect.~\ref{sec:resi}
and footnote~\ref{foot:gal} for details).}
\label{fig:corre2}
\end{figure*} 

In Fig.~\ref{fig:corre1} and \ref{fig:corre2} we show the pixel-by-pixel correlation
between dust emission in the IRAS, SPIRE and {\em Planck}-HFI 850~$\mu$m bands and 
column density of the two \ion{H}{i} components, for fields V1/V2 and V3/V4, respectively. For each
velocity channel, the surface brightness is corrected by subtracting the fitted
contribution from the other velocity channel and the offset and ecliptic latitude
gradient. Thus, for the LVC gas, the surface brightness on the y-axis is 
$I_\nu- \epsilon^\mathrm{IVC}_\nu \times N^\mathrm{IVC}_{\ion{H}{i}} - O_\nu - E_\nu \times (b-b_0)$.
The line and filled contours show the density of pixels before and after masking
the regions deviant from the model by more than $3\sigma_R$ at 250$\mu$m; the dashed and solid 
straight lines are the fitted correlation in the two cases (the slope of the solid line 
being one of the $\epsilon_\nu$ values reported in Table~\ref{table:fit}). In some
cases, masking removes regions with high gas column density where significant
excess emission shows up as an upturn of the contour plots (see, e.g.\, the case for
the LVC component at 250~$\mu$m in V2); in others, it affects regions
at lower column density. The possible nature of these residuals is discussed in Sect.~\ref{sec:resi}. 
In any case, the effect of masking on the parameter derivation is marginal:
the emissivities derived without masking would be higher by up to a maximum of 1.5$\sigma$ 
(for the LVC component in fields V2 and V3 at 100 and 250~$\mu$m), with $\sigma$ the 
uncertainty estimated from the { Monte Carlo analysis}. Nevertheless, Table~\ref{table:fit}
reports the values obtained after masking. 
The standard deviation of the residuals $\sigma_R$ is also reported in the table, together 
with $\sigma_C$, derived by removing from $\sigma_R$ the contribution of instrumental noise 
in dust emission and gas column density observation. The value $\sigma_C$ includes all 
deviations from the adopted model and should be dominated, for the low column density regions, 
by the fluctuations in the CIB \citep{PlanckEarlyXXIV}. 
{ 
Our estimates for $\sigma_R$ (and $\sigma_C$) are very close to those obtained at 100 $\mu$m
and in the {\em Planck} bands for low column density regions by \citeauthor{Planck2013XI} 
(\citeyear{Planck2013XI}; see the first line in their table C.1); they indeed confirm that 
the standard deviation of the residuals is dominated by the CIB fluctuations at all bands,
with the exception of 850$\mu$m, where the contribution of instrumental noise is larger.
}

\begin{figure*}
\sidecaption
\includegraphics[width=12cm]{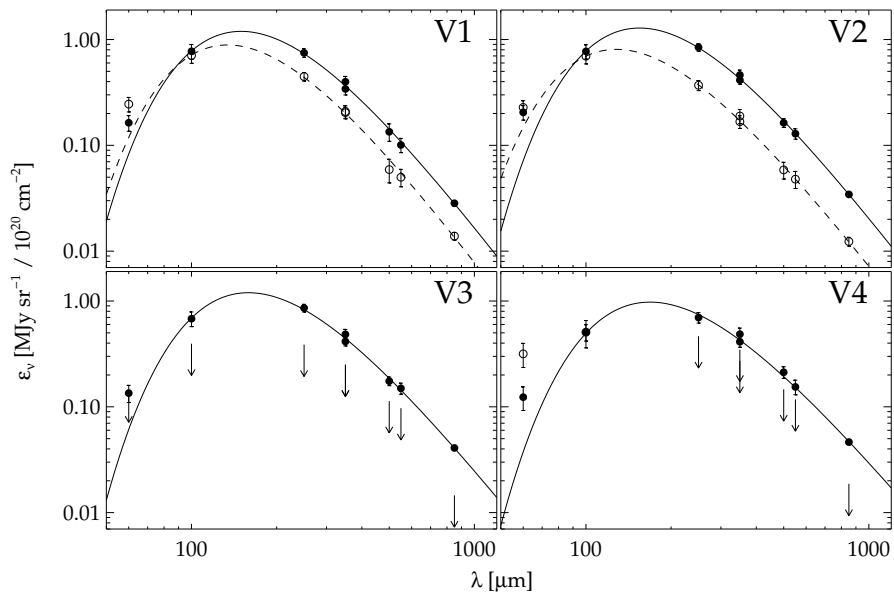}
\caption{Emissivities for the four HeViCS fields. Filled symbols and solid lines 
refer to the estimates for $\epsilon^\mathrm{LVC}_\nu$ and their (variable $\beta$) MBB
fits, respectively; open symbols, 2-$\sigma$ upper limits and dashed lines refer to 
$\epsilon^{IVC}$.
}
\label{fig:sedemi}
\end{figure*} 

In the rest of this text, we focus our attention on the results obtained for the dust emissivities
associated to the two gas components,  $\epsilon^\mathrm{LVC}_\nu$ and $\epsilon^\mathrm{IVC}_\nu$;
the discussion of the fitted offsets $O_\nu$ and ecliptic latitude gradients $E_\nu$
is presented in Appendix~\ref{app:oe}. Fig.~\ref{fig:sedemi} shows the emissivity SED
derived for each HeViCS field. The mean emissivities for the full HeViCS coverage are 
shown in Fig.~\ref{fig:sedemi_avg} and reported in Table~\ref{table:fit}. The scatter 
(as measured by the standard deviation of the emissivities of the various field) is relatively 
small, generally below 15-20\% of the mean value. This is however larger than the individual 
uncertainties of each estimate (typically in the range $\sim$ 5-10\%), pointing to real 
field-to-field variations in the emissivity values (see next section). 

It is important to remember that, in order to compare our results with those in 
\citet{Planck2013XI}, we scaled our  ALFALFA map to match the LAB column densities 
(Sect.~\ref{sec:alfalfa}). If we had not done this,
the derived emissivities would have been lower by 6\% for the LVC component, a 
difference compatible with the estimated uncertainties in the emissivity derivation,
corresponding to a maximum of 1.5$\sigma$ at 100~$\mu$m; and by a larger amount,
14\%, for the IVC component (equivalent to a maximum of 3.5$\sigma$ at 100~$\mu$m).
These differences are entirely due to the gain correction described in Sect.~\ref{sec:alfalfa},
since the offset between ALFALFA and LAB is negligible. 

{ 

We found the SPIRE and {\em Planck} emissivities in the overlapping band at 350~$\mu$m 
to be marginally consistent, with a difference of about 1.5-$\sigma$ for the LVC component;
the 350~$\mu$m SPIRE/{\em Planck} emissivity ratio is 0.86, which is entirely due to the
relative gain of the SPIRE and {\em Planck} maps used in this work. Instead, the accurate
analysis of \citet{BertincourtA&A2016} finds that the relative gain is close to unity.
The difference cannot be accounted for by including color corrections (which we neglected
at 350~$\mu$m), nor by switching to the most recent releases of {\em Planck} and {\em Herschel}
data. The use of the latest pipeline-reduced SPIRE 350~$\mu$m (downloaded from the Herschel Science 
Archive in August 2016) confirms our result. The reason for the difference is unknown. We are tempted to 
impute it to the peculiarities of the HeViCS fields. In fact, the HeViCS average surface brightness
is at least a factor two smaller than that of the {\em Spider cirrus} molecular cloud, the 
lowest-intensity field used by \citet{BertincourtA&A2016}; when our procedures for data manipulation
and analysis are applied to pipeline-reduced SPIRE data of that target, we indeed obtain a 
result which is entirely consistent with their findings.

}

\section{Absorption cross-sections of high latitude dust}
\label{sec:sed}

The SED of diffuse high latitude dust is one of the test beds for Galactic
dust models, because of two simplifying conditions: dust emission can be
reasonably assumed to be optically thin in the diffuse medium; the line of
sight at high latitude crosses a very limited range of dust environments, and
a single heating source can be assumed, that provided by the LISRF.

A very simple description of dust emission is provided by a MBB:
assuming that all dust grains share the same composition and size (and thus
attain the same temperature $T$ under the same heating conditions), the emissivity 
SED can be written as
\[
\epsilon_\nu = \frac{\tau_\nu}{N_{\mathrm{H}}} B_\nu(T).
\]
The FIR/submm absorption cross section (per unit hydrogen atom) is usually described by a 
power law; adopting 250~$\mu$m as the reference wavelength, it can be written as
\[
\frac{\tau_\nu}{N_{\mathrm{H}}}=\frac{\tau(250\mu\mathrm{m})}{N_{\mathrm{H}}} \times (250\mu\mathrm{m}/\lambda)^\beta.
\]
A fit to the emissivity SEDs can thus retrieve the dust temperature $T$ and the 
absorption cross section normalization $\tau(250\mu\mathrm{m})/ N_{\mathrm{H}} $ 
and spectral index $\beta$.

In general, a MBB results in an adequate fit to the observations, 
provided that datapoints contaminated by dust at hotter temperatures and/or
not emitting at thermal equilibrium are excluded from the analysis (for this 
reason, the $60\mu$m emissivities are not included in the fits presented here). 
However, the parameters retrieved from the fits might be different from those
of the underlying dust population. In particular, simultaneous solutions for 
$T$ and $\beta$ might be degenerate, since both quantities determines the SED shape.
An inverse correlation between these parameters is generally found and imputed 
to a bias due to standard $\chi^2$ minimization techniques, calling for more 
sophisticated Bayesian-fitting methods for a rigorous treatment of all involved 
uncertainties \citep[see, e.g., ][]{KellyApJ2012}; or to line-of-sight temperature 
mixing, even at high Galactic latitude 
\citep[see, e.g., ][]{VenezianiApJ2010,BraccoMNRAS2011}.
Even when dust is heated by a single radiation field, grains attain different 
temperatures, depending on their size and composition: this broadens the SED,
and biases the observed $\beta$ to values smaller than the intrinsic spectral
index of dust, in particular for radiation fields weaker than the LISRF 
\citep{HuntA&A2015}.
 
\begin{figure*}
\sidecaption
\includegraphics[width=12cm]{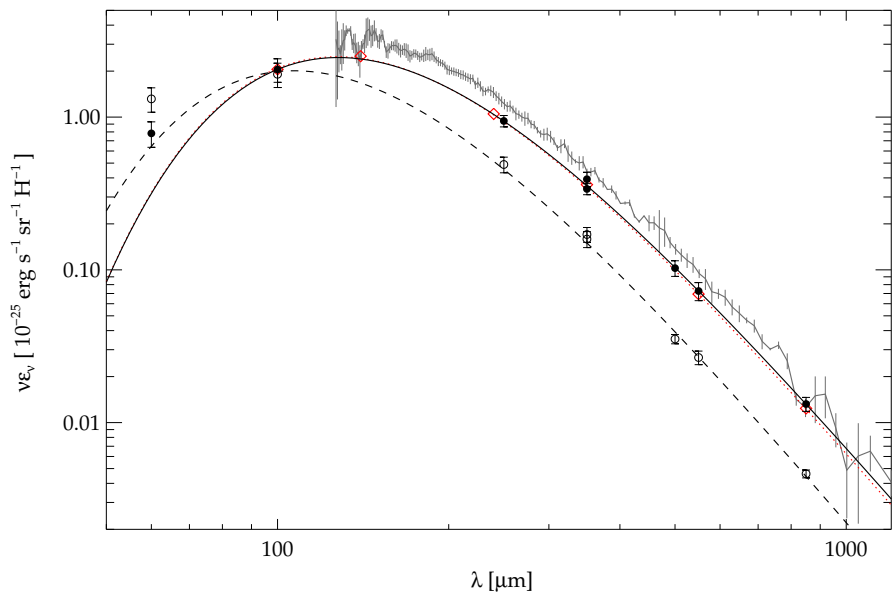}
\caption{Mean emissivities for the full HeViCS coverage. Filled circles and solid 
lines refer to the estimates for $\epsilon^\mathrm{LVC}$ and their (variable $\beta$) 
MBB fits, 
respectively; open circles and dashed lines refer to $\epsilon^\mathrm{IVC}$.  The error bars
show the standard deviation of the mean. The red squares are the COBE-DIRBE and
Planck-HFI emissivities derived in the southern Galactic pole \citep{PlanckIntermediateXVII}; 
the red dotted line is the fit to {\em Planck} high latitude emissivities \citep{Planck2013XI}; the gray 
line and error bars refer
to the high latitude emissivities from FIRAS \citep{CompiegneA&A2011}.}
\label{fig:sedemi_avg}
\end{figure*}

\begin{table*}
\caption{Parameters of single temperature MBB fits to the emissivity SEDs}
\label{table:sedfit}
\centering 
\begin{tabular}{c| c c c c c}
\hline
\hline
 & V1 & V2 & V3 & V4 & avg. \\
\hline
\multicolumn{6}{c}{}\\
\multicolumn{1}{c}{} & \multicolumn{5}{c}{LVC} \\
%fitting beta
$\frac{\tau(250\mu m)}{N\ion{H}{i}}$ / $\frac{10^{-25} \text{cm}^2}{\text{H}}$  & 0.44$\pm$0.11 & 0.54$\pm$0.12 & 0.52$\pm$0.11 & 0.49$\pm$0.13 &  0.49$\pm$0.13\\
T/K                                   & 20.7$\pm$1.4    & 20.1$\pm$1.2  & 20.3$\pm$1.2    & 20.0$\pm$1.4  & 20.4$\pm$1.5 \\
$\beta$                             & 1.69$\pm$0.14   & 1.67$\pm$0.11 & 1.51$\pm$0.12   & 1.33$\pm$0.14 & 1.53$\pm$0.17 \\
&&&&&\\
%beta=1.59
$\frac{\tau(250\mu m)}{N\ion{H}{i}}$ / $\frac{10^{-25} \text{cm}^2}{\text{H}}$  & 0.370$\pm$0.030 & 0.467$\pm$0.023 & 0.596$\pm$0.038 & 0.755$\pm$0.072 & 0.525$\pm$0.052 \\
T/K                                   & 21.6$\pm$1.1    & 20.92$\pm$0.50  & 19.61$\pm$0.52    & 17.91$\pm$0.68  & 20.00$\pm$0.70 \\
$\beta$ & \multicolumn{5}{c}{1.59 (fixed)} \\
\hline
\multicolumn{6}{c}{}\\
\multicolumn{1}{c}{} & \multicolumn{5}{c}{IVC} \\
%fitting beta
$\frac{\tau(250\mu m)}{N\ion{H}{i}}$ / $\frac{10^{-25} \text{cm}^2}{\text{H}}$ & 0.200$\pm$0.056 & 0.139$\pm$0.043 & - & - & 0.150$\pm$0.040\\
T/K                                   & 22.7$\pm$1.9    & 24.5$\pm$2.3    & - & - & 23.9$\pm$2.0\\ 
$\beta$                             & 1.77$\pm$0.16   & 1.63$\pm$0.18   & - & - & 1.65$\pm$0.15\\
&&&&&\\
%beta=1.59
$\frac{\tau(250\mu m)}{N\ion{H}{i}}$ / $\frac{10^{-25} \text{cm}^2}{\text{H}}$ & 0.153$\pm$0.015 & 0.130$\pm$0.014 & - & - & 0.137$\pm$0.011\\
T/K                                   & 24.7$\pm$1.6    & 25.1$\pm$1.2    & - & - & 24.6$\pm$1.2 \\ 
$\beta$ & \multicolumn{5}{c}{1.59 (fixed)} \\
\hline
\end{tabular}
\end{table*}

For the sake of simplicity, and in analogy to \citet{PlanckEarlyXXIV} and \citet{Planck2013XI}, 
we adopt here a single MBB model and fit it to the data for each field, and to the average emissivity,
using a standard $\chi^2$ minimization routine 
\citep[{\tt mpfit} for IDL;][]{MarkwardtProc2009}; we allow $\beta$ to vary
(these fits are plotted in Fig.~\ref{fig:sedemi} and \ref{fig:sedemi_avg})
but we also produce fits using a fixed $\beta$. The parameters 
derived from the fits are given in Table~\ref{table:sedfit}.

\subsection{LVC dust}

For dust associated with LVC gas, we find a trend of relatively higher emissivities
at longer wavelengths in the two southernmost fields. 
{
This translates to fits with
spectral index increasing from $\beta\approx 1.3$ in field V4 to $\beta\approx 1.7$
in field V1. Instead, the temperature is relatively constant in the four fields
($T=20.0-20.7$K). If instead $\beta=1.59$ is used, as derived by \citet{Planck2013XI} 
on the Diffuse High Galactic Latitude medium \citep[DHGL, defined by $b>15^\circ$, 
N$_\ion{H}{i}<5\times 10^{20}$ cm$^{-2}$; ][]{CompiegneA&A2011}, the fitted
temperature reduces to 17.9K in field V4 and increases to to 21.6K in field V1.
This behavior is a clear effect of the T-$\beta$ degeneracy \citep{ShettyApJ2009}.}

The fit to the average emissivities, as well as the emissivities themselves, 
is consistent with the previous fit to the DHGL medium (red dotted line in 
Fig.~\ref{fig:sedemi_avg}). For a variable $\beta$, we find 
$\tau(250\mu\mathrm{m})/ N_{\mathrm{H}} = (0.49\pm0.13) \times 10^{-25}$ cm$^2$ H$^{-1}$, 
$T=20.4\pm1.5$ K and $\beta=1.53\pm0.17$, while the \citet{Planck2013XI} values for the DHGL are
$\tau(250\mu\mathrm{m})/ N_{\mathrm{H}} = (0.49\pm0.14) \times 10^{-25}$ cm$^2$ H$^{-1}$, 
$T=20.3\pm1.3$ K and $\beta=1.59\pm0.12$. Analogous consistency is
obtained by fitting the COBE-DIRBE and {\em Planck}-HFI emissivities of the south 
Galactic pole \citep[red diamonds in Fig.~\ref{fig:sedemi_avg};][]{PlanckIntermediateXVII}:
$\tau(250\mu\mathrm{m})/ N_{\mathrm{H}} = (0.55\pm0.05) \times 10^{-25}$ cm$^2$ H$^{-1}$,
$T=19.8\pm1.0$ K and $\beta=1.65\pm0.10$.  
{ 
The emissivities derived in this work are thus representative of the average properties of 
high latitude dust. This is remarkable, since the DHGL and the south Galactic pole medium
cover 50\% and 20\% of the sky, respectively, while our fields are just $\sim$0.2\%.
}

\subsection{Implications for dust modelling}

The {\it Herschel}-SPIRE emissivities derived in this work also share with {\em Planck}-HFI 
the inconsistency with the dust emission spectra obtained by the FIRAS instrument
aboard the COBE satellite. The HFI team switched from a FIRAS- to a planet- (Uranus 
and Neptune) based calibration for the 350~$\mu$m and 550~$\mu$m bands, after 
recognising in the FIRAS data a bias of uncertain origin \citep{Planck2013VIII}. 
The FIRAS emissivity spectrum for the DHGL derived by \citet{CompiegneA&A2011} 
is shown in Fig.\ref{fig:sedemi_avg}. 
As it has been pointed out, the new emissivities, confirmed here by the analysis
of (Neptune-calibrated) SPIRE data, ask for a revision of the current models 
for MW dust, whose properties (grain composition and size distribution) were adjusted
to reproduce the FIRAS spectrum \citep{PlanckIntermediateXVII,PlanckIntermediateXXIX}.
For example, the MC10 dust model distributed with the {\tt DustEM} software
\citep{CompiegneA&A2011} reproduced FIRAS DHGL emissivities (after correcting them 
for the contribution of ionized gas -mostly, with a small correction for molecular gas-
to the hydrogen column density) when heated
by the LISRF. Using the new emissivities, the absorption cross-sections would need 
to be adjusted so that emission at, e.g., 250~$\mu$m, is lower by 20\%, the modelled
SED shape for LISRF heating\footnote{\citet{CompiegneA&A2011} used the standard
formulation of \citet{MathisA&A1983} for the LISRF.
\cite{DraineBook2011} proposes a few modifications to it, in order to
match COBE-DIRBE observations of diffuse stellar radiation in the near-infrared. Using the
{\tt DustEM} code and the MC10 dust model
we estimated that the modified LISRF would increase the dust emissivity by about 10\%
(the net effect being that of increasing the original LISRF by about 25\%). If this
new estimate of the LISRF is adopted, new dust models should decrease their FIR
output by 30\% with respect to FIRAS. Also, colder dust temperatures would be required.
} being still consistent with observations 
\citep{PlanckIntermediateXVII}.

{ In Appendix~\ref{app:extgal} we discuss the implications of these new absorption cross-sections
in the determination of the dust masses of external galaxies.}

\subsection{IVC dust}

Dust associated with IVC is found to have a lower emissivity than LVC dust (on average 
by about $1/2$ at 250~$\mu$m) and a lower absorption cross-section. This result confirms
the findings of \citet{PlanckEarlyXXIV}, which tentatively impute the different properties 
of IVC dust to a larger fraction of smaller grains not emitting at thermal equilibrium.
Dust processing leading to smaller grain sizes 
\citep[in particular grain shattering, dominant 
over gas-grain sputtering;][]{BocchioA&A2014} is indeed
expected in the shocked gas of Galactic fountains, which has been proposed as an explanation
for the IVC medium \citep[for a review, see, ][]{WakkerARA&A1997}. 
{
A lower opacity cross section, and a higher temperature than for LVC dust is found, for 
relatively similar $\beta$ values. Assuming
$\beta=1.59$, the emissivity is $\tau(250\mu\mathrm{m})/ N_{\mathrm{H}} = (0.14\pm0.01) 
\times 10^{-25}$ cm$^2$ H$^{-1}$ with $T\approx 25$K. 
Unfortunately, we cannot compare quantitatively our results with those obtained
on LVC and IVC dust by \citet{PlanckEarlyXXIV}, since they used an earlier
release of {\em Planck} data calibrated on FIRAS, and adopted the older FIRAS-based
value $\beta=1.8$ for both dust components. We only note that the SED is {\em bluer} 
}
because of the different dust properties, rather than because of a heating source more 
intense than the LISRF
(the IVC medium being located higher above the Galactic plane than the 
LVC; Sect.~\ref{sec:alfalfa}).

\section{Residuals}
\label{sec:resi}

\newcommand\scala{0.52735}
\newcommand\figdir{./}

\begin{figure*}
\includegraphics[scale=\scala,trim = 0 0 8.0bp 0,clip]{\figdir/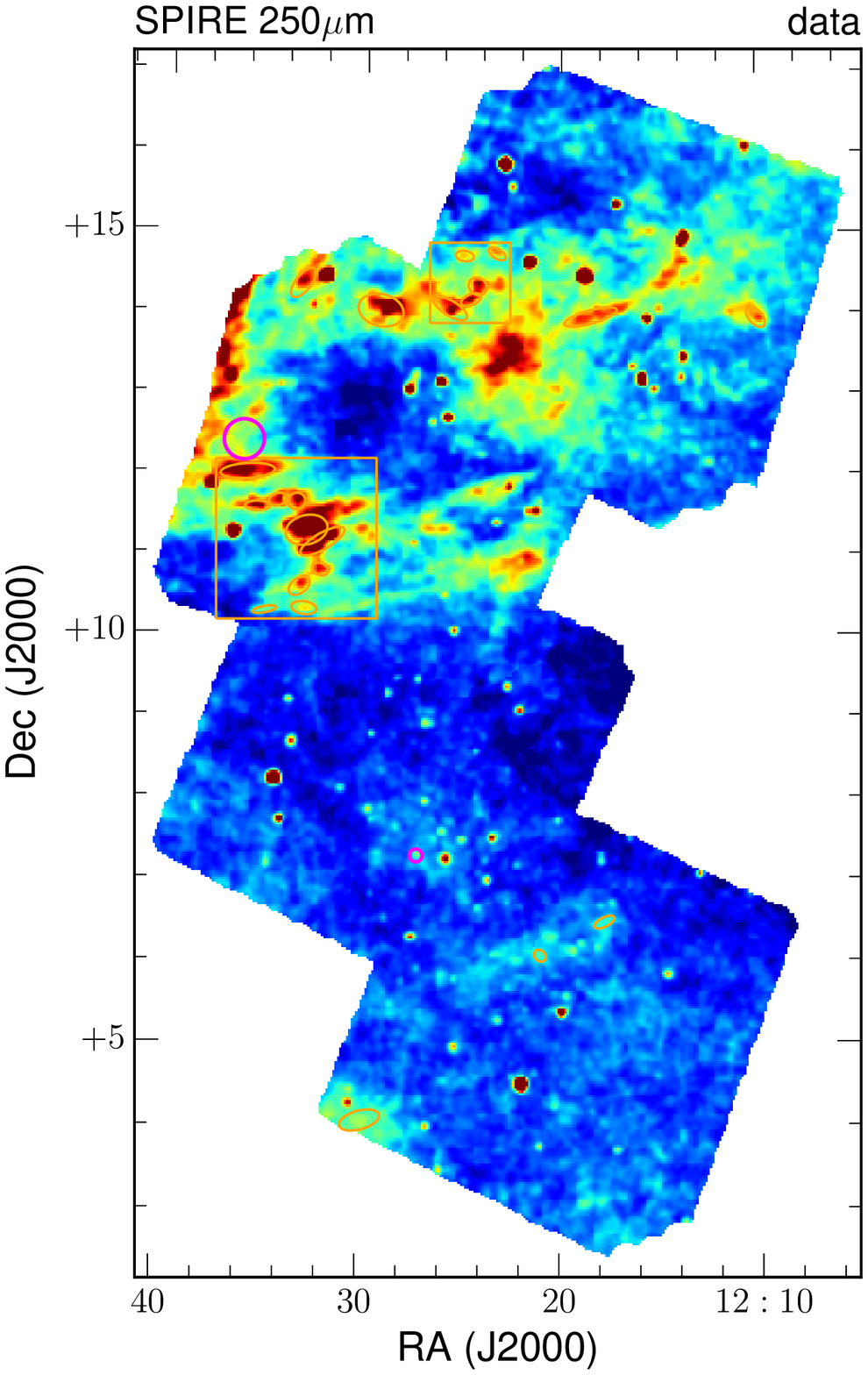}\includegraphics[scale=\scala,trim = 4bp 0 6.5bp 0,clip]{\figdir/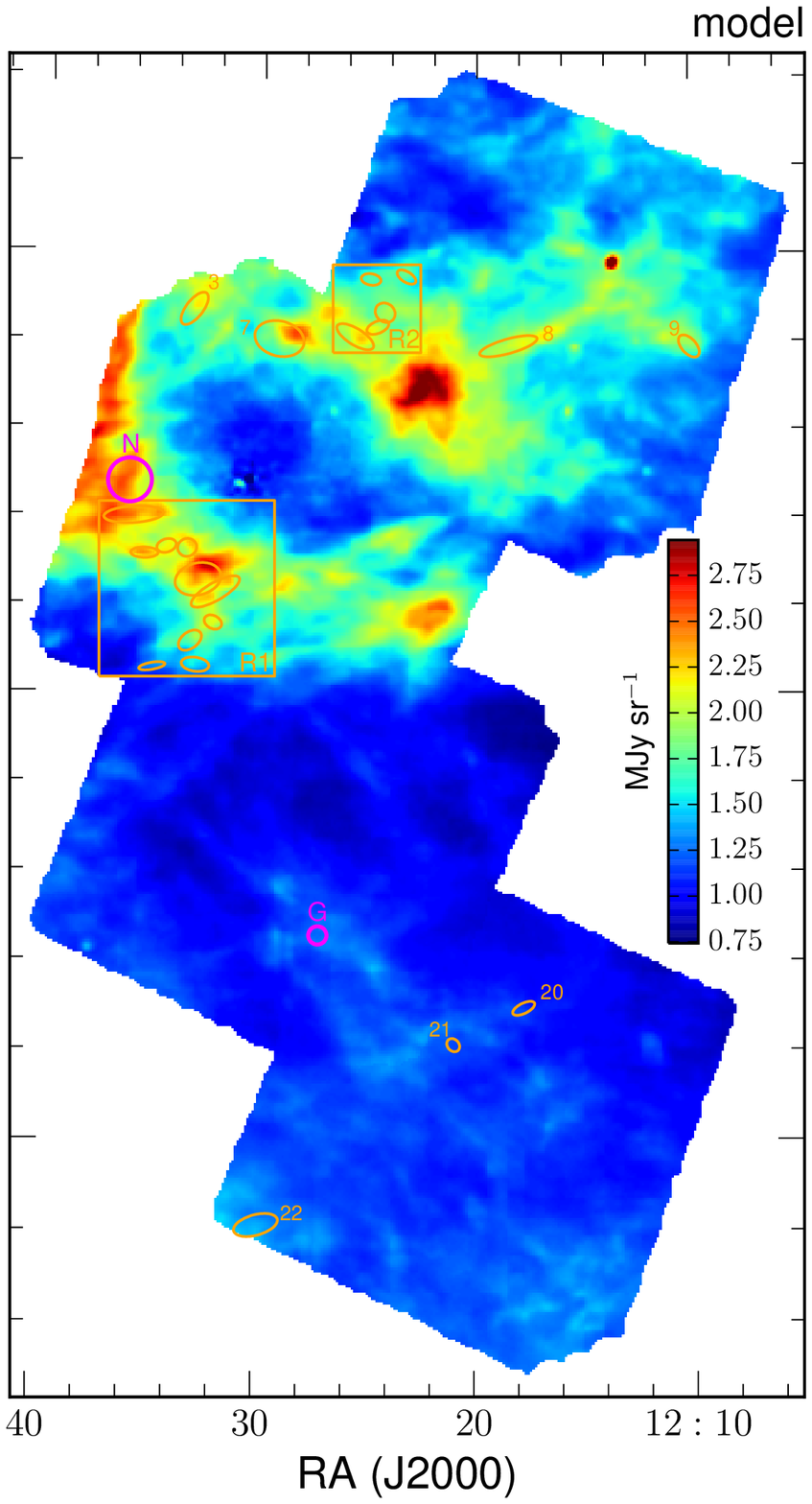}\includegraphics[scale=\scala,trim = 4bp 0 6.5bp 0,clip]{\figdir/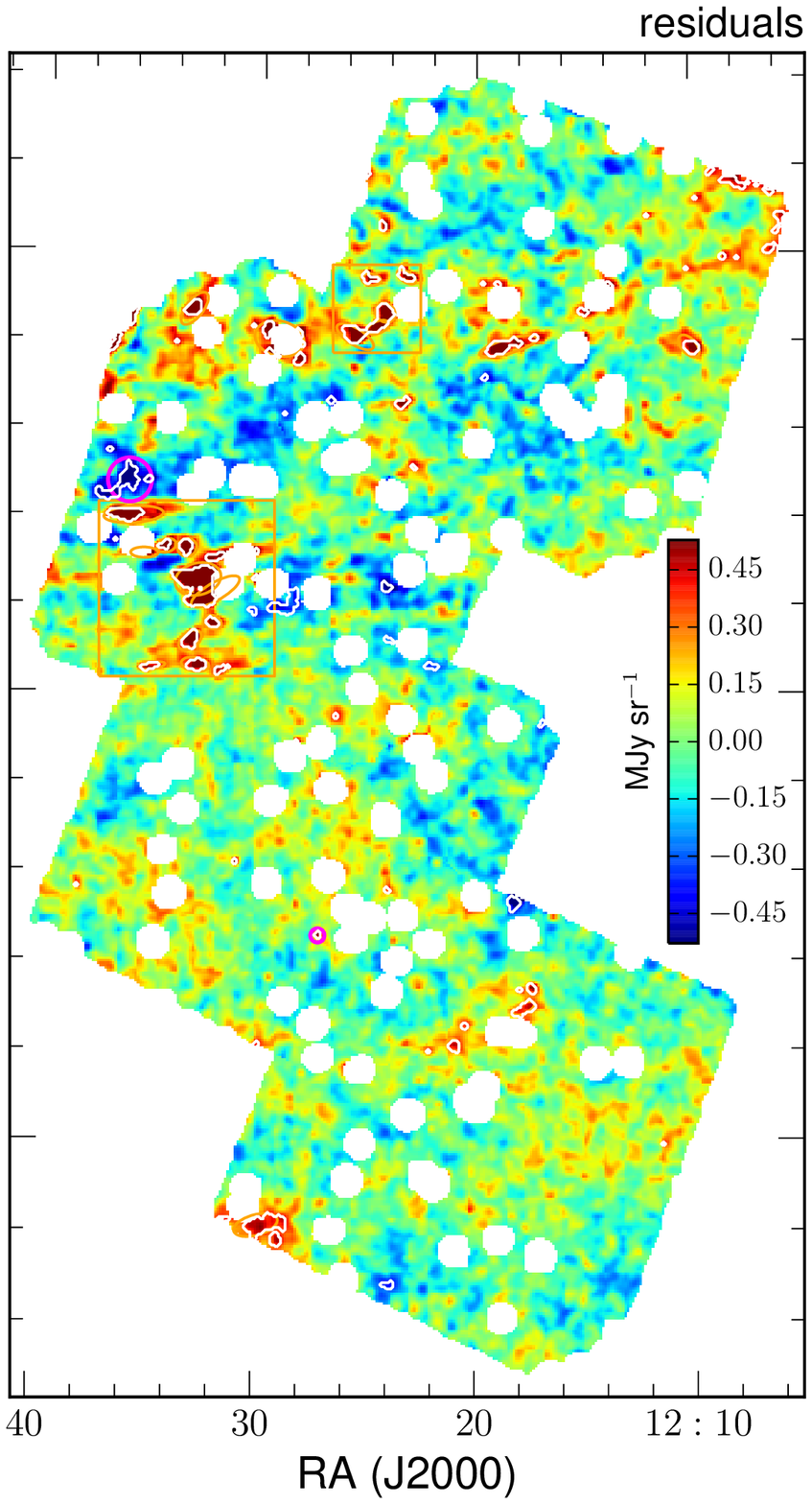}
\caption{
Observations (left column), models (middle column) and residuals (right column) 
of the HeViCS field in the SPIRE $250\mu$m band.  Observations and models have been
subtracted by the large scale terms $O_\nu + E_\nu \times (b-b_0$) and are shown on the same color scale. 
The color scale of the residuals goes from -4 to 4$\sigma_R$ (Table~\ref{table:fit}); round white areas 
show the object mask applied before the fitting, while white contours encircles the regions where 
residuals are larger than 3$\sigma_R$ (excluded during the fitting). 
The orange numbered ellipses and 
square regions refers to the high positive residual regions listed in Table~\ref{table:cata} and shown in Fig.~\ref{fig:cata}.
The magenta circle labelled 'N' highlights a region with high negative residuals; that labelled 'G' is centered
on an unmasked Virgo Cluster galaxy.
}
\label{fig:modres250}
\end{figure*} 

In Fig.~\ref{fig:modres250} we show the observed dust emission, the dust emission modelled 
from the $\ion{H}{i}$ observations using the derived emissivities (Eq.~\ref{eq:fitted}), and 
the residuals in the SPIRE $250\mu$m band. In order to show observations and models on the 
same scale, the large scale terms derived from the fit (offsets and ecliptic latitude gradients) 
have been subtracted from the images. For ease of presentations, the four HeViCS fields are 
combined in a single image { by removing the small residual offsets between the fields 
(see Appendix.~\ref{app:oe})}; the analysis, though, was conducted separately on each field,
as detailed in the previous Sections.  Analogous figures for the other IRAS, SPIRE and
HFI bands are shown in Appendix~\ref{app:otherfigs}.

Models and residuals shows consistent results in all bands. The 60~$\mu$m residuals
(Fig.~\ref{fig:modres60}) are similar to those at 100~$\mu$m (Fig.~\ref{fig:modres100}), 
both IRAS bands having a similar contribution 
from dust associated to LVC and IVC gas (at least in the V1 and V2 areas). All
SPIRE and HFI bands between 250~$\mu$m and 850~$\mu$m (Fig.~\ref{fig:modres250},
\ref{fig:modres350} to \ref{fig:modres850}), where the LVC gas contribution to dust emission is at 
least twice that of IVC, show basically the same pattern in observations, models
and residuals; the largest deviations from the models, though, more easily stand
out in shorter wavelength bands, where they dominate over the background noise.

A visual comparison between data and models shows that, on large scales,
there is a good correspondence between the features seen in dust and $\ion{H}{i}$ emission.
We have already shown the good correlation between these two ISM tracers in
Sect.~\ref{sec:res}.  However, there are significative
deviations at smaller scales: 
{ 
infrared emission appears less filamentary than $\ion{H}{i}$, mostly because
of the superimposition of the clumpier CIB to the cirrus; (some) regions
of high $\ion{H}{i}$ column density have infrared emission much in excess of
those predicted by the model.
}

\subsection{A catalog of excess regions}

We have identified the main excess regions by running the source
extraction software {\tt sextractor} \citep{BertinA&AS1996} on the 
250~$\mu$m residual maps. 
We selected
all sources with pixel values above 3$\sigma_R$
and with size larger than a beam (about 12 pixels).

To avoid fragmentation, the mask of bright background sources was 
unset. An initial source list was produced, and pruned of those objects
whose isophotal ellipses included a significant contribution from
background objects, both masked and 
unmasked\footnote{About half of the Virgo Cluster galaxies detected 
by \citet{AuldMNRAS2013} were not masked.
The brightest is VCC975, with a flux of 1.1 Jy at 250~$\mu$m. It is
shown by the magenta circle marked with 'G' in Fig.~\ref{fig:modres250}; 
the corresponding pixels are plotted as crosses on the 250~$\mu$m V3 
panels in Fig.~\ref{fig:corre2}.  Beside the obvious results
that galaxy-dominated pixels have no correlation with
foreground gas column densities, the panels show that
even for the brightest source the contribution to the
emissivity determination is limited: only pixels on the peak of 
the galaxy have been excluded from the fit.
\label{foot:gal}
}.
This contribution was estimated by making mock images of
the extragalactic sources, including the Virgo cluster galaxies
detected at $250\mu$m by \citet{AuldMNRAS2013} (assuming for
simplicity they can be described by point sources at the adopted 
resolution) and the objects in the HeViCS SPIRE point source catalog
of \citet{PappalardoA&A2015}\footnote{The few tens of thousands 
point sources measured in the catalog produce, when smoothed to the 
adopted resolution, a map characterized by fluctuations over 
an average background. By histogram fitting we estimated
values of $0.106\pm0.041$, $0.074\pm0.033$ and $0.027\pm0.020$ MJy sr$^{-1}$
for the average background and fluctuations at 250, 350 and
500~$\mu$m. When compared to estimates of the CIB \citep{GispertA&A2000},
these average values show that detected point sources contribute to only a 
fraction of it (up to 15\% at 250~$\mu$m; see also \citealt{BetherminA&A2012}).
Also, the background fluctuations inferred from resolved sources are smaller
than the $\sigma_C$ values found in this work (Table~\ref{table:fit}),
which are mostly due to undetected sources, together with the 
large residuals from the foreground cirrus modelling. When estimating
the contribution of point sources to the excess regions, we
neglected the average value, which should be accounted for in the constant 
offsets $O_\nu$.}. To keep contamination low, we only retained objects where 
the contribution from galaxies was smaller than 20\% the total residual.

\begin{table*}
\caption{Excess regions at 250~$\mu$m}
\label{table:exreg}
\centering 
\small
\begin{tabular}{l| c c c c c | c c | c c | c c | c c  }
\hline
\hline
 & RA[$^\circ$] & Dec [$^\circ$]& a [$\arcmin$] & b/a & PA [$^\circ$] & \multicolumn{2}{c |}{$R_\nu$ [ MJy sr$^{-1}$] } & 
\multicolumn{2}{c |}{$\left. \epsilon^\mathrm{LVC}_\nu\right|_\mathrm{ex} / \epsilon^\mathrm{LVC}_\nu$} & 
\multicolumn{2}{c |}{N$_{H_2}$ [cm$^{-2}$]} & \multicolumn{2}{c }{$\tau$}  \\
 &    &     &       &     & & median & peak & median & peak & median & peak & median & peak \\
\hline
 1 & 185.823746 & 14.725930 &   7.2 & 0.50 &  57 &  0.50 &  0.70 &  1.4 &  1.5 &  3.4e+19 &  4.7e+19 &  0.9 &  1.2 \\ 
 2 & 186.240067 & 14.695560 &   6.8 & 0.57 &  80 &  0.45 &  0.57 &  1.3 &  1.4 &  3.0e+19 &  3.8e+19 &  0.8 &  1.0 \\ 
 3 & 188.306244 & 14.343020 &  13.2 & 0.44 & 141 &  0.45 &  0.90 &  1.2 &  1.4 &  2.7e+19 &  5.3e+19 &  0.6 &  1.1 \\ 
 4 & 186.421265 & 14.043010 &  14.4 & 0.39 &  58 &  0.46 &  0.94 &  1.3 &  1.6 &  3.1e+19 &  6.3e+19 &  0.8 &  1.4 \\ 
 5 & 186.071487 & 14.318910 &   7.2 & 0.87 &  48 &  0.60 &  1.00 &  1.4 &  1.6 &  4.0e+19 &  6.7e+19 &  1.0 &  1.5 \\ 
 6 & 186.156967 & 14.151900 &   8.0 & 0.51 & 112 &  0.71 &  1.08 &  1.5 &  1.7 &  4.8e+19 &  7.2e+19 &  1.1 &  1.6 \\ 
 7 & 187.308731 & 14.012690 &  17.1 & 0.70 &  76 &  0.52 &  1.33 &  1.3 &  1.6 &  3.1e+19 &  7.9e+19 &  0.7 &  1.5 \\ 
 8 & 184.624084 & 13.942280 &  20.3 & 0.24 & 105 &  0.46 &  0.72 &  1.3 &  1.5 &  3.1e+19 &  4.8e+19 &  0.8 &  1.2 \\ 
 9 & 182.506973 & 13.937320 &   9.1 & 0.55 &  42 &  0.45 &  0.78 &  1.4 &  1.6 &  3.0e+19 &  5.2e+19 &  0.9 &  1.5 \\ 
10 & 188.974243 & 11.995280 &  20.4 & 0.28 &  93 &  0.50 &  0.92 &  1.2 &  1.4 &  3.0e+19 &  5.4e+19 &  0.5 &  0.9 \\ 
11 & 188.348526 & 11.628510 &   6.6 & 0.93 & 107 &  0.58 &  0.87 &  1.3 &  1.4 &  3.4e+19 &  5.1e+19 &  0.7 &  1.0 \\ 
12 & 188.587433 & 11.647190 &   6.8 & 0.66 & 106 &  0.49 &  0.78 &  1.2 &  1.4 &  2.9e+19 &  4.6e+19 &  0.6 &  0.9 \\ 
13 & 188.841675 & 11.567190 &   9.4 & 0.38 &  89 &  0.45 &  0.73 &  1.2 &  1.3 &  2.7e+19 &  4.3e+19 &  0.5 &  0.8 \\ 
14 & 188.217331 & 11.272350 &  15.9 & 0.68 & 104 &  0.77 &  1.80 &  1.3 &  1.8 &  4.5e+19 &  1.1e+20 &  0.8 &  1.8 \\ 
15 & 188.014572 & 11.132850 &  18.8 & 0.32 & 119 &  0.92 &  1.49 &  1.4 &  1.7 &  5.4e+19 &  8.8e+19 &  1.1 &  1.6 \\ 
16 & 188.038177 & 10.786410 &   6.3 & 0.71 &  62 &  0.49 &  0.69 &  1.2 &  1.3 &  2.9e+19 &  4.1e+19 &  0.6 &  0.8 \\ 
17 & 188.299667 & 10.578530 &   8.9 & 0.63 & 126 &  0.50 &  0.83 &  1.3 &  1.5 &  3.0e+19 &  4.9e+19 &  0.7 &  1.1 \\ 
18 & 188.235886 & 10.306960 &   9.5 & 0.50 &  81 &  0.44 &  0.75 &  1.3 &  1.5 &  2.6e+19 &  4.5e+19 &  0.8 &  1.3 \\ 
19 & 188.732086 & 10.282890 &   9.1 & 0.26 &  99 &  0.42 &  0.52 &  1.3 &  1.4 &  2.5e+19 &  3.1e+19 &  0.9 &  1.0 \\ 
20 & 184.445099 &  6.452250 &   8.2 & 0.42 & 116 &  0.40 &  0.48 &  1.4 &  1.6 &  2.9e+19 &  3.4e+19 &  1.1 &  1.4 \\ 
21 & 185.244553 &  6.040510 &   4.9 & 0.80 &  42 &  0.40 &  0.48 &  1.4 &  1.5 &  2.9e+19 &  3.4e+19 &  1.0 &  1.2 \\ 
22 & 187.458755 &  4.039970 &  15.4 & 0.45 & 105 &  0.41 &  0.52 &  1.4 &  1.4 &  2.9e+19 &  3.7e+19 &  0.9 &  1.1 \\ 

\hline
\end{tabular}
\label{table:cata}
\end{table*}

The final catalog is shown in Table~\ref{table:cata}, where for each
object we list the center, semi-major axis, axis ratio and position angle of 
the major axis of the isophotal ellipses, together with the median and peak residuals within
the isophote. The remaining columns will be described in the following.
Isophotal ellipses are shown in the maps of Fig.~\ref{fig:modres250}.
Individual objects or group of objects are also shown in Fig.~\ref{fig:cata}, 
overlayed on the 250~$\mu$m map at the original resolution (FWHM$\approx$18"), 
providing a better view of the morphology of diffuse emission.

\newcommand\scalac{0.428}
\newcommand\scalas{0.214}
\newcommand\figdirc{./}

\begin{figure*}
\center
\includegraphics[scale=\scalac]{\figdirc/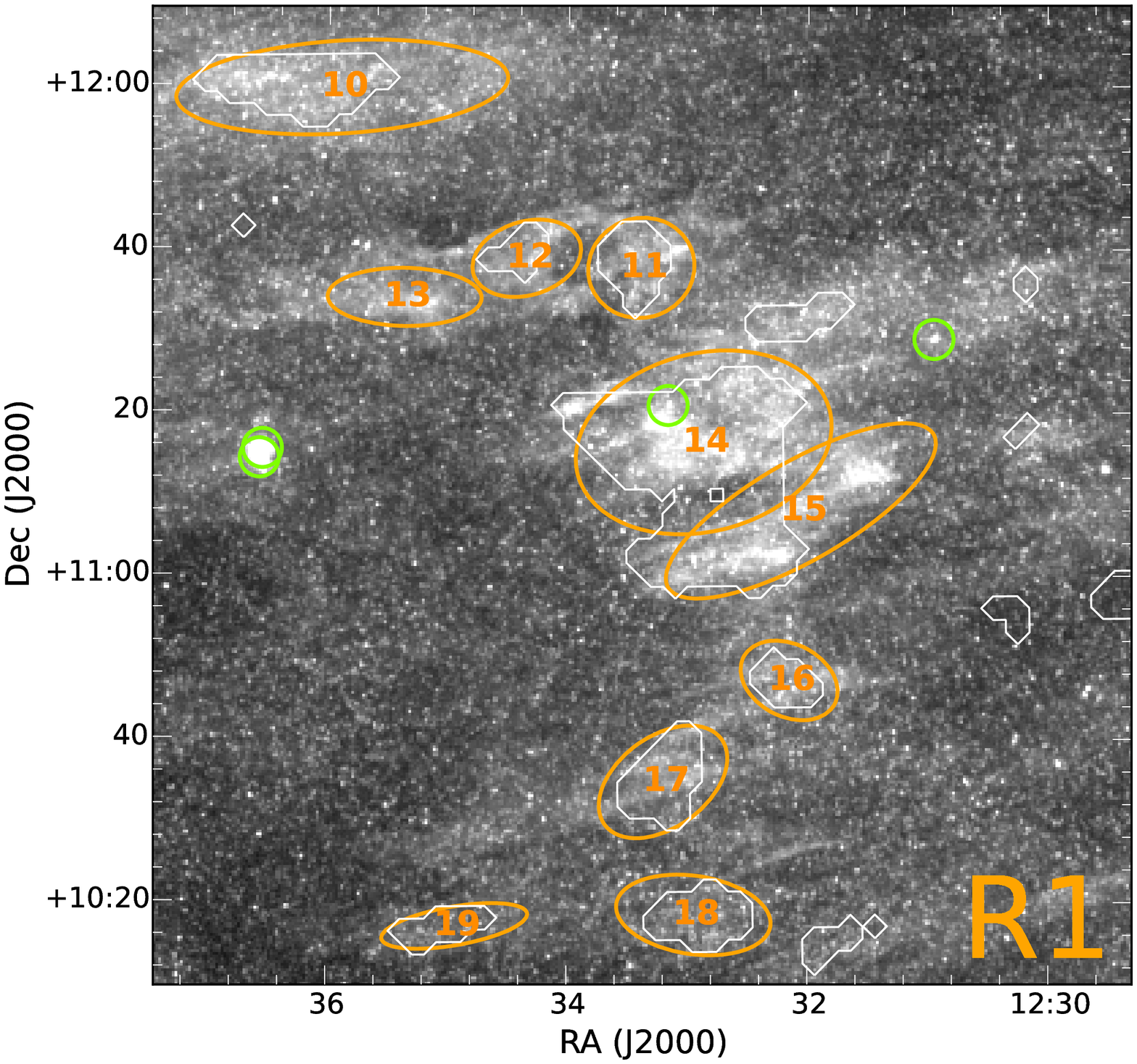}

\includegraphics[scale=\scalas]{\figdirc/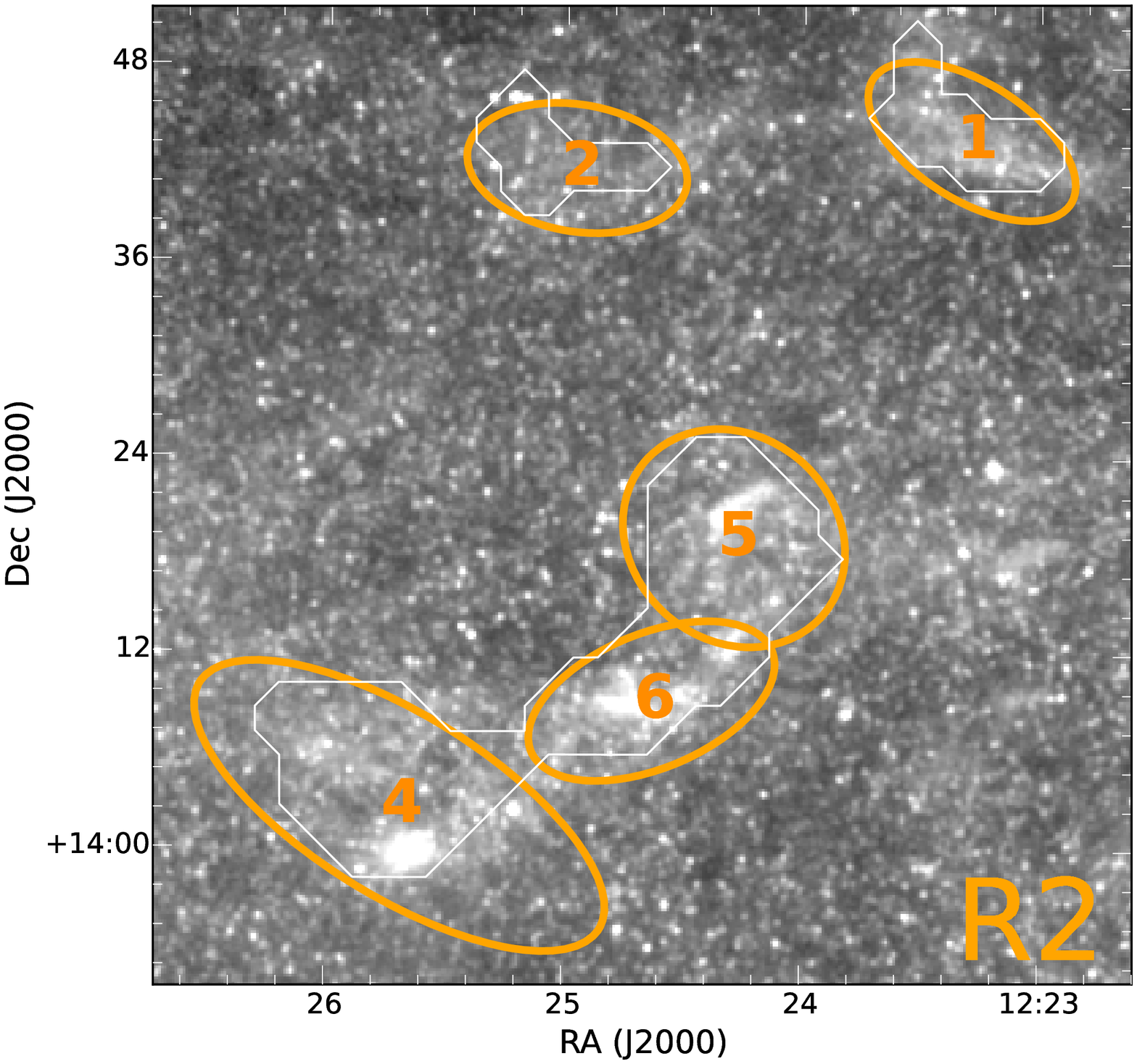}\includegraphics[scale=\scalas]{\figdirc/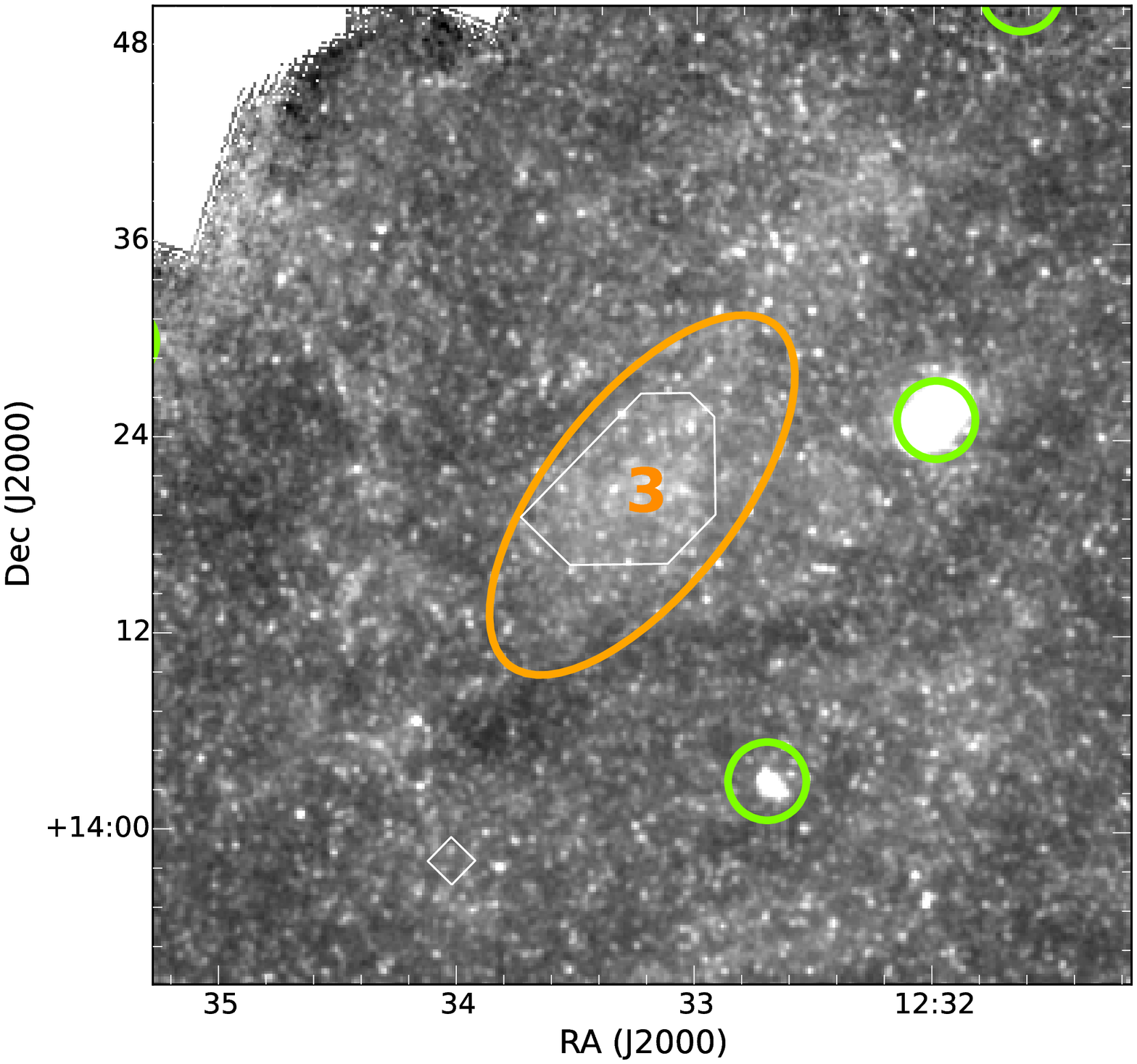}\includegraphics[scale=\scalas]{\figdirc/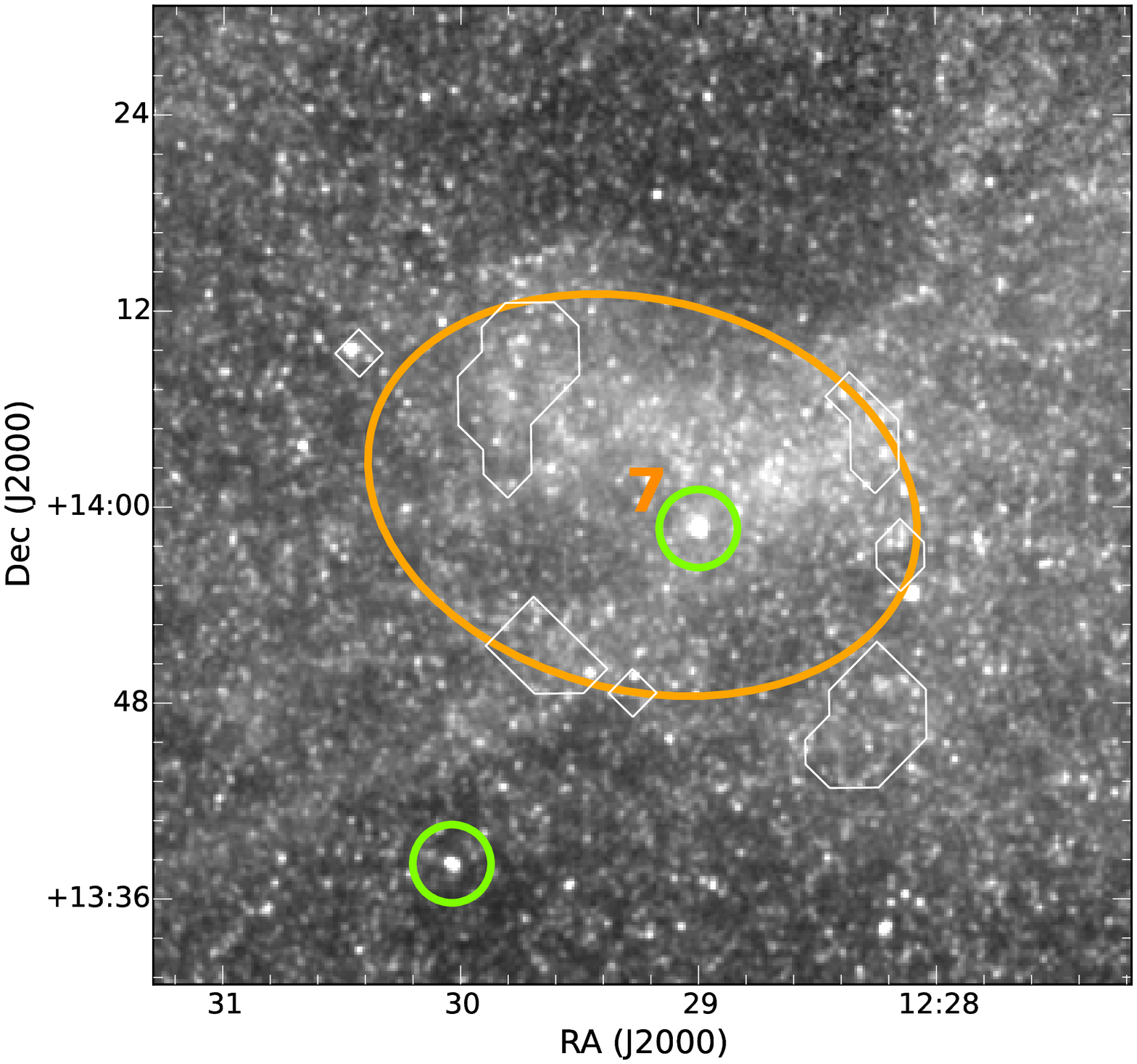}\includegraphics[scale=\scalas]{\figdirc/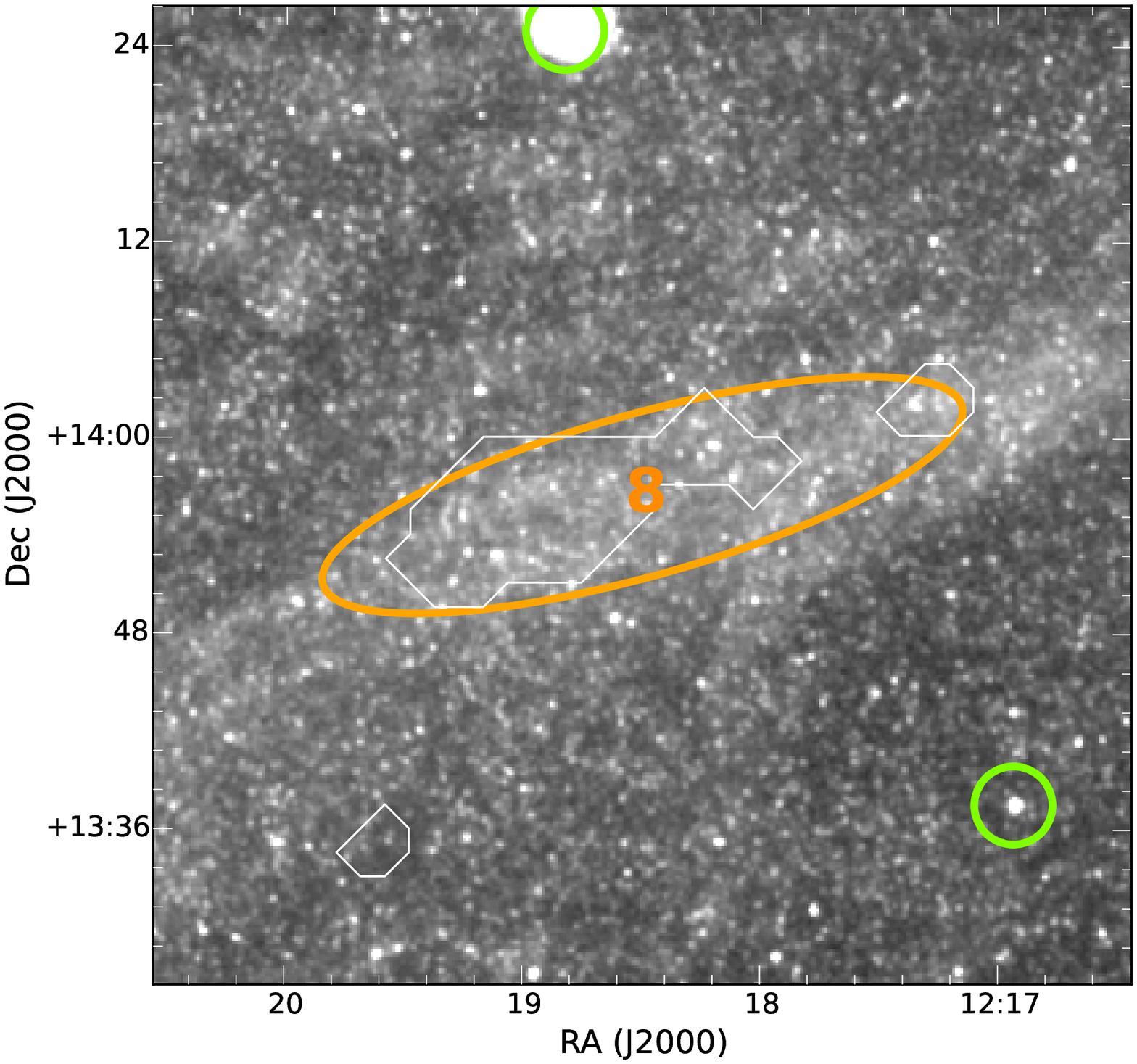}

\includegraphics[scale=\scalas]{\figdirc/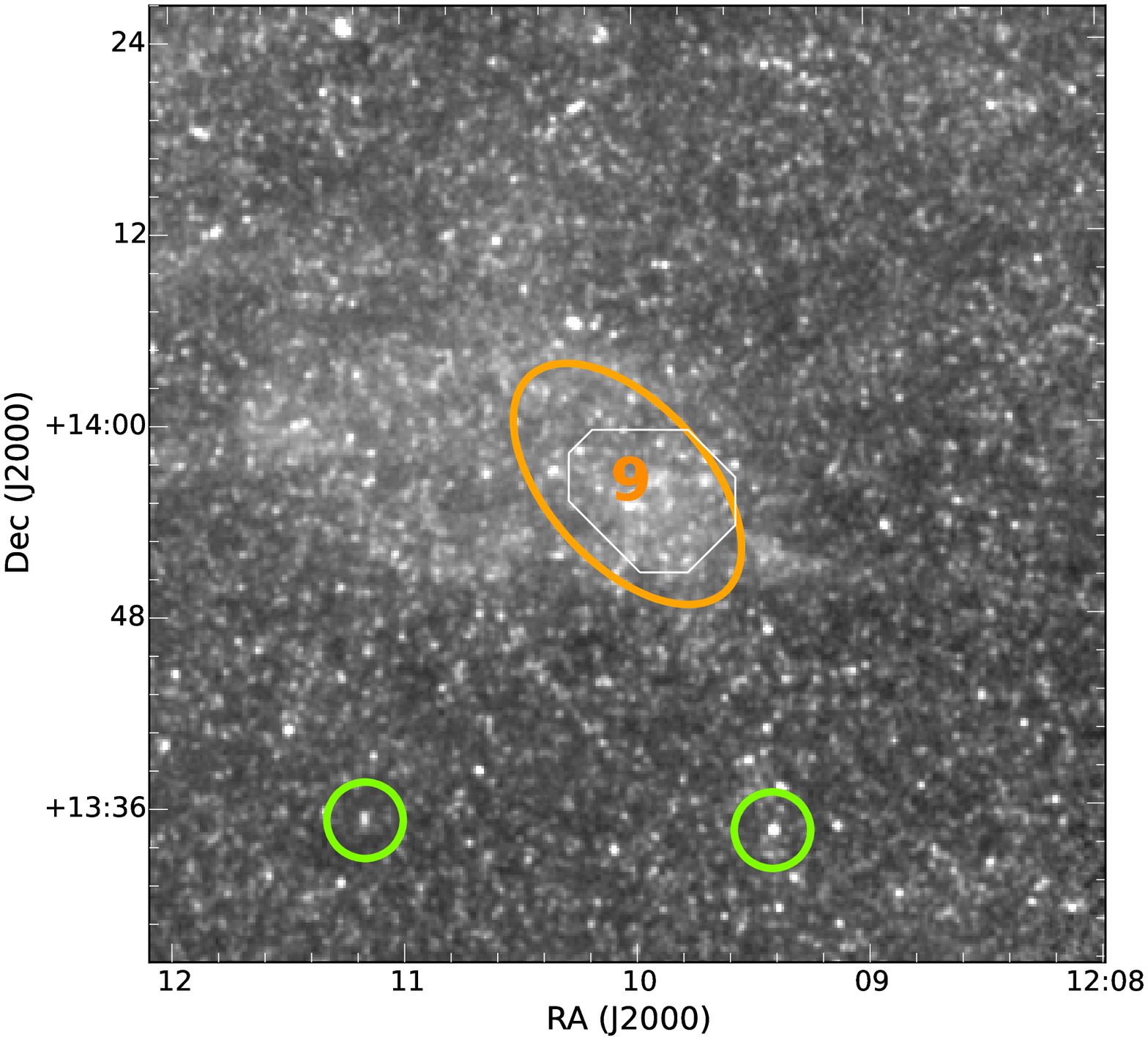}\includegraphics[scale=\scalas]{\figdirc/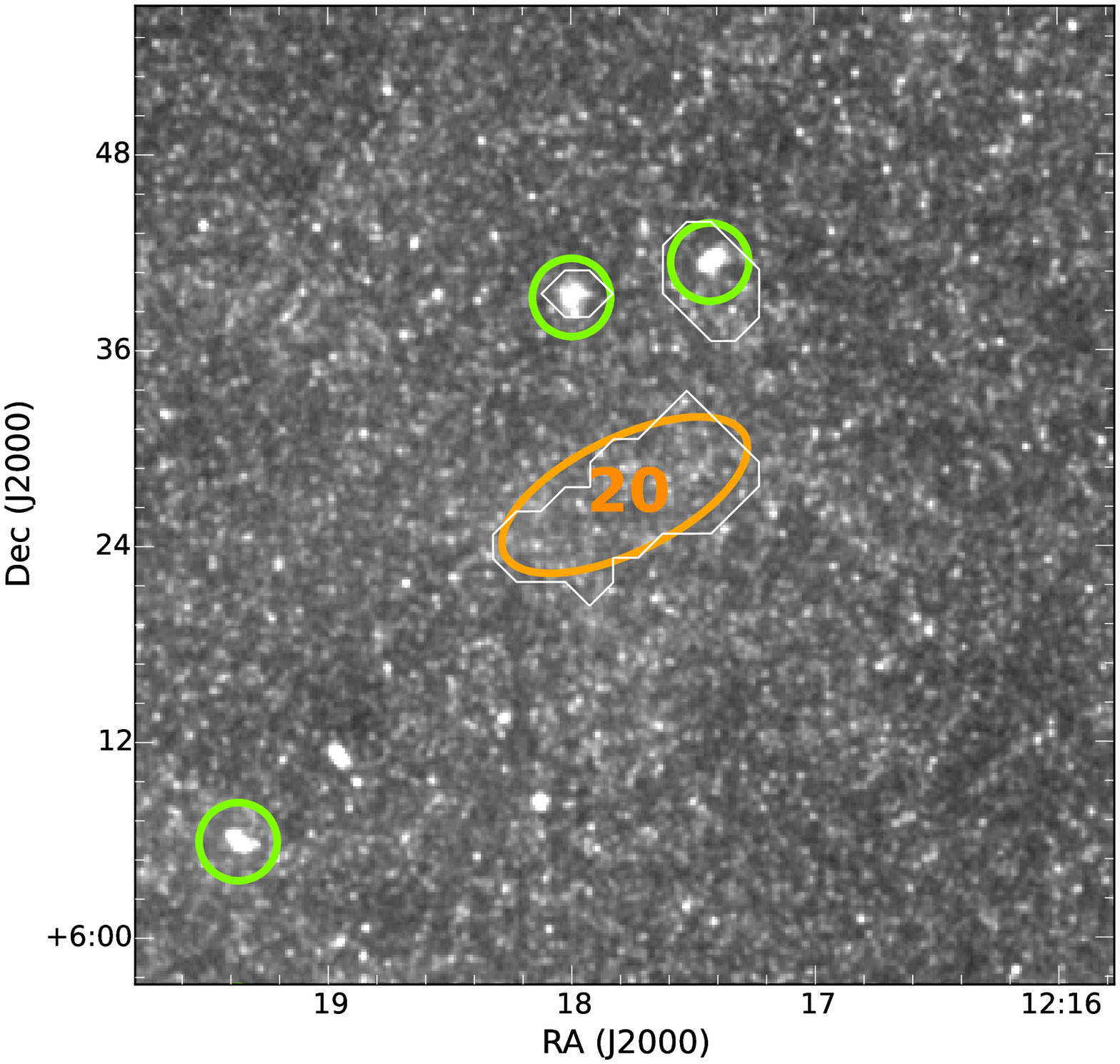}\includegraphics[scale=\scalas]{\figdirc/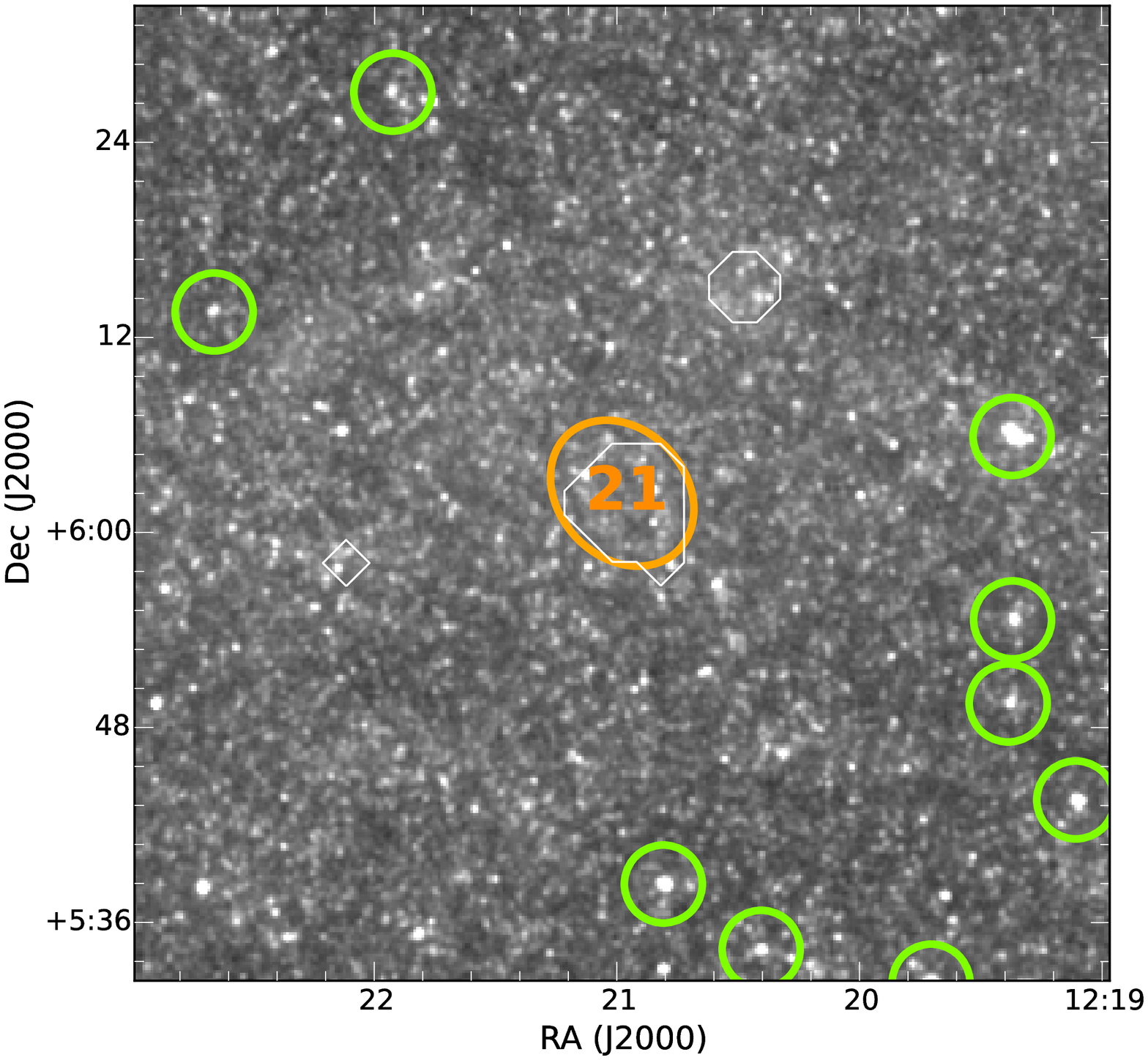}\includegraphics[scale=\scalas]{\figdirc/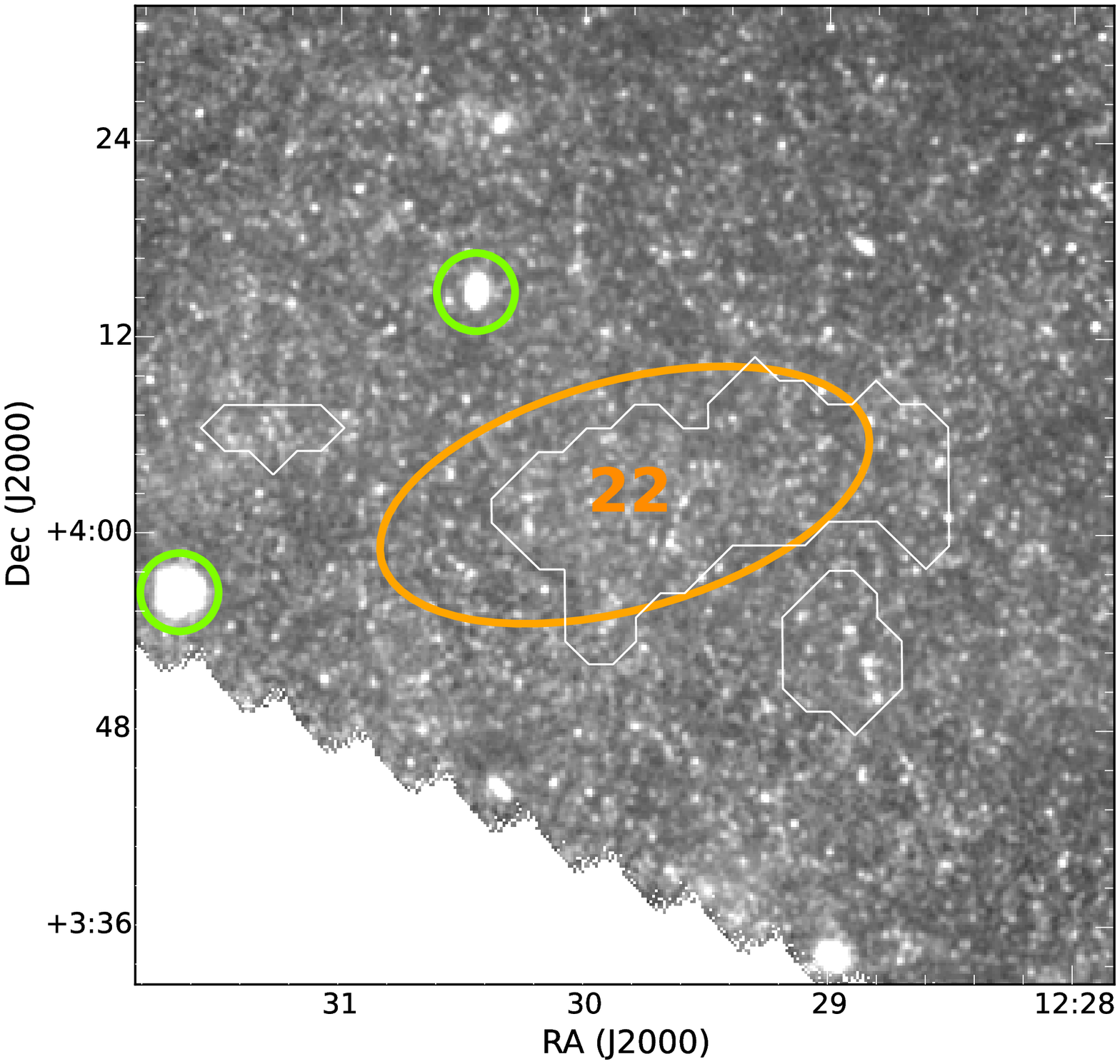}
\caption{Regions with high positive residuals, superimposed on the SPIRE 250~$\mu$m map at the original
resolution. For each object detected by {\tt sextractor}, isophotal ellipses are shown (see 
Table~\ref{table:cata}). White contours correspond to regions masked during the emissivity 
determination. Green circles (diameter = 2 FWHM) show the VCC galaxies detected by \citet{AuldMNRAS2013}.
The size of the R1 panel is $2^\circ \times 2^\circ$, $1^\circ \times 1^\circ$ for all the others.
} 
\label{fig:cata}
\end{figure*}

\subsection{A few examples}

A large group of { high-residual regions} is present in the V2 tile, near to (and
partly in correspondence with) the overlap with V3 (we define this
region R1 in Figs.~\ref{fig:modres250} and \ref{fig:cata}). It 
consists of several clouds interconnected by a filamentary structure,
whose surface brightness peaks have been masked during the emissivity
determination in all but the 850~$\mu$m image. The object with the largest
residual is \# 14 of Table~\ref{table:cata}.
The pixels within its isophotal ellipse are shown as crosses in the 
$250\mu$m dust-$\ion{H}{i}$ correlation panels for V2 (Fig.~\ref{fig:corre1}). 
The excess is clearly associated with the higher column density LVC
gas (appearing as excess emission for the lowest density IVC gas
in the field), though however the region does not lay exactly on the
local LVC peak (the maximum being at the edge of the ellipse, as
shown in the middle panel of Fig.~\ref{fig:modres250}).
Similar excesses have been found in a recent work by \citet{ReachApJ2015},
who study the correlation between $\ion{H}{i}$ column density from the Arecibo 
GALFA survey and dust emission from {\em Planck}. Our work is analogous to 
theirs, though the column densities of the HeViCS fields are smaller
than those of the high latitude clouds studied by \citet{ReachApJ2015}.
They argue that the residuals could be produced by: different local dust
properties; a local increase of the ISM density not traced by $N_\ion{H}{i}$,
due to the presence of molecular gas; an underestimation
of $N_\mathrm{H}$ due to the neglect of $\ion{H}{i}$ self-absorption.
We examine these alternatives below.

\subsubsection{Different dust properties}

\begin{figure}
\includegraphics[width=\hsize]{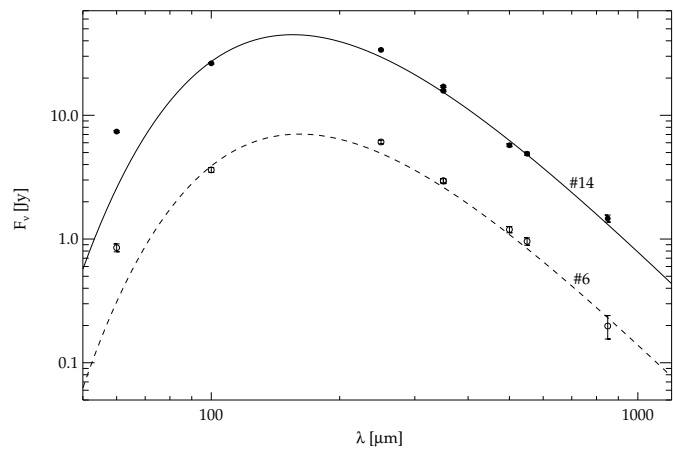}
\caption{The SED of object 14 in field V2 and 6 in V1.
Errorbars for each aperture have been estimated from 
$\sigma_R$ (Table~\ref{table:fit}). Modified blackbody fits assume
$\beta=1.59$ and exclude the 60~$\mu$m data.}
\label{fig:sedhr}
\end{figure}

If one assumes that the LVC $\ion{H}{i}$ column density is a faithful tracer of 
the full ISM density, the pixel-by-pixel correlation of the excess emission 
could be explained by the contribution of dust at higher emissivity
than that for the diffuse medium. However, it is not straightforward
to derive the increase in emissivity without knowing the fraction of
the total gas column density that is associated with the more emissive
dust. If the whole column density under the excess area is associated
with dust at a higher emissivity $\epsilon^\mathrm{LVC}_\nu|_\mathrm{ex}$, 
the residuals can be written as the difference between the emission
of this dust and the emission that is predicted from the whole field,
\[
R_\nu=\epsilon^\mathrm{LVC}_\nu|_\mathrm{ex} N_\ion{H}{i} - \epsilon^\mathrm{LVC}_\nu N_\ion{H}{i},
\]
which leads to
\[
\frac{\left. \epsilon^\mathrm{LVC}_\nu\right|_\mathrm{ex}}{\epsilon^\mathrm{LVC}_\nu} = 
1+\frac{R_\nu}{\epsilon^\mathrm{LVC}_\nu N_\ion{H}{i}}.
\]
At 250~$\mu$m, the excess area has residuals values $R_\nu=$ 0.77 and 
1.8 MJy sr$^{-1}$ for the average and for the peak, respectively,
that would correspond to a moderate increase in emissivity by a factor of
1.3 to 1.8.  In principle, this higher emissivity can be simply due to
a larger dust temperature in the excess region with respect to the 
average temperature derived from the full field. However, the SED of
the residuals within the area is consistent with a temperature close
to the average temperature of the field (T=20.4K for $\beta=1.59$, to be compared with
T=20.5$\pm$0.5 for field V2; see Fig.~\ref{fig:sedhr}). 

For the
same temperature, a higher emissivity corresponds to a higher dust
absorption cross-section.  Larger dust cross-sections are found in 
dense Milky Way clouds: for example \citet{JuvelaA&A2015} measured 
the absorption cross-sections in Galactic cold cores using {\em Planck} 
and {\it Herschel} data, finding values up to more than three times 
those of the diffuse medium.
{
The larger cross-section can result from grain growth due to accretion of
gas atoms (C, H) into mantles \citep{YsardA&A2015,KoehlerA&A2015},
to the formation of ice mantles, and to the coagulation of grains
\citep{OssenkopfA&A1994,KoehlerA&A2015}.
}
However, our excess regions lack the main characteristic defining
dense Galactic cold clouds, i.e.\ the lower temperature with 
respect to the surrounding medium. Also, Galactic cold clumps
have on average higher column densities \citep{PlanckEarlyXXIII};
none of the sources of the {\em Planck}
Catalogue of Galactic Cold Clumps \citep{Planck2015XXVIII}
falls within the HeViCS footprint.

\subsubsection{Molecular gas}

Conversely, if one assumes that the dust properties do not change 
significantly in the limited column density range of the HeViCS fields,
the excess could be explained by untraced molecular gas. 
Indeed, excess emission over the dust/$\ion{H}{i}$ correlation has been used
to detect molecular clouds \citep[see, e.g., ][]{DesertApJ1988,ReachApJ1998}.
Due to the limited resolution of available surveys, these searches only 
detected larger excesses than those measured here: for instance, the
region we are considering is marked by low surface brightness contours
in the 100~$\mu$m excess map of the north Galactic pole in \citet{ReachApJ1998}
and has not been included in their final catalog of detected clouds.

{
Adopting the same emissivity as for atomic gas, the residuals could be 
used to derive the column density of untraced H nucleons, and thus to
estimate the $\mathrm{H}_2$ column density,
}
\[
N_{\mathrm{H}_2} = \frac{1}{2} \frac{R_\nu}{\epsilon^\mathrm{LVC}_\nu}.
\]
For the average and peak column densities of LVC gas in the excess area 
one should expect a molecular gas column density ranging from
$N_{\mathrm{H}_2}= 0.4$ to $1.1\times 10^{20}$ cm$^{-2}$.
Unfortunately, there are no molecular gas surveys at high Galactic
latitude sensitive enough to detect these column density levels.
The HeViCS area has been covered by the North Galactic Hemisphere 
CO(1-0) survey of \citet{HartmannApJ1998}, with a sparsely sampled
$\sim 1^\circ$ x 1$^\circ$ grid. However, in none of the pointings was
CO emission detected, with a 2-$\sigma$ upper limit corresponding 
to $N_{\mathrm{H}_2}=1.2\times 10^{20}$ cm$^{-2}$  (assuming the typical velocity 
width of the clouds detected in the survey and the standard 
$X_{CO} = 2\times 10^{20}$ cm$^{-2}$ (K km s$^{-1}$)$^{-1}$ conversion 
factor for Galactic molecular clouds; \citealt{BolattoARA&A2013}).
Besides having a detection limit close to that inferred from our
highest excess region, the grid of pointings (with beam FWHM = 8$\farcm$4) 
misses all regions with high residuals. The fully sampled CO(1-0) map 
derived from {\em Planck} observations cannot be of help either, being featureless 
within the HeViCS footprint \citep[we used the type-3 map of ][]{Planck2013XIII};
it also has a 2-$\sigma$ upper limit 4 times higher than that of 
\citet{HartmannApJ1998}.

{
\citet{GillmonApJ2006} studied the correlation between IRAS surface 
brightnesses and molecular gas column densities derived from UV absorption lines
in a sample of high-latitude AGNs; they concluded that significant fractions
of $\mathrm{H}_2$ are present in most diffuse cirrus clouds. 
Unfortunately, the limited dynamic range of our residuals does not help in 
understanding if the excess regions are just the tip of diffuse molecular
clouds or isolated cores (3-$\sigma_R$ corresponding to about $0.3\times 10^{20}$ 
cm$^{-2}$, a third of the estimate for the peak value). Molecular gas with
$N_{\mathrm{H}_2}\approx2-3\times 10^{20}$ cm$^{-2}$ was detected by 
\citet{CorteseMNRAS2010} in correspondence to a 5$\arcmin$ plume to the north 
of the interacting pair NGC4435/NGC4438 and interpreted as a compact cirrus cloud
(the plume is detected in HeViCS images, but it is masked in our analysis due to 
the proximity to the bright galaxies).
Compact molecular clouds with $N_{\mathrm{H}_2}\la 1\times 10^{20}$ cm$^{-2}$
have been found in a few limited fields at high Galactic latitude 
\citep[Small Area Molecular Structures - SAMS; ][]{HeithausenA&A2002,HeithausenProc2007}.
}
Dust emission associated with SAMS has been detected in SPIRE images
\citep{HeithausenA&A2012}, and indeed for one of them dust emissivity
in excess of that for the diffuse medium has been reported 
\citep{DaviesMNRAS2010}. The SAMS studied so far are however smaller,
with typical sizes of 1-2$\arcmin$; yet they tend to be grouped in clusters
\citep{HeithausenProc2007}. Thus, our excess region, which appears
more extended and filamentary also at the original 250~$\mu$m resolution
could be a web of connected SAMS. \citet{HeithausenA&A2002} claims
that SAMS are transitory objects, as they are not sufficiently shielded from UV
radiation to survive for long time \citep{HeithausenA&A2002,HeithausenProc2007}.
Yet, \citet{GillmonApJ2006} found that H$_2$ self-shielding can already occur for
N$_\mathrm{H_2} \ga $ 10$^{18}$ cm$^{-2}$ in regions with $I_\nu \ga$ 2 MJy sr$^{-1}$
at 100~$\mu$m. These surface brightness levels are indeed consistent with
those found here.

\subsubsection{$\ion{H}{i}$ self-absorption}

Instead, if the excess is due to the neglect of $\ion{H}{i}$ self-absorption 
in the estimate of the gas column density, the total $\ion{H}{i}$ column density 
should be 30 to 80\% larger than what is actually measured in the optically
thin limit for LVC gas (the estimate depending on the use of average or peak
values, respectively). The optical depth at the line center would need to range 
between 0.8 and 1.8 to explain the excess \citep[Eq. 15 in][]{ReachApJ2015}. 
These optical depths are larger than those derived from \ion{H}{i} absorption 
(see Sect.~\ref{sec:alfalfa}); yet, none of the probed lines-of-sight falls on 
a high residual region. The values estimated here are smaller than those 
derived in the higher density regions analyzed by \citet{ReachApJ2015};
{
even if they are seldom encountered at high Galactic latitude \citep[less than
4\% of $b > 30^\circ$ lines of sight in][have higher optical depths]{HeilesApJS2003}
the coverage fraction of the residuals makes this hypothesis compatible with
the absorption line observations. However, it is not clear why
the excess in object \# 14 should be due to $\ion{H}{i}$ self-absorption, while the
peak of $N^\mathrm{LVC}_\ion{H}{i}$ (at the northern edge of the
selected area) does not show a similarly high residual.}
\newline

{
\noindent Summarizing, we have no elements to favour one of the proposed explanations for
the excesses over the other. We are tempted to exclude $\ion{H}{i}$ self-absorption,
because of the mismatch between high column densities and high optical depth
estimates; and to favour the molecular gas explanation, because of the detection 
of similar column densities in other fields. Also, the latter hypothesis could
be more readily tested with future dedicated observations of molecular lines 
over the areas of the excess regions.
}

As another example, we have considered a region in tile V1,  
(R2 in Figs.~\ref{fig:modres250} and \ref{fig:cata}). It consists
of a few clumps, three of which are aligned over a filament. 
Among them we have chosen source 6 of Table~\ref{table:cata}.
As for the previous example, the SED (dashed line in Fig.~\ref{fig:sedhr})
is compatible with dust at a temperature (T=19.7K) very close to the field 
average. The pixel-by-pixel correlation is shown by the symbols in the 
250~$\mu$m panel for field V1 in Fig.~\ref{fig:corre1}. The excess is again 
associated with LVC gas, though not with the highest column densities available in 
the field. The residuals range from $R_\nu$ = 0.7 (median) to 1.1 MJy sr$^{-1}$
(peak), which could result from either a dust emissivity a factor 1.5 to 1.7
higher than the average for the V1 field, to the presence of molecular gas with 
column density $N_{\mathrm{H}_2}=$ 0.5 to 0.7 $\times 10^{20}$ cm$^{-2}$, or
to a $\ion{H}{i}$ line central optical depth between 1.1 and 1.6. 

The properties of the other excess regions in the field are similar. 
For all the sources we have computed the change in dust emissivity,
the H$_2$ column density, or the \ion{H}{i} opacity which could explain
the residuals: these calculations are included in  Table~\ref{table:cata}. 

\subsection{Negative residuals}
\label{sec:negative}

So far, we discussed only the case of high positive residuals.
There are, however, regions with negative residuals, in some cases
so large (in absolute value) that they have been masked during the 
emissivity estimate. One such case is indicated by the red circle
labelled 'N'
in Fig.~\ref{fig:modres250} (RA=189.02458$^\circ$, dec=12.389513$^\circ$), whose pixels are shown as diamonds 
in the V2 $250\mu$m panel of Fig.~\ref{fig:corre1}. The negative
residual corresponds to a 15$\arcmin$x30$\arcmin$ cloud in the LVC map,
{ sharing the same velocity range as the neighborhood}. The structure
is a moderate enhancement of the neighboring gas column density,
with $N_{\mathrm{H}}\approx 0.5 \times 10^{20}$ cm$^{-2}$ 
on top of a "background"
column density $N_{\mathrm{H}}=$ 2.3 $\times 10^{20}$ cm$^{-2}$.
Most of the diffuse emission in the area comes from dust associated
with this background, and this explains why the selected pixels
in Fig.~\ref{fig:corre1} still show the same correlation as for the
diffuse field. However, the emissivity is significantly lower
than the mean: these residuals could be either explained by 
the absence of dust in the cloud (the {\em missing} emission
indeed corresponds to the amount of dust one would expect from the 
$\ion{H}{i}$ density enhancement for the average field emissivity)
or to a much smaller emissivity than the field, resulting from a colder 
dust temperature in the cloud. Although both interpretation are
heuristically similar, a colder dust temperature seems to be
excluded, since the negative residuals in this area can be seen also at
the longer wavelengths.

\begin{figure}
\includegraphics[width=\hsize]{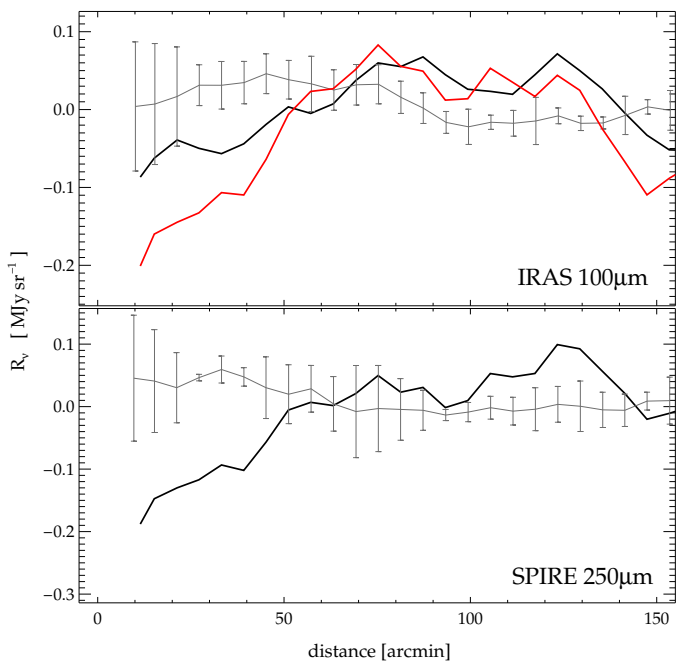}
\caption{Profile of the 100 and $250\mu$m residuals as a function of the
projected distance from M87 for the V2 field (black solid line). The gray
lines and error bars are the average and standard deviation of the profiles
from the center of each of the other three HeViCS tiles.
The red solid line is the profile at
100~$\mu$m predicted from the $250\mu$m data and assuming the average
100/250 color ratio (see text for details). Each profile have been obtained by
azimuthally averaging the residuals over circular coronas of width 6$\arcmin$.
}
\label{fig:profm87}
\end{figure}

At a lower level, negative (and positive) residuals can be found in all
fields and bands, as expected from the scatter in the $I_\nu$ / $N_{\mathrm{H}}$
correlation and from the anisotropies of the CIB 
\citep[see, e.g.\ ][]{Planck2013XXX}. A curious case is, however,
that of the V2 field, where a large area of negative residuals is present,
coincidentally, around the center of the background Virgo cluster { and
inside of the cirrus ring feature}. 
{
This is shown in Fig.~\ref{fig:profm87}, where 
we plot the profiles of the 100 and $250\mu$m residuals as a function of the 
projected distance from the galaxy M87, at the center of the cluster (black lines), 
and we compare them with the flatter profiles obtained with the respect to the center 
in each of the other three HeViCS tiles (the gray lines and error bars show their
average and standard deviation).
}
The lower surface brightness inside the ring feature, when compared
to the average level outside of it (in V2 and in the southern HeViCS tiles),
appears as a "hole". Its presence
was recognised from the first HeViCS analysis, and was 
later evident also in the full sky IRAS and {\em Planck} surveys,
from 100~$\mu$m to 550~$\mu$m. 

It is not straightforward to understand whether this residual is due to a
{ local large-scale} underdensity in the CIB, or to different dust properties:
if the reason is the latter, the residual could result from a lower
emissivity (and thus lower cross-section, since the temperature estimated within
1$^\circ$ from M87, T=20.8K for $\beta=1.59$, is very close to that for the 
LVC component) in a region with low gas column density 
($N_\ion{H}{i}\approx 10^{20}$ cm$^{-2}$ for the LVC). This is
in contrast, however, with the {\em excess} of emissivity seen at IRAS and {\em Planck}
wavelelength for $N_{\mathrm{H}} < 10^{20}$ cm$^{-2}$, which could be due
to variations in the dust composition and size distribution at the lower
column densities \citep{YsardA&A2015}, or to the contribution of dust associated 
with diffused ionised gas \citep{Planck2013XI}.
{
On the contrary, the "hole" could be due to \ion{H}{ii}-associated dust
(which we did not account for in our analysis; Sect.~\ref{sec:ana}) {\em only}
if the column density of ionised gas is smaller inside of the feature than outside 
of it. However, the feature would appear as a significant dearth of H$\alpha$ 
emission even in the available large-scale surveys such as WHAM \citep{HaffnerApJS2003},
which is not the case.
}

\section{Upper limits on intra-cluster dust emission}
\label{sec:icd}

{
Despite the topic might seem out of the scope of this paper,
one of the aim of HeViCS (and of the cirrus analysis) was
the detection of emission from dust in the ICM.
Dust grains could be removed from cluster galaxies and transported to the
ICM by the same environmental processes (such as ram pressure stripping 
and tidal interaction{; for a review, see \citealt{BoselliPASP2006}}) 
which remove their gas content, or by internal process, such as 
global galactic winds \citep{PopescuA&A2000a}.  The dust
passed to the ICM could in principle be detected by its reddening
of background galaxies and QSOs. There is a long history of attempts
at detecting such cluster reddening, but the results are still
controversial and inconclusive 
(for instance, compare the detection claimed by \citealt{McGeeMNRAS2010} 
with the absence of significative excess reddening in 
\citealt{GutierrezA&A2014}).

A key issue is the ability of dust to survive destruction in the hot ICM gas 
\citep{DwekApJ1990}. For the Virgo cluster, when only environmental processes 
are considered, FIR emission would come predominantly from dust produced in 
the winds of evolved stars \citep{PopescuA&A2000a}.
In this case, the predicted surface brightness is very low, 0.01 MJy sr$^{-1}$ 
at 100~$\mu$m. However, in a less prohibitive scenario, in
which dust is injected by galactic winds, a larger FIR 
contribution could come from galaxies originally on the outskirt of
the cluster and falling towards the center.
If the dust is able to infall within 1$^\circ$ from M87,
it could produce levels up to $\la$ 0.1 MJy sr$^{-1}$ at 
175~$\mu$m (case B of \citealt{PopescuA&A2000a}).

The complex structure of the residuals around M87, and in particular the 
large scale negative feature discussed in Sect.~\ref{sec:negative}, makes 
it complicate to detect ICM dust in Virgo. In fact,
contamination from the foreground cirrus has long been recognised to be the 
limiting factor in detecting diffuse FIR emission from clusters 
\citep{WiseApJ1993,StickelA&A2002,BaiApJL2007,KitayamaApJ2009}.

Following \citet{StickelA&A1998}, we have attempted to measure
an excess in Virgo at 100~$\mu$m with respect to 250~$\mu$m.
We used the 250~$\mu$m V2 map to estimate the 100~$\mu$m
emission that would be expected on the basis of the average 100/250
color ratio (1.025, measured on field V2). This estimate is shown by the
red line in Fig.~\ref{fig:profm87}. Indeed, the profile of the
{\em true} residuals reveals a 100 $\mu$m emission in excess of the 
predicted one.
The excess is $\approx$0.1 MJy sr$^{-1}$
in the cluster center, a value close to the detection claimed by
\citet{StickelA&A1998} in Coma; to the models of \citet{DwekApJ1990} 
for the Coma cluster; and to the more optimistic predictions of 
intracluster dust emission in Virgo by \citet{PopescuA&A2000a}.  
It is also close, however, to the upper limits of 0.06 MJy sr$^{-1}$
estimated using {\em Spitzer} data at 70 and 160~$\mu$m by 
\citet{KitayamaApJ2009}, who dismissed the dust detection in the
Coma cluster by \citet{StickelA&A1998} as due to cirrus contamination.
We are not able therefore to confirm that
the excess is due to cluster emission, since it is of the same
order as the noise in the residuals and in the random profiles
centered on the other fields (gray lines and error bars
in Fig.~\ref{fig:profm87})
which are caused by both variations in the dust emissivity
and by the variance of the CIB \citep{Planck2013XI}. The
issue is further complicated by the decorrelation between
the 100$\mu$m CIB power spectrum and that at longer
wavelengths \citep{Planck2013XXX}.

Notably, most of the truly intergalactic features (optically
highlighted by \citealt{MihosApJL2005})
cannot be seen in dust emission. The only exception is the plume
to the north of NGC4435/NGC4438 which \citet{CorteseMNRAS2010}
found to be most likely a foreground cirrus cloud rather than a
tidal tail in the cluster.  In fact, the only
features of the deep optical survey of Virgo presented 
by \citet{MihosProc2015} clearly associated with emission
in our maps are those corresponding to the highest column densities
of the Galactic $\ion{H}{i}$.
}

\section{Summary and conclusions}
\label{sec:sum}

We have studied the correlation between FIR emission and
atomic gas column density for the low density, high Galactic
latitude cirrus in the foreground of the Virgo Cluster.
For the dust emission, we used the HeViCS maps at 250, 350 
and 500~$\mu$m obtained with the SPIRE instrument aboard the 
{\it Herschel} Space Observatory \citep{DaviesA&A2010,DaviesMNRAS2012,AuldMNRAS2013}; 
maps at 60 and 100~$\mu$m obtained from the IRAS-IRIS dataset 
\citep{MivilleDeschenesApJS2005} and HFI maps at 350, 550 and 
850~$\mu$m from the 2013 {\em Planck} data release \citep{Planck2013I}. 
Two $\ion{H}{i}$ column density maps were obtained from the 
ALFALFA survey \citep{GiovanelliAJ2005}, separating an IVC
component with emission at velocities in the range
$-100 < v_\mathrm{LSR}/$ km s$^{-1} < -20$ and a LVC
component with $-20 < v_\mathrm{LSR}/$ km s$^{-1} < 100$.
For each band and for each of the four HeViCS fields, we
have derived the dust emissivity associated with the two
gas components. Our results can be summarised as follows:

\begin{itemize}

\item The values of the emissivities we found for dust
associated with the LVC are consistent with the determinations 
by \citet{Planck2013XI} and \citet{PlanckIntermediateXVII}. 
In the SPIRE bands we measure
$\epsilon^\mathrm{LVC}_\nu = (0.787\pm0.077)$, $(0.395\pm0.034)$ and
$(0.171\pm0.032) \times 10^{-20}$ MJy sr$^{-1}$ cm$^2$ at 250, 350
and 500~$\mu$m, respectively. The values represent the average
and the standard deviation of the emissivities derived for each
of the $\approx 4^\circ \times 4^\circ$ HeViCS fields. The scatter 
{ in the emissivity values} is generally larger than the 
uncertainties estimated for each individual field and points
toward intrinsic field-to-field variations. Indeed, the scatter
measured in the analysis of {\em Planck} data (covering a much larger
range of column density than HeViCS) suggests variation of
the dust properties along different lines of sight and environments
\citep{YsardA&A2015,FanciulloA&A2015}. Yet it is remarkable that 
the average values derived for the HeViCS footprint, covering just
0.2\% of the sky, are in excellent agreement (see 
Fig.~\ref{fig:sedemi_avg}) with the analysis of high latitude dust 
emission by the {\em Planck} team, covering from 20\% 
\citep{PlanckIntermediateXVII} to 50\% \citep{Planck2013XI} of the sky.

\item By fitting the average emissivity SED with a MBB,
we derived a dust absorption cross-section
\[
\frac{\tau^\mathrm{LVC}_\nu}{N_{\ion{H}{i}}} = (0.49\pm0.13)
\times
\left(\frac{250\mu\mathrm{m}}{\lambda}\right)^{1.53\pm0.17} 
\times 10^{-25} \text{cm}^2 \, \text{H}^{-1}
\]
with an average dust temperature $T=20.4\pm1.5$K. Again, this
is consistent with the values derived by 
\citet{Planck2013XI} and \citet{PlanckIntermediateXVII}. In particular
the planet-calibrated SPIRE data confirms the lower emissivity,
absorption cross-section (and spectral index) with respect to the FIRAS spectrum
and calls for a revisitation of the FIRAS-calibrated dust models
\citep[like, e.g., ][]{DraineARA&A2003,ZubkoApJS2004,CompiegneA&A2011,JonesA&A2013}.
Nevertheless, we found that the dust absorption cross-section used in most of 
the HeViCS papers results in only a slight underestimate of 
the dust mass in a galaxy with respect to those obtained with
the value above ({ see Appendix~\ref{app:extgal}}).

\item Dust associated with the IVC component has lower emissivity
than the LVC component, 
$\epsilon^\mathrm{IVC}_\nu = (0.408\pm0.054)$, $(0.186\pm0.026)$ and
$(0.0588\pm0.00023) \times 10^{-20}$ MJy sr$^{-1}$ cm$^2$ at 250, 350
and 500~$\mu$m; it also has a bluer spectrum, resulting in a hotter
dust temperature when fitted with a MBB, and in a lower dust cross-section.
These results confirm the early analysis of {\em Planck} data for high latitude
clouds, and point towards different properties of dust in the
IVC with respect to the LVC \citep{PlanckEarlyXXIV}.

\item The high resolution of the ALFALFA data allowed us to study
the departure of the $\ion{H}{i}$-dust correlation at scales smaller
that $\sim 20\arcmin$. { This corresponds to (uncertain) linear 
sizes of $0.6-0.9$~pc, assuming as a distance the scaleheight of 
the global MW emission from cold dust \citep[100~pc][]{MisiriotisA&A2006}
or the scaleheight of the \ion{H}{i} disk \citep[150~pc][]{KalberlaARA&A2009},
respectively.
}
We found regions with residuals significantly higher than the average scatter 
of the correlation.
The high residual regions are most
probably the result of local variation in the dust properties (increase 
in the dust emissivity by 20 to 80\%) or excess emission associated with 
the small molecular clouds of average column density
$N_{\mathrm{H}_2} \lesssim 10^{20}$ cm$^{-2}$. We also found regions
of negative residuals, and in particular a $\ion{H}{i}$ feature 
of the LVC component with no associated dust emission. This parcel of 
gas devoid of dust seems to suggest that, besides changes in the grain 
composition and size distribution, the scatter in the emissivity 
measurements could be due to an imperfect spatial correspondence 
between the ISM tracers at these smaller scales; on the other
hand, dedicated observations of molecular gas tracers are needed to 
see if the excess emission can be due to (so far) untraced gas or to 
dust clouds devoid of a gas component.

\item Due to the cirrus contamination and large residuals in the
proximity of the center of the the Virgo cluster, we  
{
find no strong evidence for dust emission from the ICM.
In fact, even for the most optimistic prediction for the Virgo cluster
\citep{PopescuA&A2000a} the surface brightness is not likely to exceed
$\approx$0.1 MJy sr$^{-1}$ at 100~$\mu$m, a level which is the
same as the average scatter of the residuals, dominated in most
cases by CIB fluctuations.
}

\end{itemize}
{
This work was dedicated to the study of the cirrus absorption properties in
the FIR/submm. The availability of deep UV \citep{BoissierA&A2015},
optical \citep{FerrareseApJS2012,MihosProc2015} and atomic gas 
\citep{GiovanelliAJ2005,PeekApJS2011} surveys of the Virgo cluster
will also allow in the future a study of the emissivity of scattered starlight, 
thus posing further constraints on the dust grain models for the high
Galactic latitude cirrus.
}

\begin{acknowledgements}

We are grateful to Edvige Corbelli, Luca Cortese, Emmanuel Xilouris
and an anonymous referee
for comments that improved the quality of the paper.

S. B., J. I. D., M. Baes, M. Bocchio, V. C., C. J. R. C., A. J., S. M. acknowledge
support from the European Research Council (ERC) in the form of the FP7 
project DustPedia (P.I. Jonathan Davies, proposal 606824).

S. B., C. G., L. K. H.  are grateful for support from PRIN-INAF 2012/13.

The ALFALFA team at Cornell was supported by the NSF grants AST-0607007 and AST-1107390 to
R. G. and M. H. and by grants of the Brinson Foundation to M.H.

S.Z. has been supported by the EU Marie Curie Career Integration Grant "SteMaGE" Nr. PCIG12-GA-2012-326466  (Call Identifier: FP7-PEOPLE-2012 CIG).

Based on observations obtained with {\em Planck} (http://www.esa.int/Planck), an ESA science mission with instruments and contributions directly funded by ESA Member States, NASA, and Canada.

This research has made use of: APLpy, an open-source plotting package for Python hosted at http://aplpy.github.com;
the NASA/IPAC Extragalactic Database (NED) which is operated by the Jet Propulsion Laboratory, 
California Institute of Technology, under contract with the National Aeronautics and Space Administration;
NASA's Astrophysics Data System.

\end{acknowledgements}

\bibliographystyle{aa}
\bibliography{/Users/sbianchi/Documents/tex/DUST}

\begin{thebibliography}{111}
\expandafter\ifx\csname natexlab\endcsname\relax\def\natexlab#1{#1}\fi

\bibitem[{{Auld} {et~al.}(2013){Auld}, {Bianchi}, {Smith}, {Davies}, {Bendo},
  {di Serego}, {Cortese}, {Baes}, {Bomans}, {Boquien}, {Boselli}, {Ciesla},
  {Clemens}, {Corbelli}, {De Looze}, {Fritz}, {Gavazzi}, {Pappalardo},
  {Grossi}, {Hunt}, {Madden}, {Magrini}, {Pohlen}, {Verstappen}, {Vlahakis},
  {Xilouris}, \& {Zibetti}}]{AuldMNRAS2013}
{Auld}, R., {Bianchi}, S., {Smith}, M.~W.~L., {et~al.} 2013, MNRAS, 428, 1880

\bibitem[{{Bai} {et~al.}(2007){Bai}, {Rieke}, \& {Rieke}}]{BaiApJL2007}
{Bai}, L., {Rieke}, G.~H., \& {Rieke}, M.~J. 2007, ApJL, 668, L5

\bibitem[{{Bertin} \& {Arnouts}(1996)}]{BertinA&AS1996}
{Bertin}, E. \& {Arnouts}, S. 1996, A\&AS, 117, 393

\bibitem[{{Bertin} {et~al.}(2002){Bertin}, {Mellier}, {Radovich}, {Missonnier},
  {Didelon}, \& {Morin}}]{swarp}
{Bertin}, E., {Mellier}, Y., {Radovich}, M., {et~al.} 2002, in Astronomical
  Society of the Pacific Conference Series, Vol. 281, Astronomical Data
  Analysis Software and Systems XI, ed. D.~A. {Bohlender}, D.~{Durand}, \&
  T.~H. {Handley}, 228

\bibitem[{{Bertincourt} {et~al.}(2016){Bertincourt}, {Lagache}, {Martin},
  {Schulz}, {Conversi}, {Dassas}, {Maurin}, {Abergel}, {Beelen}, {Bernard},
  {Crill}, {Dole}, {Eales}, {Gudmundsson}, {Lellouch}, {Moreno}, \&
  {Perdereau}}]{BertincourtA&A2016}
{Bertincourt}, B., {Lagache}, G., {Martin}, P.~G., {et~al.} 2016, A\&A, 588,
  A107

\bibitem[{{B{\'e}thermin} {et~al.}(2012){B{\'e}thermin}, {Le Floc'h}, {Ilbert},
  {Conley}, {Lagache}, {Amblard}, {Arumugam}, {Aussel}, {Berta}, {Bock},
  {Boselli}, {Buat}, {Casey}, {Castro-Rodr{\'{\i}}guez}, {Cava}, {Clements},
  {Cooray}, {Dowell}, {Eales}, {Farrah}, {Franceschini}, {Glenn}, {Griffin},
  {Hatziminaoglou}, {Heinis}, {Ibar}, {Ivison}, {Kartaltepe}, {Levenson},
  {Magdis}, {Marchetti}, {Marsden}, {Nguyen}, {O'Halloran}, {Oliver}, {Omont},
  {Page}, {Panuzzo}, {Papageorgiou}, {Pearson}, {P{\'e}rez-Fournon}, {Pohlen},
  {Rigopoulou}, {Roseboom}, {Rowan-Robinson}, {Salvato}, {Schulz}, {Scott},
  {Seymour}, {Shupe}, {Smith}, {Symeonidis}, {Trichas}, {Tugwell}, {Vaccari},
  {Valtchanov}, {Vieira}, {Viero}, {Wang}, {Xu}, \&
  {Zemcov}}]{BetherminA&A2012}
{B{\'e}thermin}, M., {Le Floc'h}, E., {Ilbert}, O., {et~al.} 2012, A\&A, 542,
  A58

\bibitem[{{Bianchi}(2013)}]{BianchiA&A2013}
{Bianchi}, S. 2013, A\&A, 552, A89

\bibitem[{{Bianchi} {et~al.}(2000){Bianchi}, {Davies}, \&
  {Alton}}]{BianchiA&A2000b}
{Bianchi}, S., {Davies}, J.~I., \& {Alton}, P.~B. 2000, A\&A, 359, 65

\bibitem[{{Bocchio} {et~al.}(2014){Bocchio}, {Jones}, \&
  {Slavin}}]{BocchioA&A2014}
{Bocchio}, M., {Jones}, A.~P., \& {Slavin}, J.~D. 2014, A\&A, 570, A32

\bibitem[{{Bocchio} {et~al.}(2013){Bocchio}, {Jones}, {Verstraete}, {Xilouris},
  {Micelotta}, \& {Bianchi}}]{BocchioA&A2013}
{Bocchio}, M., {Jones}, A.~P., {Verstraete}, L., {et~al.} 2013, A\&A, 556, A6

\bibitem[{{Boissier} {et~al.}(2015){Boissier}, {Boselli}, {Voyer}, {Bianchi},
  {Pappalardo}, {Guhathakurta}, {Heinis}, {Cortese}, {Duc}, {Cuillandre},
  {Davies}, \& {Smith}}]{BoissierA&A2015}
{Boissier}, S., {Boselli}, A., {Voyer}, E., {et~al.} 2015, A\&A, 579, A29

\bibitem[{{Bolatto} {et~al.}(2013){Bolatto}, {Wolfire}, \&
  {Leroy}}]{BolattoARA&A2013}
{Bolatto}, A.~D., {Wolfire}, M., \& {Leroy}, A.~K. 2013, ARA\&A, 51, 207

\bibitem[{{Boselli} {et~al.}(2010{\natexlab{a}}){Boselli}, {Ciesla}, {Buat},
  {Cortese}, {Auld}, {Baes}, {Bendo}, {Bianchi}, {Bock}, {Bomans}, {Bradford},
  {Castro-Rodriguez}, {Chanial}, {Charlot}, {Clemens}, {Clements}, {Corbelli},
  {Cooray}, {Cormier}, {Dariush}, {Davies}, {de Looze}, {di Serego Alighieri},
  {Dwek}, {Eales}, {Elbaz}, {Fadda}, {Fritz}, {Galametz}, {Galliano},
  {Garcia-Appadoo}, {Gavazzi}, {Gear}, {Giovanardi}, {Glenn}, {Gomez},
  {Griffin}, {Grossi}, {Hony}, {Hughes}, {Hunt}, {Isaak}, {Jones}, {Levenson},
  {Lu}, {Madden}, {O'Halloran}, {Okumura}, {Oliver}, {Page}, {Panuzzo},
  {Papageorgiou}, {Parkin}, {Perez-Fournon}, {Pierini}, {Pohlen}, {Rangwala},
  {Rigby}, {Roussel}, {Rykala}, {Sabatini}, {Sacchi}, {Sauvage}, {Schulz},
  {Schirm}, {Smith}, {Spinoglio}, {Stevens}, {Sundar}, {Symeonidis}, {Trichas},
  {Vaccari}, {Verstappen}, {Vigroux}, {Vlahakis}, {Wilson}, {Wozniak},
  {Wright}, {Xilouris}, {Zeilinger}, \& {Zibetti}}]{BoselliA&A2010}
{Boselli}, A., {Ciesla}, L., {Buat}, V., {et~al.} 2010{\natexlab{a}}, A\&A,
  518, L61

\bibitem[{{Boselli} {et~al.}(2012){Boselli}, {Ciesla}, {Cortese}, {Buat},
  {Boquien}, {Bendo}, {Boissier}, {Eales}, {Gavazzi}, {Hughes}, {Pohlen},
  {Smith}, {Baes}, {Bianchi}, {Clements}, {Cooray}, {Davies}, {Gear}, {Madden},
  {Magrini}, {Panuzzo}, {Remy}, {Spinoglio}, \& {Zibetti}}]{BoselliA&A2012}
{Boselli}, A., {Ciesla}, L., {Cortese}, L., {et~al.} 2012, A\&A, 540, A54

\bibitem[{{Boselli} {et~al.}(2010{\natexlab{b}}){Boselli}, {Eales}, {Cortese},
  {Bendo}, {Chanial}, {Buat}, {Davies}, {Auld}, {Rigby}, {Baes}, {Barlow},
  {Bock}, {Bradford}, {Castro-Rodriguez}, {Charlot}, {Clements}, {Cormier},
  {Dwek}, {Elbaz}, {Galametz}, {Galliano}, {Gear}, {Glenn}, {Gomez}, {Griffin},
  {Hony}, {Isaak}, {Levenson}, {Lu}, {Madden}, {O'Halloran}, {Okamura},
  {Oliver}, {Page}, {Panuzzo}, {Papageorgiou}, {Parkin}, {Perez-Fournon},
  {Pohlen}, {Rangwala}, {Roussel}, {Rykala}, {Sacchi}, {Sauvage}, {Schulz},
  {Schirm}, {Smith}, {Spinoglio}, {Stevens}, {Symeonidis}, {Vaccari},
  {Vigroux}, {Wilson}, {Wozniak}, {Wright}, \& {Zeilinger}}]{BoselliPASP2010}
{Boselli}, A., {Eales}, S., {Cortese}, L., {et~al.} 2010{\natexlab{b}}, PASP,
  122, 261

\bibitem[{{Boselli} \& {Gavazzi}(2006)}]{BoselliPASP2006}
{Boselli}, A. \& {Gavazzi}, G. 2006, PASP, 118, 517

\bibitem[{{Boulanger} {et~al.}(1996){Boulanger}, {Abergel}, {Bernard},
  {Burton}, {D\'esert}, {Hartmann}, {Lagache}, \& {Puget}}]{BoulangerA&A1996}
{Boulanger}, F., {Abergel}, A., {Bernard}, J.-P., {et~al.} 1996, A\&A, 312, 256

\bibitem[{{Boulanger} \& {Perault}(1988)}]{BoulangerApJ1988}
{Boulanger}, F. \& {Perault}, M. 1988, ApJ, 330, 964

\bibitem[{{Bracco} {et~al.}(2011){Bracco}, {Cooray}, {Veneziani}, {Amblard},
  {Serra}, {Wardlow}, {Thompson}, {White}, {Auld}, {Baes}, {Bertoldi},
  {Buttiglione}, {Cava}, {Clements}, {Dariush}, {de Zotti}, {Dunne}, {Dye},
  {Eales}, {Fritz}, {Gomez}, {Hopwood}, {Ibar}, {Ivison}, {Jarvis}, {Lagache},
  {Lee}, {Leeuw}, {Maddox}, {Micha{\l}owski}, {Pearson}, {Pohlen}, {Rigby},
  {Rodighiero}, {Smith}, {Temi}, {Vaccari}, \& {van der
  Werf}}]{BraccoMNRAS2011}
{Bracco}, A., {Cooray}, A., {Veneziani}, M., {et~al.} 2011, MNRAS, 412, 1151

\bibitem[{{Clark} {et~al.}(2016){Clark}, {Schofield}, {Gomez}, \&
  {Davies}}]{ClarkMNRAS2016}
{Clark}, C.~J.~R., {Schofield}, S.~P., {Gomez}, H.~L., \& {Davies}, J.~I. 2016,
  MNRAS, 459, 1646

\bibitem[{{Compi{\`e}gne} {et~al.}(2011){Compi{\`e}gne}, {Verstraete}, {Jones},
  {Bernard}, {Boulanger}, {Flagey}, {Le Bourlot}, {Paradis}, \&
  {Ysard}}]{CompiegneA&A2011}
{Compi{\`e}gne}, M., {Verstraete}, L., {Jones}, A., {et~al.} 2011, A\&A, 525,
  A103

\bibitem[{{Condon} {et~al.}(1998){Condon}, {Cotton}, {Greisen}, {Yin},
  {Perley}, {Taylor}, \& {Broderick}}]{CondonAJ1998}
{Condon}, J.~J., {Cotton}, W.~D., {Greisen}, E.~W., {et~al.} 1998, AJ, 115,
  1693

\bibitem[{{Cortese} {et~al.}(2010){Cortese}, {Bendo}, {Isaak}, {Davies}, \&
  {Kent}}]{CorteseMNRAS2010}
{Cortese}, L., {Bendo}, G.~J., {Isaak}, K.~G., {Davies}, J.~I., \& {Kent},
  B.~R. 2010, MNRAS, 403, L26

\bibitem[{{Dale} {et~al.}(2012){Dale}, {Aniano}, {Engelbracht}, {Hinz},
  {Krause}, {Montiel}, {Roussel}, {Appleton}, {Armus}, {Beir{\~a}o}, {Bolatto},
  {Brandl}, {Calzetti}, {Crocker}, {Croxall}, {Draine}, {Galametz}, {Gordon},
  {Groves}, {Hao}, {Helou}, {Hunt}, {Johnson}, {Kennicutt}, {Koda}, {Leroy},
  {Li}, {Meidt}, {Miller}, {Murphy}, {Rahman}, {Rix}, {Sandstrom}, {Sauvage},
  {Schinnerer}, {Skibba}, {Smith}, {Tabatabaei}, {Walter}, {Wilson}, {Wolfire},
  \& {Zibetti}}]{DaleApJ2012}
{Dale}, D.~A., {Aniano}, G., {Engelbracht}, C.~W., {et~al.} 2012, ApJ, 745, 95

\bibitem[{{Davies} {et~al.}(2010{\natexlab{a}}){Davies}, {Baes}, {Bendo},
  {Bianchi}, {Bomans}, {Boselli}, {Clemens}, {Corbelli}, {Cortese}, {Dariush},
  {de Looze}, {di Serego Alighieri}, {Fadda}, {Fritz}, {Garcia-Appadoo},
  {Gavazzi}, {Giovanardi}, {Grossi}, {Hughes}, {Hunt}, {Jones}, {Madden},
  {Pierini}, {Pohlen}, {Sabatini}, {Smith}, {Verstappen}, {Vlahakis},
  {Xilouris}, \& {Zibetti}}]{DaviesA&A2010}
{Davies}, J.~I., {Baes}, M., {Bendo}, G.~J., {et~al.} 2010{\natexlab{a}}, A\&A,
  518, L48

\bibitem[{{Davies} {et~al.}(2012){Davies}, {Bianchi}, {Cortese}, {Auld},
  {Baes}, {Bendo}, {Boselli}, {Ciesla}, {Clemens}, {Corbelli}, {de Looze},
  {Alighieri}, {Fritz}, {Gavazzi}, {Pappalardo}, {Grossi}, {Hunt}, {Madden},
  {Magrini}, {Pohlen}, {Smith}, {Verstappen}, \& {Vlahakis}}]{DaviesMNRAS2012}
{Davies}, J.~I., {Bianchi}, S., {Cortese}, L., {et~al.} 2012, MNRAS, 419, 3505

\bibitem[{{Davies} {et~al.}(2010{\natexlab{b}}){Davies}, {Wilson}, {Auld},
  {Baes}, {Barlow}, {Bendo}, {Bock}, {Boselli}, {Bradford}, {Buat},
  {Castro-Rodriguez}, {Chanial}, {Charlot}, {Ciesla}, {Clements}, {Cooray},
  {Cormier}, {Cortese}, {Dwek}, {Eales}, {Elbaz}, {Galametz}, {Galliano},
  {Gear}, {Glenn}, {Gomez}, {Griffin}, {Hony}, {Isaak}, {Levenson}, {Lu},
  {Madden}, {O'Halloran}, {Okumura}, {Oliver}, {Page}, {Panuzzo},
  {Papageorgiou}, {Parkin}, {Perez-Fournon}, {Pohlen}, {Rangwala}, {Rigby},
  {Roussel}, {Rykala}, {Sacchi}, {Sauvage}, {Schulz}, {Schirm}, {Smith},
  {Spinoglio}, {Stevens}, {Srinivasan}, {Symeonidis}, {Trichas}, {Vaccari},
  {Vigroux}, {Wozniak}, {Wright}, \& {Zeilinger}}]{DaviesMNRAS2010}
{Davies}, J.~I., {Wilson}, C.~D., {Auld}, R., {et~al.} 2010{\natexlab{b}},
  MNRAS, 409, 102

\bibitem[{{Desert} {et~al.}(1988){Desert}, {Bazell}, \&
  {Boulanger}}]{DesertApJ1988}
{Desert}, F.~X., {Bazell}, D., \& {Boulanger}, F. 1988, ApJ, 334, 815

\bibitem[{{D\'esert} {et~al.}(1990){D\'esert}, {Boulanger}, \&
  {Puget}}]{DesertA&A1990}
{D\'esert}, F.~X., {Boulanger}, F., \& {Puget}, J.~L. 1990, A\&A, 237, 215

\bibitem[{{Draine}(2003)}]{DraineARA&A2003}
{Draine}, B.~T. 2003, ARA\&A, 41, 241

\bibitem[{{Draine}(2011)}]{DraineBook2011}
{Draine}, B.~T. 2011, {Physics of the Interstellar and Intergalactic Medium}
  ({Princeton University Press})

\bibitem[{{Draine} {et~al.}(2007){Draine}, {Dale}, {Bendo}, {Gordon}, {Smith},
  {Armus}, {Engelbracht}, {Helou}, {Kennicutt}, {Li}, {Roussel}, {Walter},
  {Calzetti}, {Moustakas}, {Murphy}, {Rieke}, {Bot}, {Hollenbach}, {Sheth}, \&
  {Teplitz}}]{DraineApJ2007}
{Draine}, B.~T., {Dale}, D.~A., {Bendo}, G., {et~al.} 2007, ApJ, 663, 866

\bibitem[{{Draine} \& {Lee}(1984)}]{DraineApJ1984}
{Draine}, B.~T. \& {Lee}, H.~M. 1984, ApJ, 285, 89

\bibitem[{{Dwek} {et~al.}(1990){Dwek}, {Rephaeli}, \& {Mather}}]{DwekApJ1990}
{Dwek}, E., {Rephaeli}, Y., \& {Mather}, J.~C. 1990, ApJ, 350, 104

\bibitem[{{Fanciullo} {et~al.}(2015){Fanciullo}, {Guillet}, {Aniano}, {Jones},
  {Ysard}, {Miville-Desch{\^e}nes}, {Boulanger}, \&
  {K{\"o}hler}}]{FanciulloA&A2015}
{Fanciullo}, L., {Guillet}, V., {Aniano}, G., {et~al.} 2015, A\&A, 580, A136

\bibitem[{{Ferrarese} {et~al.}(2012){Ferrarese}, {C{\^o}t{\'e}}, {Cuillandre},
  {Gwyn}, {Peng}, {MacArthur}, {Duc}, {Boselli}, {Mei}, {Erben}, {McConnachie},
  {Durrell}, {Mihos}, {Jord{\'a}n}, {Lan{\c c}on}, {Puzia}, {Emsellem},
  {Balogh}, {Blakeslee}, {van Waerbeke}, {Gavazzi}, {Vollmer}, {Kavelaars},
  {Woods}, {Ball}, {Boissier}, {Courteau}, {Ferriere}, {Gavazzi},
  {Hildebrandt}, {Hudelot}, {Huertas-Company}, {Liu}, {McLaughlin}, {Mellier},
  {Milkeraitis}, {Schade}, {Balkowski}, {Bournaud}, {Carlberg}, {Chapman},
  {Hoekstra}, {Peng}, {Sawicki}, {Simard}, {Taylor}, {Tully}, {van Driel},
  {Wilson}, {Burdullis}, {Mahoney}, \& {Manset}}]{FerrareseApJS2012}
{Ferrarese}, L., {C{\^o}t{\'e}}, P., {Cuillandre}, J.-C., {et~al.} 2012, ApJS,
  200, 4

\bibitem[{{Finkbeiner}(2003)}]{FinkbeinerApJS2003}
{Finkbeiner}, D.~P. 2003, ApJS, 146, 407

\bibitem[{{Gillmon} \& {Shull}(2006)}]{GillmonApJ2006}
{Gillmon}, K. \& {Shull}, J.~M. 2006, ApJ, 636, 908

\bibitem[{{Giovanelli} {et~al.}(2005){Giovanelli}, {Haynes}, {Kent},
  {Perillat}, {Saintonge}, {Brosch}, {Catinella}, {Hoffman}, {Stierwalt},
  {Spekkens}, {Lerner}, {Masters}, {Momjian}, {Rosenberg}, {Springob},
  {Boselli}, {Charmandaris}, {Darling}, {Davies}, {Garcia Lambas}, {Gavazzi},
  {Giovanardi}, {Hardy}, {Hunt}, {Iovino}, {Karachentsev}, {Karachentseva},
  {Koopmann}, {Marinoni}, {Minchin}, {Muller}, {Putman}, {Pantoja}, {Salzer},
  {Scodeggio}, {Skillman}, {Solanes}, {Valotto}, {van Driel}, \& {van
  Zee}}]{GiovanelliAJ2005}
{Giovanelli}, R., {Haynes}, M.~P., {Kent}, B.~R., {et~al.} 2005, AJ, 130, 2598

\bibitem[{{Giovanelli} {et~al.}(2007){Giovanelli}, {Haynes}, {Kent},
  {Saintonge}, {Stierwalt}, {Altaf}, {Balonek}, {Brosch}, {Brown}, {Catinella},
  {Furniss}, {Goldstein}, {Hoffman}, {Koopmann}, {Kornreich}, {Mahmood},
  {Martin}, {Masters}, {Mitschang}, {Momjian}, {Nair}, {Rosenberg}, \&
  {Walsh}}]{GiovanelliAJ2007}
{Giovanelli}, R., {Haynes}, M.~P., {Kent}, B.~R., {et~al.} 2007, AJ, 133, 2569

\bibitem[{{Gispert} {et~al.}(2000){Gispert}, {Lagache}, \&
  {Puget}}]{GispertA&A2000}
{Gispert}, R., {Lagache}, G., \& {Puget}, J.~L. 2000, A\&A, 360, 1

\bibitem[{{Griffin} {et~al.}(2010){Griffin}, {Abergel}, {Abreu}, {Ade},
  {Andr{\'e}}, {Augueres}, {Babbedge}, {Bae}, {Baillie}, {Baluteau}, {Barlow},
  {Bendo}, {Benielli}, {Bock}, {Bonhomme}, {Brisbin}, {Brockley-Blatt},
  {Caldwell}, {Cara}, {Castro-Rodriguez}, {Cerulli}, {Chanial}, {Chen},
  {Clark}, {Clements}, {Clerc}, {Coker}, {Communal}, {Conversi}, {Cox},
  {Crumb}, {Cunningham}, {Daly}, {Davis}, {de Antoni}, {Delderfield}, {Devin},
  {di Giorgio}, {Didschuns}, {Dohlen}, {Donati}, {Dowell}, {Dowell}, {Duband},
  {Dumaye}, {Emery}, {Ferlet}, {Ferrand}, {Fontignie}, {Fox}, {Franceschini},
  {Frerking}, {Fulton}, {Garcia}, {Gastaud}, {Gear}, {Glenn}, {Goizel},
  {Griffin}, {Grundy}, {Guest}, {Guillemet}, {Hargrave}, {Harwit}, {Hastings},
  {Hatziminaoglou}, {Herman}, {Hinde}, {Hristov}, {Huang}, {Imhof}, {Isaak},
  {Israelsson}, {Ivison}, {Jennings}, {Kiernan}, {King}, {Lange}, {Latter},
  {Laurent}, {Laurent}, {Leeks}, {Lellouch}, {Levenson}, {Li}, {Li},
  {Lilienthal}, {Lim}, {Liu}, {Lu}, {Madden}, {Mainetti}, {Marliani}, {McKay},
  {Mercier}, {Molinari}, {Morris}, {Moseley}, {Mulder}, {Mur}, {Naylor},
  {Nguyen}, {O'Halloran}, {Oliver}, {Olofsson}, {Olofsson}, {Orfei}, {Page},
  {Pain}, {Panuzzo}, {Papageorgiou}, {Parks}, {Parr-Burman}, {Pearce},
  {Pearson}, {P{\'e}rez-Fournon}, {Pinsard}, {Pisano}, {Podosek}, {Pohlen},
  {Polehampton}, {Pouliquen}, {Rigopoulou}, {Rizzo}, {Roseboom}, {Roussel},
  {Rowan-Robinson}, {Rownd}, {Saraceno}, {Sauvage}, {Savage}, {Savini},
  {Sawyer}, {Scharmberg}, {Schmitt}, {Schneider}, {Schulz}, {Schwartz},
  {Shafer}, {Shupe}, {Sibthorpe}, {Sidher}, {Smith}, {Smith}, {Smith},
  {Spencer}, {Stobie}, {Sudiwala}, {Sukhatme}, {Surace}, {Stevens}, {Swinyard},
  {Trichas}, {Tourette}, {Triou}, {Tseng}, {Tucker}, {Turner}, {Vaccari},
  {Valtchanov}, {Vigroux}, {Virique}, {Voellmer}, {Walker}, {Ward}, {Waskett},
  {Weilert}, {Wesson}, {White}, {Whitehouse}, {Wilson}, {Winter}, {Woodcraft},
  {Wright}, {Xu}, {Zavagno}, {Zemcov}, {Zhang}, \& {Zonca}}]{GriffinA&A2010}
{Griffin}, M.~J., {Abergel}, A., {Abreu}, A., {et~al.} 2010, A\&A, 518, L3

\bibitem[{{Guti{\'e}rrez} \& {L{\'o}pez-Corredoira}(2014)}]{GutierrezA&A2014}
{Guti{\'e}rrez}, C.~M. \& {L{\'o}pez-Corredoira}, M. 2014, A\&A, 571, A66

\bibitem[{{Haffner} {et~al.}(2003){Haffner}, {Reynolds}, {Tufte}, {Madsen},
  {Jaehnig}, \& {Percival}}]{HaffnerApJS2003}
{Haffner}, L.~M., {Reynolds}, R.~J., {Tufte}, S.~L., {et~al.} 2003, ApJS, 149,
  405

\bibitem[{{Hartmann} {et~al.}(1998){Hartmann}, {Magnani}, \&
  {Thaddeus}}]{HartmannApJ1998}
{Hartmann}, D., {Magnani}, L., \& {Thaddeus}, P. 1998, ApJ, 492, 205

\bibitem[{{Heiles} \& {Troland}(2003)}]{HeilesApJS2003}
{Heiles}, C. \& {Troland}, T.~H. 2003, ApJS, 145, 329

\bibitem[{{Heithausen}(2002)}]{HeithausenA&A2002}
{Heithausen}, A. 2002, A\&A, 393, L41

\bibitem[{{Heithausen}(2007)}]{HeithausenProc2007}
{Heithausen}, A. 2007, in Astronomical Society of the Pacific Conference
  Series, Vol. 365, SINS - Small Ionized and Neutral Structures in the Diffuse
  Interstellar Medium, ed. M.~{Haverkorn} \& W.~M. {Goss}, 177

\bibitem[{{Heithausen}(2012)}]{HeithausenA&A2012}
{Heithausen}, A. 2012, A\&A, 543, A21

\bibitem[{{Helou} \& {Walker}(1988)}]{IrasProc1988}
{Helou}, G. \& {Walker}, D.~W., eds. 1988, {Infrared astronomical satellite
  (IRAS) catalogs and atlases. Volume 7: The small scale structure catalog},
  Vol.~7

\bibitem[{{Hunt} {et~al.}(2015){Hunt}, {Draine}, {Bianchi}, {Gordon}, {Aniano},
  {Calzetti}, {Dale}, {Helou}, {Hinz}, {Kennicutt}, {Roussel}, {Wilson},
  {Bolatto}, {Boquien}, {Croxall}, {Galametz}, {Gil de Paz}, {Koda},
  {Mu{\~n}oz-Mateos}, {Sandstrom}, {Sauvage}, {Vigroux}, \&
  {Zibetti}}]{HuntA&A2015}
{Hunt}, L.~K., {Draine}, B.~T., {Bianchi}, S., {et~al.} 2015, A\&A, 576, A33

\bibitem[{{James} {et~al.}(2002){James}, {Dunne}, {Eales}, \&
  {Edmunds}}]{JamesMNRAS2002}
{James}, A., {Dunne}, L., {Eales}, S., \& {Edmunds}, M.~G. 2002, MNRAS, 335,
  753

\bibitem[{{Jones} {et~al.}(2013){Jones}, {Fanciullo}, {K{\"o}hler},
  {Verstraete}, {Guillet}, {Bocchio}, \& {Ysard}}]{JonesA&A2013}
{Jones}, A.~P., {Fanciullo}, L., {K{\"o}hler}, M., {et~al.} 2013, A\&A, 558,
  A62

\bibitem[{{Juvela} {et~al.}(2015){Juvela}, {Ristorcelli}, {Marshall},
  {Montillaud}, {Pelkonen}, {Ysard}, {McGehee}, {Paladini}, {Pagani},
  {Malinen}, {Rivera-Ingraham}, {Lef{\`e}vre}, {T{\'o}th}, {Montier},
  {Bernard}, \& {Martin}}]{JuvelaA&A2015}
{Juvela}, M., {Ristorcelli}, I., {Marshall}, D.~J., {et~al.} 2015, A\&A, 584,
  A93

\bibitem[{{Kalberla} {et~al.}(2005){Kalberla}, {Burton}, {Hartmann}, {Arnal},
  {Bajaja}, {Morras}, \& {P{\"o}ppel}}]{KalberlaA&A2005}
{Kalberla}, P.~M.~W., {Burton}, W.~B., {Hartmann}, D., {et~al.} 2005, A\&A,
  440, 775

\bibitem[{{Kalberla} \& {Kerp}(2009)}]{KalberlaARA&A2009}
{Kalberla}, P.~M.~W. \& {Kerp}, J. 2009, ARA\&A, 47, 27

\bibitem[{{Kelly} {et~al.}(2012){Kelly}, {Shetty}, {Stutz}, {Kauffmann},
  {Goodman}, \& {Launhardt}}]{KellyApJ2012}
{Kelly}, B.~C., {Shetty}, R., {Stutz}, A.~M., {et~al.} 2012, ApJ, 752, 55

\bibitem[{{Kitayama} {et~al.}(2009){Kitayama}, {Ito}, {Okada}, {Kaneda},
  {Takahashi}, {Ota}, {Onaka}, {Tajiri}, {Nagata}, \&
  {Yamada}}]{KitayamaApJ2009}
{Kitayama}, T., {Ito}, Y., {Okada}, Y., {et~al.} 2009, ApJ, 695, 1191

\bibitem[{{K{\"o}hler} {et~al.}(2015){K{\"o}hler}, {Ysard}, \&
  {Jones}}]{KoehlerA&A2015}
{K{\"o}hler}, M., {Ysard}, N., \& {Jones}, A.~P. 2015, A\&A, 579, A15

\bibitem[{{Kuntz} \& {Danly}(1996)}]{KuntzApJ1996}
{Kuntz}, K.~D. \& {Danly}, L. 1996, ApJ, 457, 703

\bibitem[{{Lagache} {et~al.}(1999){Lagache}, {Abergel}, {Boulanger},
  {D\'esert}, \& {Puget}}]{LagacheA&A1999}
{Lagache}, G., {Abergel}, A., {Boulanger}, F., {D\'esert}, F.~X., \& {Puget},
  J.~L. 1999, A\&A, 344, 322

\bibitem[{{Lagache} {et~al.}(2000){Lagache}, {Haffner}, {Reynolds}, \&
  {Tufte}}]{LagacheA&A2000}
{Lagache}, G., {Haffner}, L.~M., {Reynolds}, R.~J., \& {Tufte}, S.~L. 2000,
  A\&A, 354, 247

\bibitem[{{Low} {et~al.}(1984){Low}, {Young}, {Beintema}, {Gautier},
  {Beichman}, {Aumann}, {Gillett}, {Neugebauer}, {Boggess}, \&
  {Emerson}}]{LowApJL1984}
{Low}, F.~J., {Young}, E., {Beintema}, D.~A., {et~al.} 1984, ApJL, 278, L19

\bibitem[{{Magrini} {et~al.}(2011){Magrini}, {Bianchi}, {Corbelli}, {Cortese},
  {Hunt}, {Smith}, {Vlahakis}, {Davies}, {Bendo}, {Baes}, {Boselli}, {Clemens},
  {Casasola}, {de Looze}, {Fritz}, {Giovanardi}, {Grossi}, {Hughes}, {Madden},
  {Pappalardo}, {Pohlen}, {di Serego Alighieri}, \&
  {Verstappen}}]{MagriniA&A2011}
{Magrini}, L., {Bianchi}, S., {Corbelli}, E., {et~al.} 2011, A\&A, 535, A13

\bibitem[{{Markwardt}(2009)}]{MarkwardtProc2009}
{Markwardt}, C.~B. 2009, in Astronomical Society of the Pacific Conference
  Series, Vol. 411, Astronomical Data Analysis Software and Systems XVIII, ed.
  D.~A. {Bohlender}, D.~{Durand}, \& P.~{Dowler}, 251

\bibitem[{{Mathis} {et~al.}(1983){Mathis}, {Mezger}, \&
  {Panagia}}]{MathisA&A1983}
{Mathis}, J.~S., {Mezger}, P.~G., \& {Panagia}, N. 1983, A\&A, 128, 212

\bibitem[{{McGee} \& {Balogh}(2010)}]{McGeeMNRAS2010}
{McGee}, S.~L. \& {Balogh}, M.~L. 2010, MNRAS, 405, 2069

\bibitem[{{Mihos}(2015)}]{MihosProc2015}
{Mihos}, C. 2015, in IAU Symposium No. 317, 2015, Vol.~22, The General Assembly
  of Galaxy Halos

\bibitem[{{Mihos} {et~al.}(2005){Mihos}, {Harding}, {Feldmeier}, \&
  {Morrison}}]{MihosApJL2005}
{Mihos}, J.~C., {Harding}, P., {Feldmeier}, J., \& {Morrison}, H. 2005, ApJL,
  631, L41

\bibitem[{{Misiriotis} {et~al.}(2006){Misiriotis}, {Xilouris},
  {Papamastorakis}, {Boumis}, \& {Goudis}}]{MisiriotisA&A2006}
{Misiriotis}, A., {Xilouris}, E.~M., {Papamastorakis}, J., {Boumis}, P., \&
  {Goudis}, C.~D. 2006, A\&A, 459, 113

\bibitem[{{Miville-Desch{\^e}nes} \&
  {Lagache}(2005)}]{MivilleDeschenesApJS2005}
{Miville-Desch{\^e}nes}, M. \& {Lagache}, G. 2005, ApJS, 157, 302

\bibitem[{{Neugebauer} {et~al.}(1984){Neugebauer}, {Habing}, {van Duinen},
  {Aumann}, {Baud}, {Beichman}, {Beintema}, {Boggess}, {Clegg}, {de Jong},
  {Emerson}, {Gautier}, {Gillett}, {Harris}, {Hauser}, {Houck}, {Jennings},
  {Low}, {Marsden}, {Miley}, {Olnon}, {Pottasch}, {Raimond}, {Rowan-Robinson},
  {Soifer}, {Walker}, {Wesselius}, \& {Young}}]{NeugebauerApJL1984}
{Neugebauer}, G., {Habing}, H.~J., {van Duinen}, R., {et~al.} 1984, ApJ, 278,
  L1

\bibitem[{{Ossenkopf} \& {Henning}(1994)}]{OssenkopfA&A1994}
{Ossenkopf}, V. \& {Henning}, T. 1994, A\&A, 291, 943

\bibitem[{{Ott}(2010)}]{OttProc2010}
{Ott}, S. 2010, in ASP Conference Series, Vol. 434, 139

\bibitem[{{Pappalardo} {et~al.}(2015){Pappalardo}, {Bendo}, {Bianchi}, {Hunt},
  {Zibetti}, {Corbelli}, {di Serego Alighieri}, {Grossi}, {Davies}, {Baes}, {De
  Looze}, {Fritz}, {Pohlen}, {Smith}, {Verstappen}, {Boquien}, {Boselli},
  {Cortese}, {Hughes}, {Viaene}, {Bizzocchi}, \& {Clemens}}]{PappalardoA&A2015}
{Pappalardo}, C., {Bendo}, G.~J., {Bianchi}, S., {et~al.} 2015, A\&A, 573, A129

\bibitem[{{Peek} {et~al.}(2011){Peek}, {Heiles}, {Douglas}, {Lee}, {Grcevich},
  {Stanimirovi{\'c}}, {Putman}, {Korpela}, {Gibson}, {Begum}, {Saul},
  {Robishaw}, \& {Kr{\v c}o}}]{PeekApJS2011}
{Peek}, J.~E.~G., {Heiles}, C., {Douglas}, K.~A., {et~al.} 2011, ApJS, 194, 20

\bibitem[{{Pilbratt} {et~al.}(2010){Pilbratt}, {Riedinger}, {Passvogel},
  {Crone}, {Doyle}, {Gageur}, {Heras}, {Jewell}, {Metcalfe}, {Ott}, \&
  {Schmidt}}]{PilbrattA&A2010}
{Pilbratt}, G.~L., {Riedinger}, J.~R., {Passvogel}, T., {et~al.} 2010, A\&A,
  518, L1

\bibitem[{{Planck Collaboration I}(2014)}]{Planck2013I}
{Planck Collaboration I}. 2014, A\&A, 571, A1

\bibitem[{{Planck Collaboration Int. XVII}(2014)}]{PlanckIntermediateXVII}
{Planck Collaboration Int. XVII}. 2014, A\&A, 566, A55

\bibitem[{{Planck Collaboration Int. XXIX}(2016)}]{PlanckIntermediateXXIX}
{Planck Collaboration Int. XXIX}. 2016, A\&A, 586, A132

\bibitem[{{Planck Collaboration Int. XXV}(2015)}]{PlanckA&A2015}
{Planck Collaboration Int. XXV}. 2015, A\&A, 582, A28

\bibitem[{{Planck Collaboration IX}(2014)}]{Planck2013IX}
{Planck Collaboration IX}. 2014, A\&A, 571, A9

\bibitem[{{Planck Collaboration VIII}(2014)}]{Planck2013VIII}
{Planck Collaboration VIII}. 2014, A\&A, 571, A8

\bibitem[{{Planck Collaboration XI}(2014)}]{Planck2013XI}
{Planck Collaboration XI}. 2014, A\&A, 571, A11

\bibitem[{{Planck Collaboration XII}(2014)}]{Planck2013XII}
{Planck Collaboration XII}. 2014, A\&A, 571, A12

\bibitem[{{Planck Collaboration XIII}(2014)}]{Planck2013XIII}
{Planck Collaboration XIII}. 2014, A\&A, 571, A13

\bibitem[{{Planck Collaboration XIV}(2014)}]{Planck2013XIV}
{Planck Collaboration XIV}. 2014, A\&A, 571, A14

\bibitem[{{Planck Collaboration XXIII}(2011)}]{PlanckEarlyXXIII}
{Planck Collaboration XXIII}. 2011, A\&A, 536, A23

\bibitem[{{Planck Collaboration XXIV}(2011)}]{PlanckEarlyXXIV}
{Planck Collaboration XXIV}. 2011, A\&A, 536, A24

\bibitem[{{Planck Collaboration XXVIII}(2016)}]{Planck2015XXVIII}
{Planck Collaboration XXVIII}. 2016, A\&A, in press

\bibitem[{{Planck Collaboration XXX}(2014)}]{Planck2013XXX}
{Planck Collaboration XXX}. 2014, A\&A, 571, A30

\bibitem[{{Poglitsch} {et~al.}(2010){Poglitsch}, {Waelkens}, {Geis},
  {Feuchtgruber}, {Vandenbussche}, {Rodriguez}, {Krause}, {Renotte}, {van
  Hoof}, {Saraceno}, {Cepa}, {Kerschbaum}, {Agn{\`e}se}, {Ali}, {Altieri},
  {Andreani}, {Augueres}, {Balog}, {Barl}, {Bauer}, {Belbachir}, {Benedettini},
  {Billot}, {Boulade}, {Bischof}, {Blommaert}, {Callut}, {Cara}, {Cerulli},
  {Cesarsky}, {Contursi}, {Creten}, {De Meester}, {Doublier}, {Doumayrou},
  {Duband}, {Exter}, {Genzel}, {Gillis}, {Gr{\"o}zinger}, {Henning},
  {Herreros}, {Huygen}, {Inguscio}, {Jakob}, {Jamar}, {Jean}, {de Jong},
  {Katterloher}, {Kiss}, {Klaas}, {Lemke}, {Lutz}, {Madden}, {Marquet},
  {Martignac}, {Mazy}, {Merken}, {Montfort}, {Morbidelli}, {M{\"u}ller},
  {Nielbock}, {Okumura}, {Orfei}, {Ottensamer}, {Pezzuto}, {Popesso},
  {Putzeys}, {Regibo}, {Reveret}, {Royer}, {Sauvage}, {Schreiber}, {Stegmaier},
  {Schmitt}, {Schubert}, {Sturm}, {Thiel}, {Tofani}, {Vavrek}, {Wetzstein},
  {Wieprecht}, \& {Wiezorrek}}]{PoglitschA&A2010}
{Poglitsch}, A., {Waelkens}, C., {Geis}, N., {et~al.} 2010, \aap, 518, L2

\bibitem[{{Popescu} \& {Tuffs}(2013)}]{PopescuMNRAS2013}
{Popescu}, C.~C. \& {Tuffs}, R.~J. 2013, \mnras, 436, 1302

\bibitem[{{Popescu} {et~al.}(2000){Popescu}, {Tuffs}, {Fischera}, \&
  {V{\"o}lk}}]{PopescuA&A2000a}
{Popescu}, C.~C., {Tuffs}, R.~J., {Fischera}, J., \& {V{\"o}lk}, H. 2000, A\&A,
  354, 480

\bibitem[{{Reach} {et~al.}(2015){Reach}, {Heiles}, \& {Bernard}}]{ReachApJ2015}
{Reach}, W.~T., {Heiles}, C., \& {Bernard}, J.-P. 2015, ApJ, 811, 118

\bibitem[{{Reach} {et~al.}(1998){Reach}, {Wall}, \& {Odegard}}]{ReachApJ1998}
{Reach}, W.~T., {Wall}, W.~F., \& {Odegard}, N. 1998, ApJ, 507, 507

\bibitem[{{Rudick} {et~al.}(2010){Rudick}, {Mihos}, {Harding}, {Feldmeier},
  {Janowiecki}, \& {Morrison}}]{RudickApJ2010}
{Rudick}, C.~S., {Mihos}, J.~C., {Harding}, P., {et~al.} 2010, ApJ, 720, 569

\bibitem[{{Schlegel} {et~al.}(1998){Schlegel}, {Finkbeiner}, \&
  {Davis}}]{SchlegelApJ1998}
{Schlegel}, D.~J., {Finkbeiner}, D.~P., \& {Davis}, M. 1998, ApJ, 500, 525

\bibitem[{{Shetty} {et~al.}(2009){Shetty}, {Kauffmann}, {Schnee}, \&
  {Goodman}}]{ShettyApJ2009}
{Shetty}, R., {Kauffmann}, J., {Schnee}, S., \& {Goodman}, A.~A. 2009, ApJ,
  696, 676

\bibitem[{{Smith}(2012)}]{SmithThesis2012}
{Smith}, M. W.~L. 2012, PhD thesis, Cardiff University

\bibitem[{{Smith} {et~al.}(2010){Smith}, {Vlahakis}, {Baes}, {Bendo},
  {Bianchi}, {Bomans}, {Boselli}, {Clemens}, {Corbelli}, {Cortese}, {Dariush},
  {Davies}, {de Looze}, {di Serego Alighieri}, {Fadda}, {Fritz},
  {Garcia-Appadoo}, {Gavazzi}, {Giovanardi}, {Grossi}, {Hughes}, {Hunt},
  {Jones}, {Madden}, {Pierini}, {Pohlen}, {Sabatini}, {Verstappen}, {Xilouris},
  \& {Zibetti}}]{SmithA&A2010}
{Smith}, M.~W.~L., {Vlahakis}, C., {Baes}, M., {et~al.} 2010, A\&A, 518, L51

\bibitem[{{Spitzer}(1978)}]{SpitzerBook1978}
{Spitzer}, L. 1978, {Physical Processes in the Interstellar Medium} (New York:
  {Wiley})

\bibitem[{{Stickel} {et~al.}(2002){Stickel}, {Klaas}, {Lemke}, \&
  {Mattila}}]{StickelA&A2002}
{Stickel}, M., {Klaas}, U., {Lemke}, D., \& {Mattila}, K. 2002, A\&A, 383, 367

\bibitem[{{Stickel} {et~al.}(1998){Stickel}, {Lemke}, {Mattila}, {Haikala}, \&
  {Haas}}]{StickelA&A1998}
{Stickel}, M., {Lemke}, D., {Mattila}, K., {Haikala}, L.~K., \& {Haas}, M.
  1998, A\&A, 329, 55

\bibitem[{{Veneziani} {et~al.}(2010){Veneziani}, {Ade}, {Bock}, {Boscaleri},
  {Crill}, {de Bernardis}, {De Gasperis}, {de Oliveira-Costa}, {De Troia}, {Di
  Stefano}, {Ganga}, {Jones}, {Kisner}, {Lange}, {MacTavish}, {Masi},
  {Mauskopf}, {Montroy}, {Natoli}, {Netterfield}, {Pascale}, {Piacentini},
  {Pietrobon}, {Polenta}, {Ricciardi}, {Romeo}, \& {Ruhl}}]{VenezianiApJ2010}
{Veneziani}, M., {Ade}, P.~A.~R., {Bock}, J.~J., {et~al.} 2010, ApJ, 713, 959

\bibitem[{{Wakker}(2001)}]{WakkerApJS2001}
{Wakker}, B.~P. 2001, ApJS, 136, 463

\bibitem[{{Wakker} \& {van Woerden}(1997)}]{WakkerARA&A1997}
{Wakker}, B.~P. \& {van Woerden}, H. 1997, ARA\&A, 35, 217

\bibitem[{{Wise} {et~al.}(1993){Wise}, {O'Connell}, {Bregman}, \&
  {Roberts}}]{WiseApJ1993}
{Wise}, M.~W., {O'Connell}, R.~W., {Bregman}, J.~N., \& {Roberts}, M.~S. 1993,
  ApJ, 405, 94

\bibitem[{{Witt} {et~al.}(2010){Witt}, {Gold}, {Barnes}, {DeRoo}, {Vijh}, \&
  {Madsen}}]{WittApJ2010}
{Witt}, A.~N., {Gold}, B., {Barnes}, III, F.~S., {et~al.} 2010, ApJ, 724, 1551

\bibitem[{{Ysard} {et~al.}(2015){Ysard}, {K{\"o}hler}, {Jones},
  {Miville-Desch{\^e}nes}, {Abergel}, \& {Fanciullo}}]{YsardA&A2015}
{Ysard}, N., {K{\"o}hler}, M., {Jones}, A., {et~al.} 2015, A\&A, 577, A110

\bibitem[{{Zubko} {et~al.}(2004){Zubko}, {Dwek}, \& {Arendt}}]{ZubkoApJS2004}
{Zubko}, V., {Dwek}, E., \& {Arendt}, R.~G. 2004, ApJS, 152, 211

\end{thebibliography}

\begin{appendix}

\section{Offsets and Zodiacal light gradients}
\label{app:oe}

We discuss here the offsets $O_\nu$ and the ecliptic latitude gradient
$E_\nu$ resulting from the fitting described in Sect.~\ref{sec:ana} and \ref{sec:res}.

\begin{figure*}
\includegraphics[width=9cm]{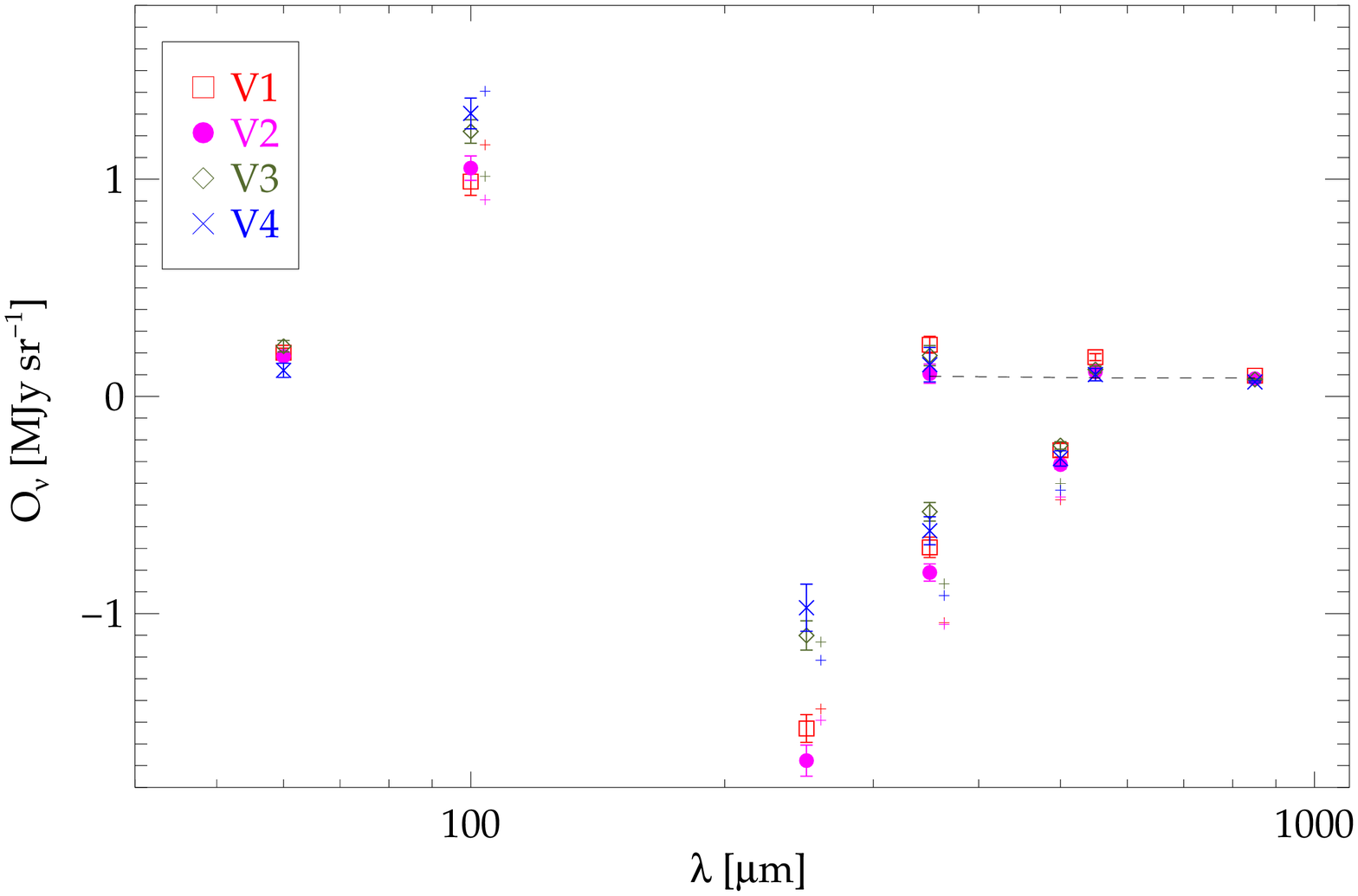}
\includegraphics[width=9cm]{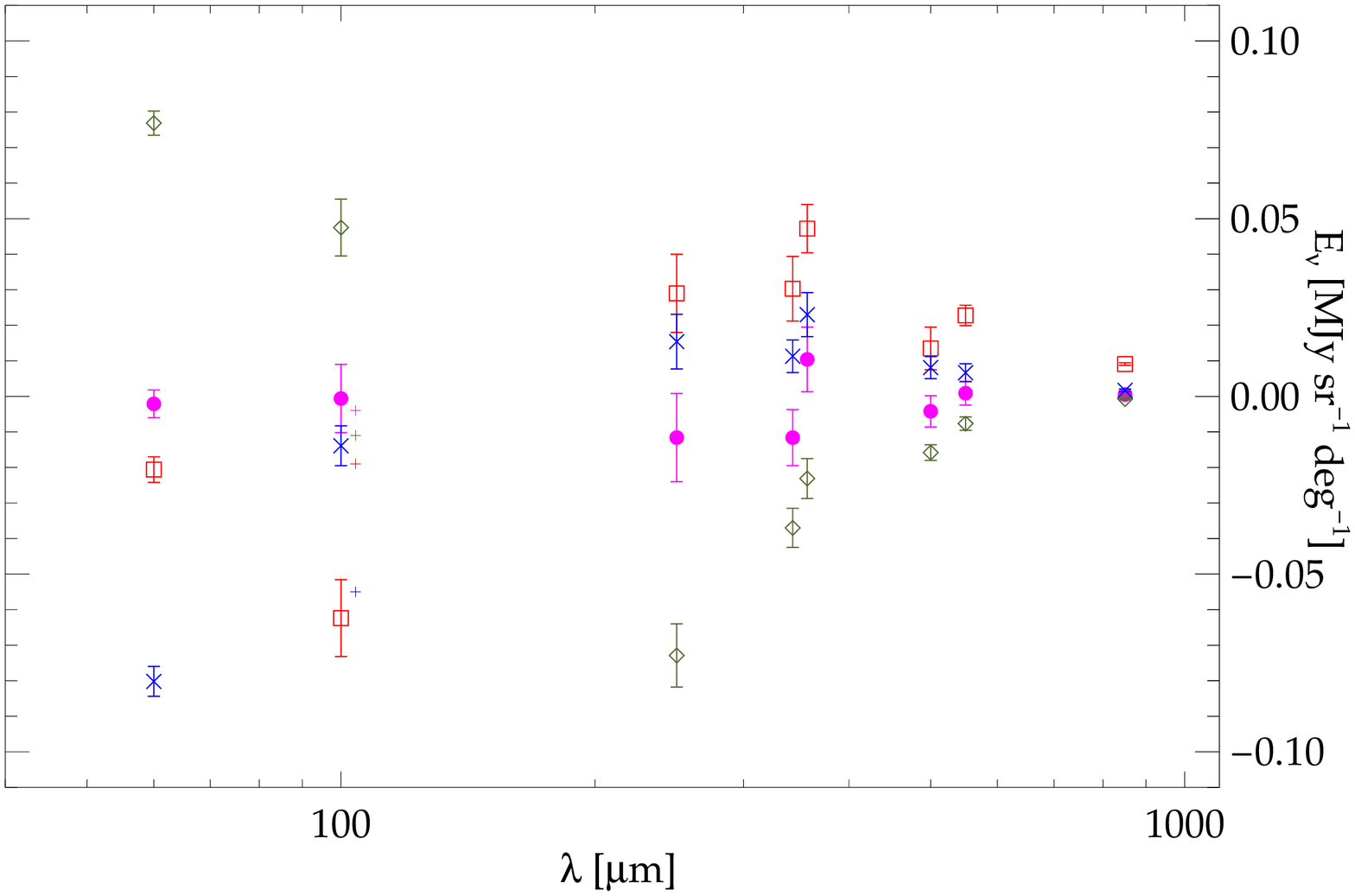}
\caption{Left panel: offsets $O_\nu$. With crosses
we show also: estimates at 100~$\mu$m obtained summing the offset with LAB data
and the residual zodiacal light contamination in IRIS maps as derived by
\citet{Planck2013XI}; estimates for SPIRE data using the SPIRE/HFI offsets
from the {\tt zeroPointCorrection}
procedure in HIPE. The dashed line shows the offset derived by the {\em Planck} team
on zodi-corrected maps \citep{Planck2013VIII}. Right panel: fitted gradient
in ecliptic latitude $E_\nu$. Crosses show the gradient as derived from the
correction described in \citet{Planck2013XI}. To avoid cluttering, we have 
slightly displaced some symbols from their exact wavelength.}
\label{fig:zero}
\end{figure*} 

The derived $O_\nu$ are shown in the left panel of Fig.~\ref{fig:zero}. For
the IRIS 60~$\mu$m, the offset is  positive, small, and very similar for all fields. 
At 100~$\mu$m, it is larger and likely dominated by uncertainties in the subtraction
of the zodiacal light contribution from IRIS maps: this is discussed in \citet{Planck2013XI},
where a correction is found by comparing smoothed IRIS images with the earlier
100~$\mu$m maps from \citet{SchlegelApJ1998}, which implemented a different, empirical, 
zodiacal light subtraction. When this correction is taken into account, our values
of $O_\nu$ are compatible with those derived by \citet{Planck2013XI} by correlating
dust emission and atomic gas data on their {\em Low N$_{\ion{H}{i}}$} mask.
The offsets we found for {\em Planck} data are also 
compatible with the values derived on the same sky area by \citet{Planck2013VIII}. The
offsets derived for SPIRE data are instead negative, since the maps by construction have 
their median level set to zero: fields with larger emission (V1 and V2 at 250~$\mu$m) have a 
median level biased to more positive values, and the offset is more negative. As a crosscheck,
we also derived the  offset between SPIRE and {\em Planck}-HFI maps using the {\tt zeroPointCorrection} 
task within HIPE. The {\em Planck} maps used by the task have been corrected so that they also 
include the emission from the CIB. Thus our offset can be compared
with the difference between the CIB estimate used by the {\em Planck} team \citep{Planck2013VIII}
and the HIPE offset. These values (shown as crosses at 250, 350 and 500~$\mu$m in the
left panel of Fig.~\ref{fig:zero}) are compatible with those derived here at 250~$\mu$m
(the HIPE offset resulting from a spectrum extrapolation) but show some differences at
the longer wavelengths. A possible explanation might be in the uncertainties in the zero-level 
determination or in the 
spectral baseline subtraction of $\ion{H}{i}$ maps, that might result in different correlations
between dust emission and gas and between two dust emission channels, as noted also by
\citet{Planck2013VIII}. For the same reason, intensity levels estimated using 
Eq.~\ref{eq:fitted} in two adjacent HeViCS fields might not overlap. 
{ However, in all cases these residual offsets are $\la 1 \sigma$}.

The right panel of Fig.~\ref{fig:zero} shows the fitted ecliptic latitude gradient,
$E_\nu$. In half of the fits (see Table~\ref{table:fit}), the derived $E_\nu$ is found to 
be insignificant ($< 3\sigma$). When measured at $> 3\sigma$, it is generally
consistent between SPIRE and HFI bands (i.e.\ always positive for fields V1 and V4,
always negative for V3). However, we found it to be present also in HFI maps, which
were already corrected for zodiacal light; and the V1 field has a larger gradient than V2
(where it is almost always consistent with zero), despite their common
extent in ecliptic latitude. Either our na\"ive choice for the gradient is not
able to capture the complexity of a full zodiacal light model \citep[see ][ for the
modelling of Zodiacal light in HFI maps]{Planck2013XIV} or it could possibly hide, as 
we discussed for $O_\nu$, spatial gradients due to uncertainties in the $\ion{H}{i}$ spectra
baseline subtraction. Gradients are also found for the IRIS data: at 100~$\mu$m,
gradients are found to be of the same order of what can be derived from the
zodiacal light subtraction of \citet{Planck2013XI} (crosses in the 
right panel of Fig.~\ref{fig:zero}). However, the largest (negative) gradient derived 
here is for field V1, while using the \citet{Planck2013XI} correction it is for
field V4. Whatever is the true nature of $E_\nu$, we nevertheless retained the term 
in the fit because we found that it greatly reduced the scatter in the determination 
of the dust emissivities associated with the IVC component, without modifying significantly
the results for the LVC component.

\section{Implications for the estimate of dust mass in external galaxies}
\label{app:extgal}

{ 
The absorption cross-section of high latitude MW dust is typically used as a proxy for
the properties of dust in external galaxies. We evaluate here the impact of the new
MW LVC cross-section found in {\em Planck} papers and in the current one, when
these are used to derive the dust masses.

\citet{BianchiA&A2013} argued that, under the common assumption that
the FIR SED is due to dust heated by an average ISRF \citep{DraineApJ2007,DaleApJ2012},
fits using a single temperature MBB (essentially, using a single
dust grain of average absorption cross section) do not perform much worse than fits using more
realistic dust models (with a distribution of grain sizes of different materials).
Assuming the same dust-to-gas mass ratio ($\sim 0.01$) as derived for the 
\citet{CompiegneA&A2011} dust model, and adopting their correction to account
for hydrogen atoms in the ionised and molecular gas (increasing the hydrogen 
column density by 23\%), we can derive from the current data an average dust absorption cross 
section per unit of mass $\kappa_{\mathrm{abs}}(250\mu\mathrm{m}) = 2.2 \pm 0.6 \,\mathrm{cm}^2 \,\mathrm{g}^{-1}$
(with $\beta=1.5$). On the same sample of galaxies used in \citet{BianchiA&A2013},
the dust mass derived using this cross section would be on average a factor 1.55
larger than that obtained using the average properties of the FIRAS-constrained
\citet{CompiegneA&A2011} dust model, with $\kappa_{\mathrm{abs}}(250\mu\mathrm{m}) = 
5.1\,\mathrm{cm}^2 \,\mathrm{g}^{-1}$ ($\beta=1.91$). In several papers of the HeViCS 
series we instead used $\kappa_{\mathrm{abs}}(350\mu\mathrm{m}) = 1.92
\, \mathrm{cm}^2 \,\mathrm{g}^{-1}$ ($\beta=2$). Accidentally\footnote{
The average cross section used in most HeViCS papers 
\citep[see, e.g., ][]{SmithA&A2010,MagriniA&A2011,DaviesMNRAS2012,AuldMNRAS2013}
was taken from the MW dust model
of \citet{DraineARA&A2003}. Even though the model was constrained on the FIRAS
spectrum, its emission under the LISRF heating was lower than the (corrected)
DHGL SED of \citet{CompiegneA&A2011}. \citet{BianchiA&A2013} maintained that this could
be due to a different correction for the contribution of ionised (and molecular) gas to
the hydrogen column density. 
{ Inconsistencies between {\em Planck} observations and the
\citet{DraineARA&A2003} models were also noted by \citet{PlanckIntermediateXVII}
and \citet{PlanckIntermediateXXIX}.
}}, the dust masses obtained with this value are very close 
to those obtained with the new absorption cross-section, being on average lower by 5\%. 
Because of the lower spectral index of the new cross section, the fitted temperature 
would be larger, the average for the sample being $\sim$25K, instead of 
$\sim$21 and 22K obtained with the \citet{CompiegneA&A2011} model spectral index and
the $\beta=2$ HeViCS choice, respectively.

{ 

\citet{ClarkMNRAS2016} estimated the dust absorption cross section from integrated 
FIR/submm SEDs and properties for 22 galaxies in the Herschel Reference Survey 
\citep[HRS;][]{BoselliPASP2010}. They used the mass of metals
as a proxy of the dust mass and assumed a power law for the cross-section spectrum, 
following the method by \citet{JamesMNRAS2002}. If their analysis is repeated using 
$\beta=1.5$, as suggested by this work, it is found that $\kappa_{\mathrm{abs}}(250\mu\mathrm{m}) 
= 1.02^{+1.09}_{-0.49}  \,\mathrm{cm}^2 \,\mathrm{g}^{-1}$. The average value is
lower than the estimates based on DHGL emission by a factor two, and thus the dust
masses would be higher by the same factor. The discrepancy
might suggest that dust in the DHGL medium has different average properties than the bulk of
the dust mass in a galaxy. If this is true, dust masses might be better estimated from 
global SEDs using the cross sections from \citet{ClarkMNRAS2016}; yet, the DHGL value
is still within the large scatter of their estimates, which might point to
galaxy-to-galaxy variations of the average dust properties.

Finally, a value $\beta\approx 1.5$ is compatible with the SPIRE colors of the HRS
galaxies \citep{BoselliA&A2010,BoselliA&A2012}. The trend in SPIRE
colors could however be produced by subtle effects, such as deviations from a pure
power-law in a restricted wavelength range, or by the mixing of the dust temperatures
for grains of different material and size \citep{HuntA&A2015}. This reminds us that
the result of the present work might just be an apparent $\beta$ value providing
a good description of the emissivity SED, but hiding the complexity of the dust 
grain mixture producing it.
}
}

\section{Data, models and residuals}
\label{app:otherfigs}

\begin{figure*}
\includegraphics[scale=\scala,trim = 0 0 8.0bp 0,clip]{\figdir/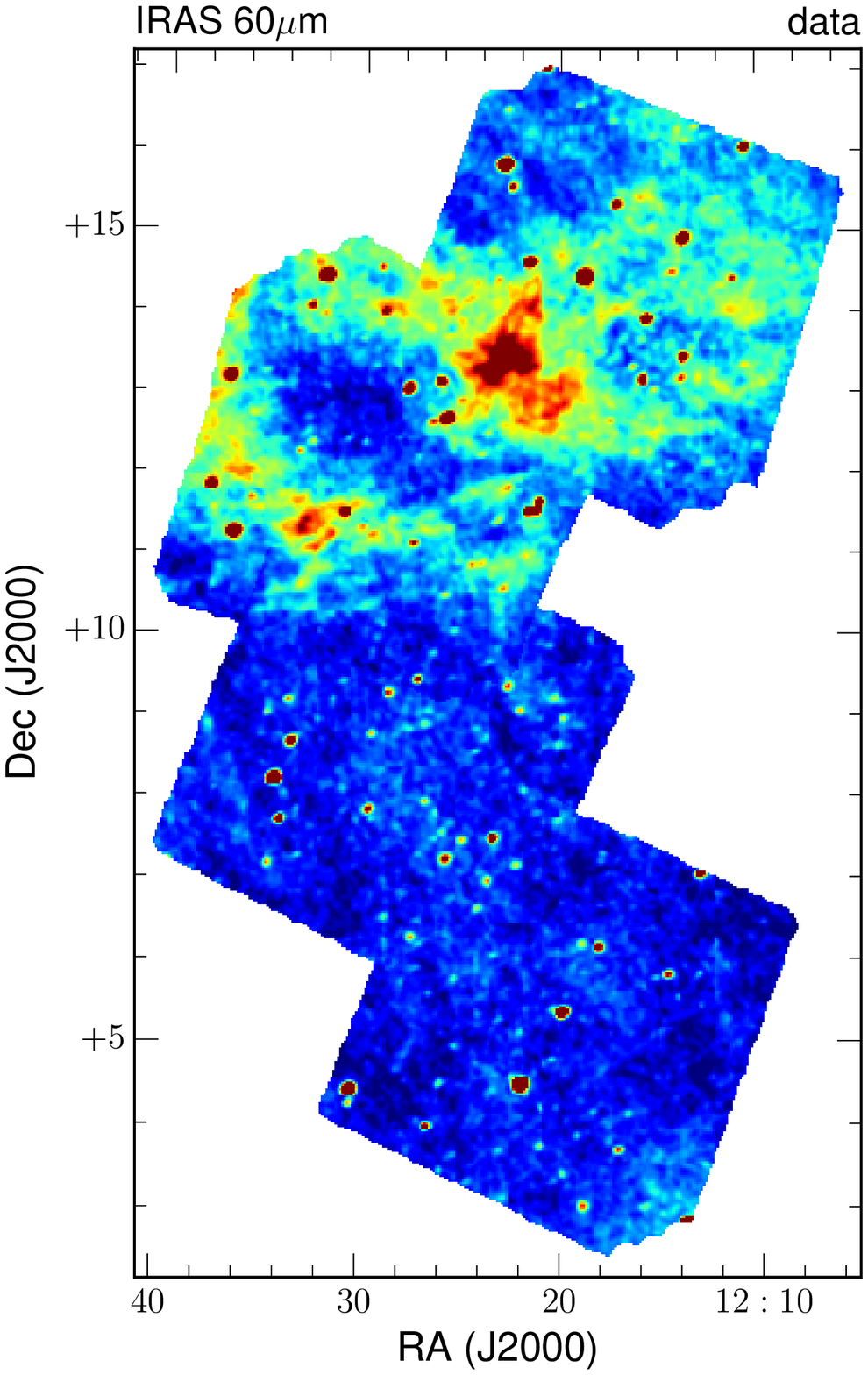}\includegraphics[scale=\scala,trim = 4bp 0 6.5bp 0,clip]{\figdir/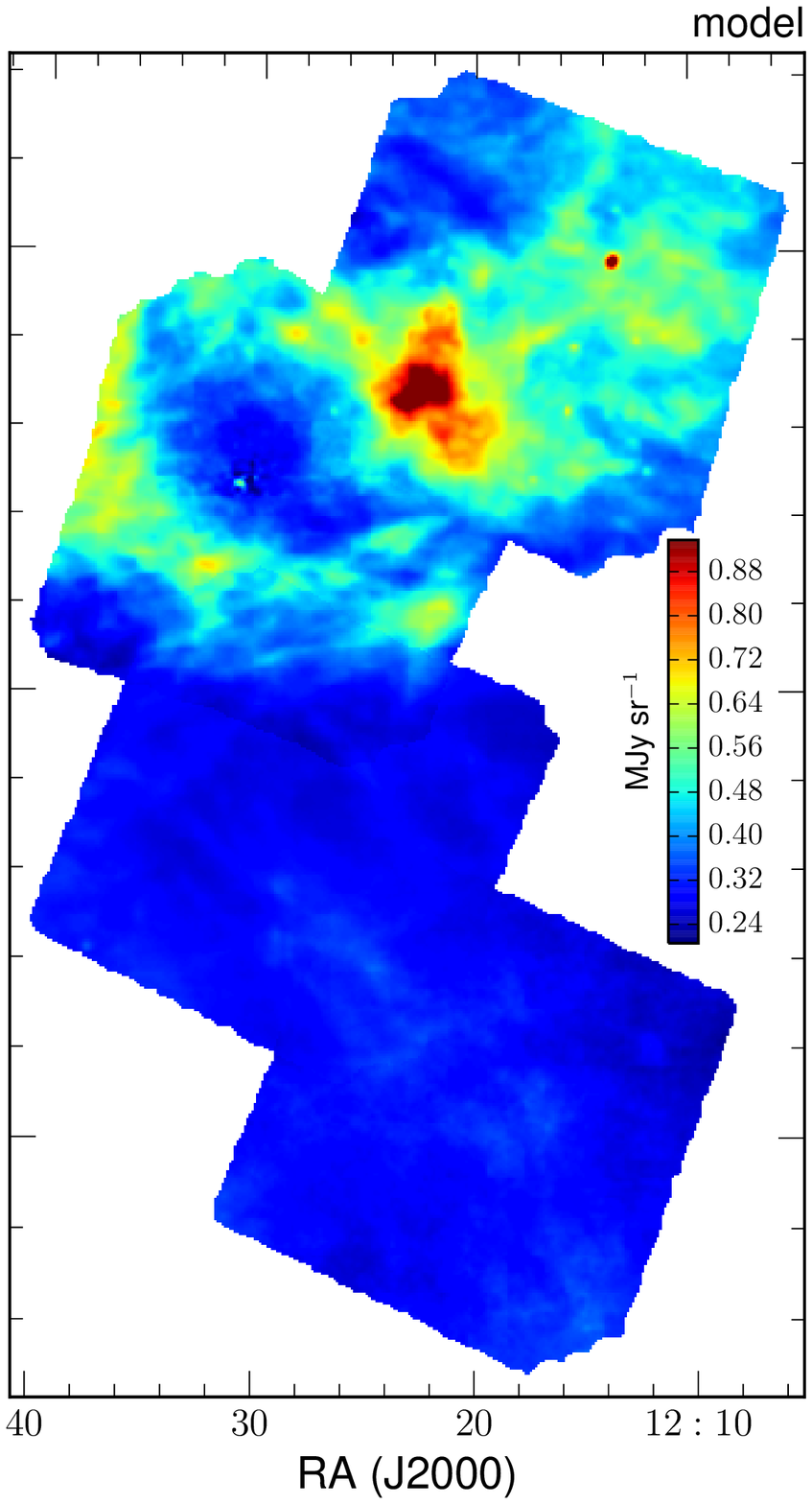}\includegraphics[scale=\scala,trim = 4bp 0 6.5bp 0,clip]{\figdir/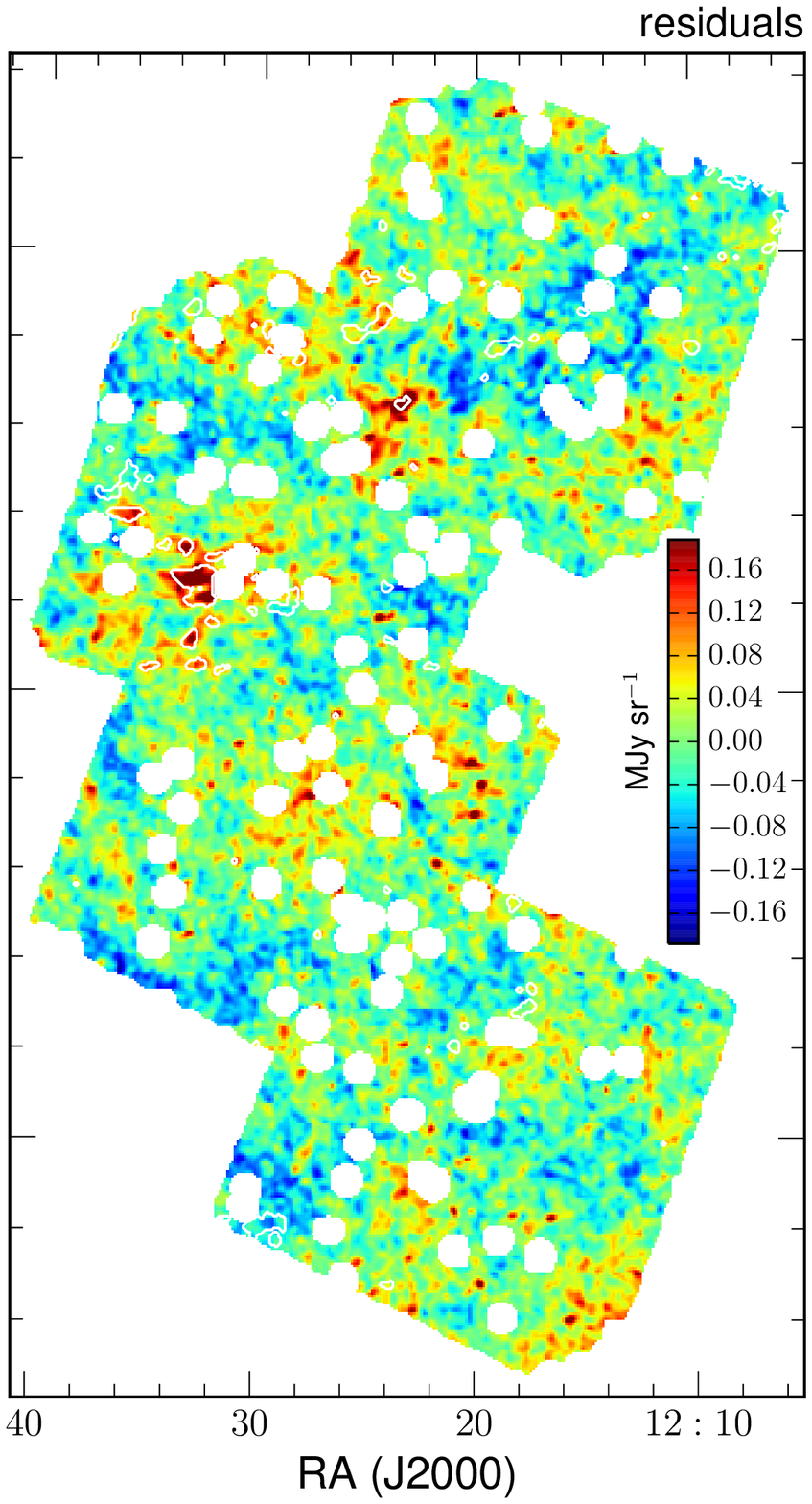}
\caption{Same as Fig.~\ref{fig:modres250}, but for the IRAS $60\mu$m band}
\label{fig:modres60}
\end{figure*}

\begin{figure*}
\includegraphics[scale=\scala,trim = 0 0 8.0bp 0,clip]{\figdir/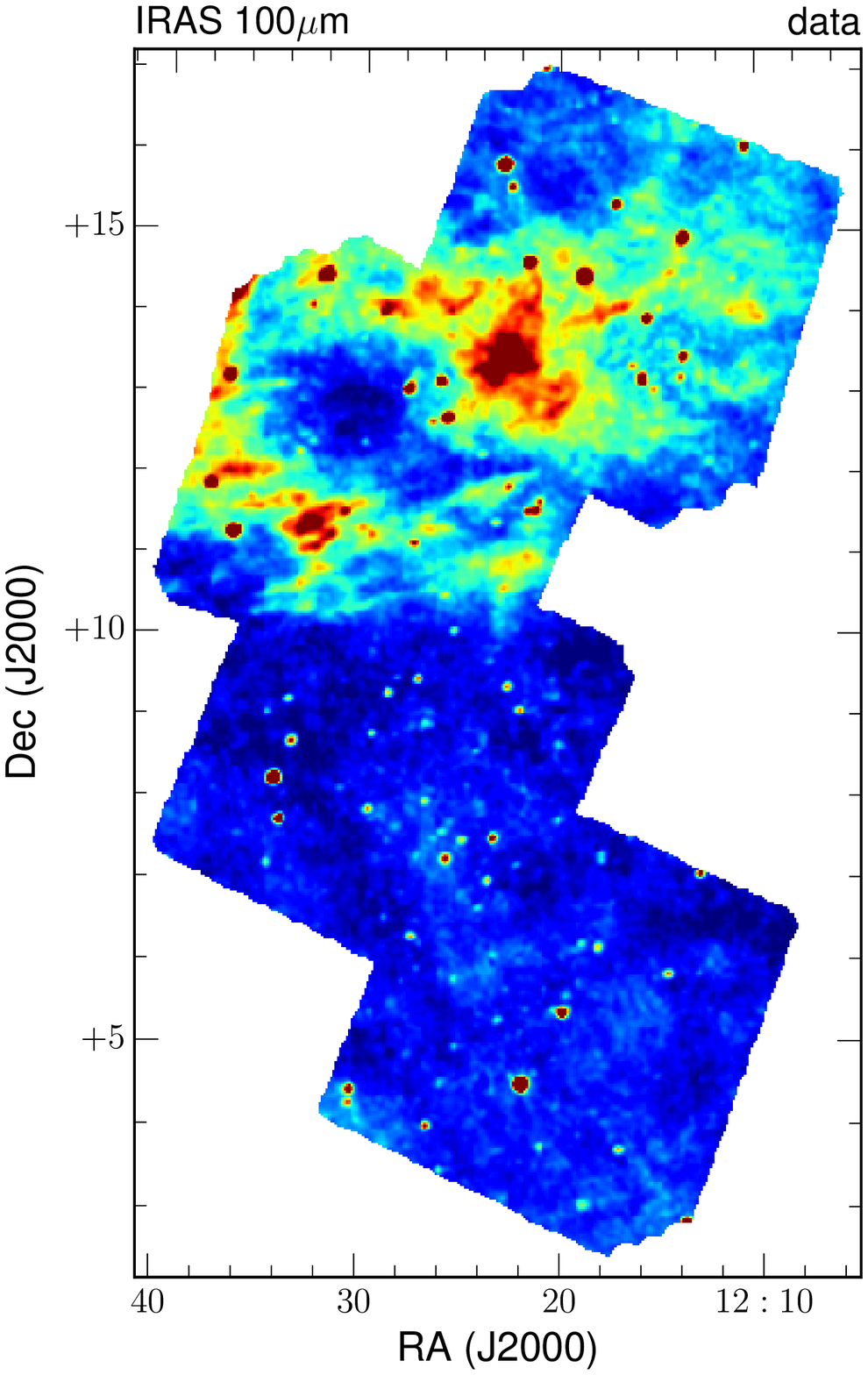}\includegraphics[scale=\scala,trim = 4bp 0 6.5bp 0,clip]{\figdir/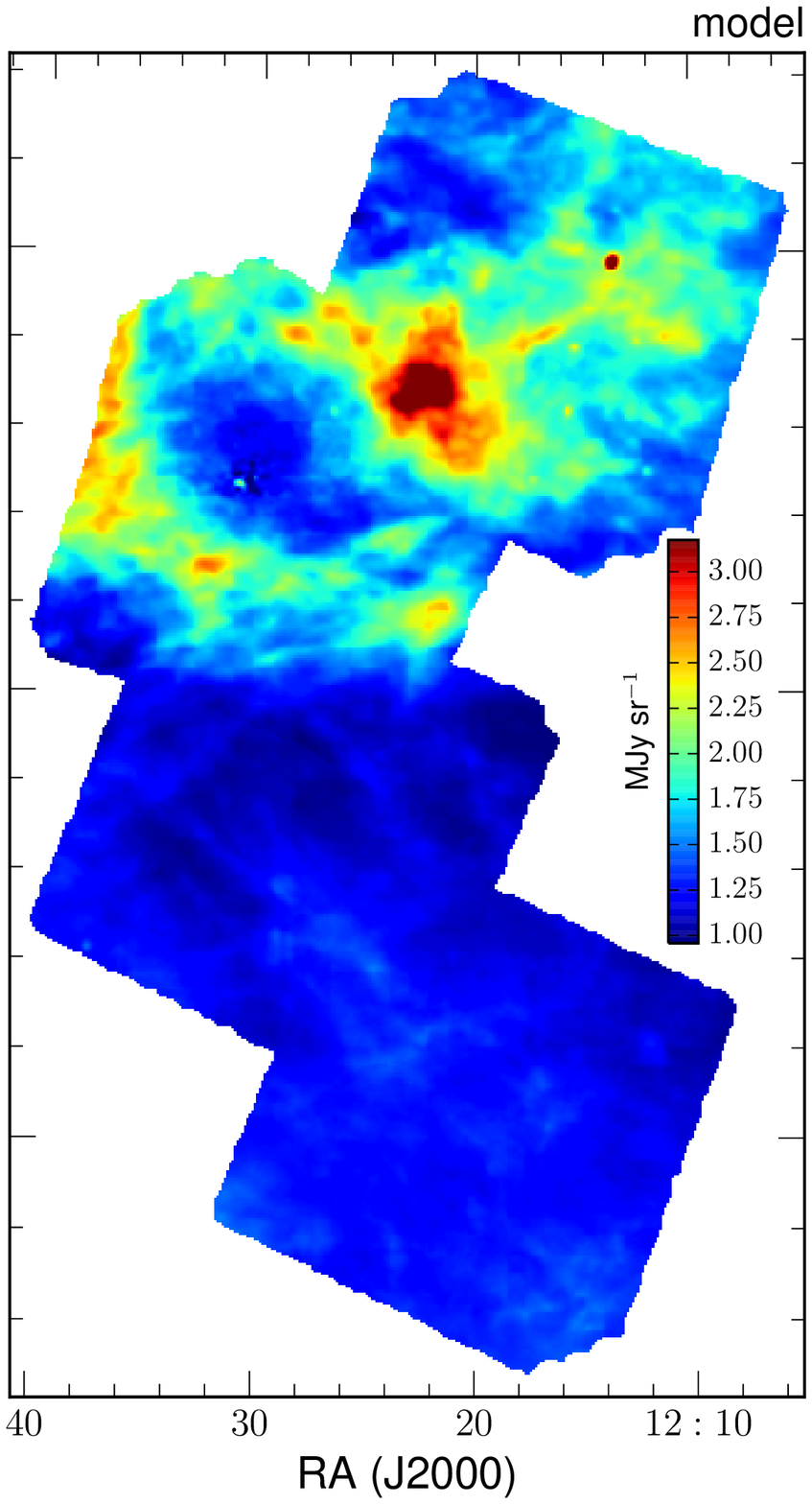}\includegraphics[scale=\scala,trim = 4bp 0 6.5bp 0,clip]{\figdir/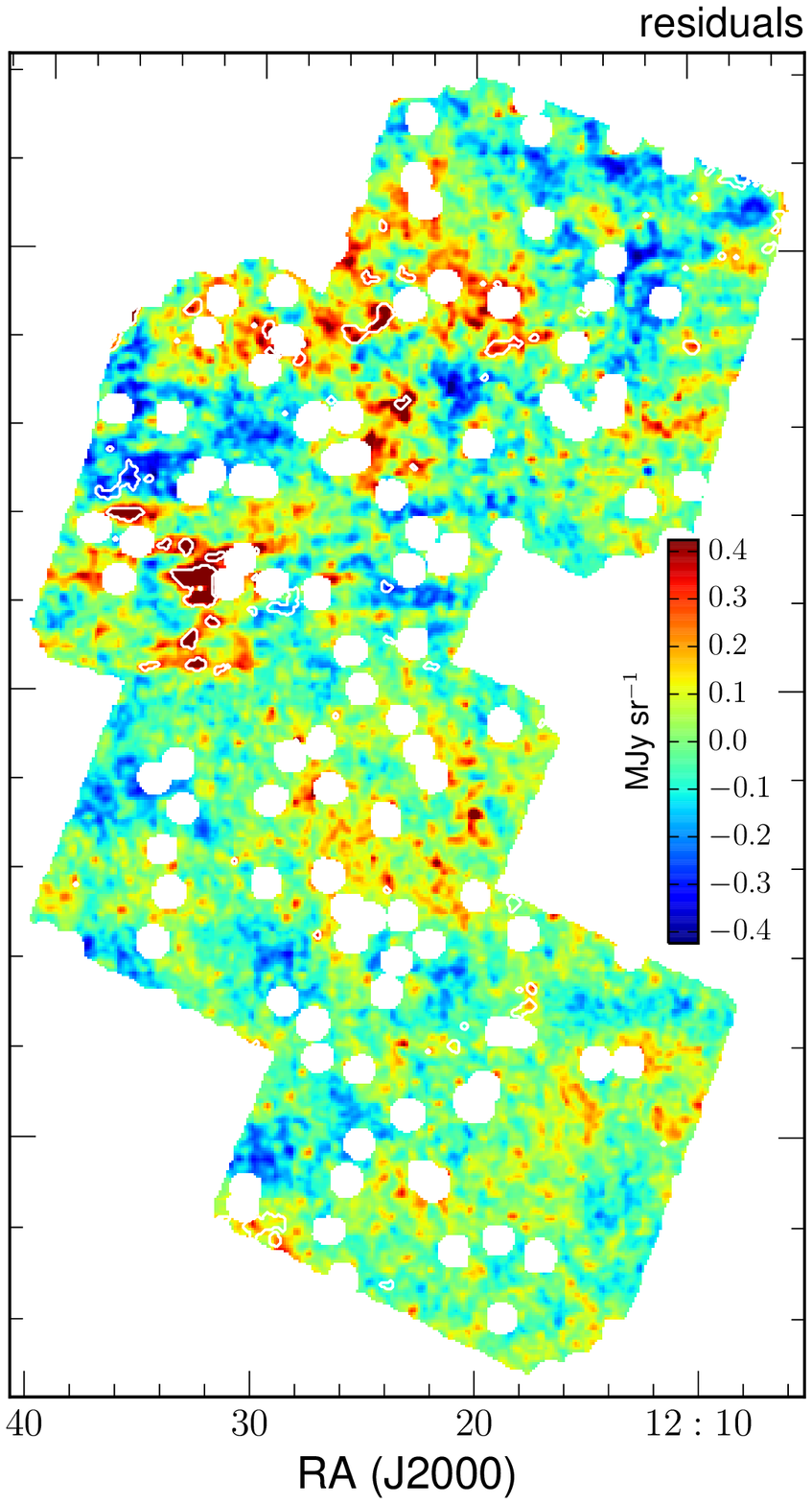}
\caption{Same as Fig.~\ref{fig:modres250}, but for the IRAS $100\mu$m band}
\label{fig:modres100}
\end{figure*}

\begin{figure*}
\includegraphics[scale=\scala,trim = 0 0 8.0bp 0,clip]{\figdir/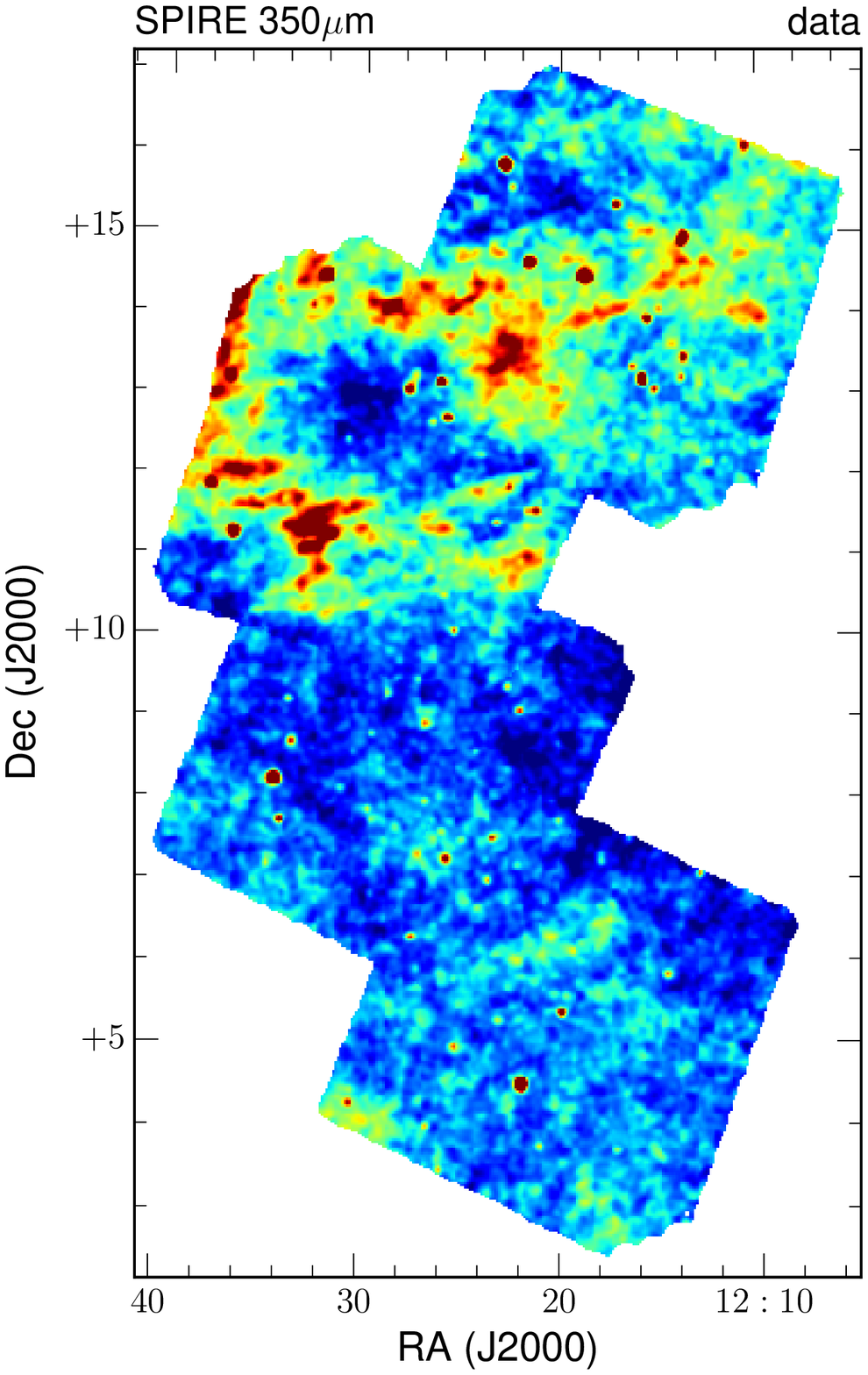}\includegraphics[scale=\scala,trim = 4bp 0 6.5bp 0,clip]{\figdir/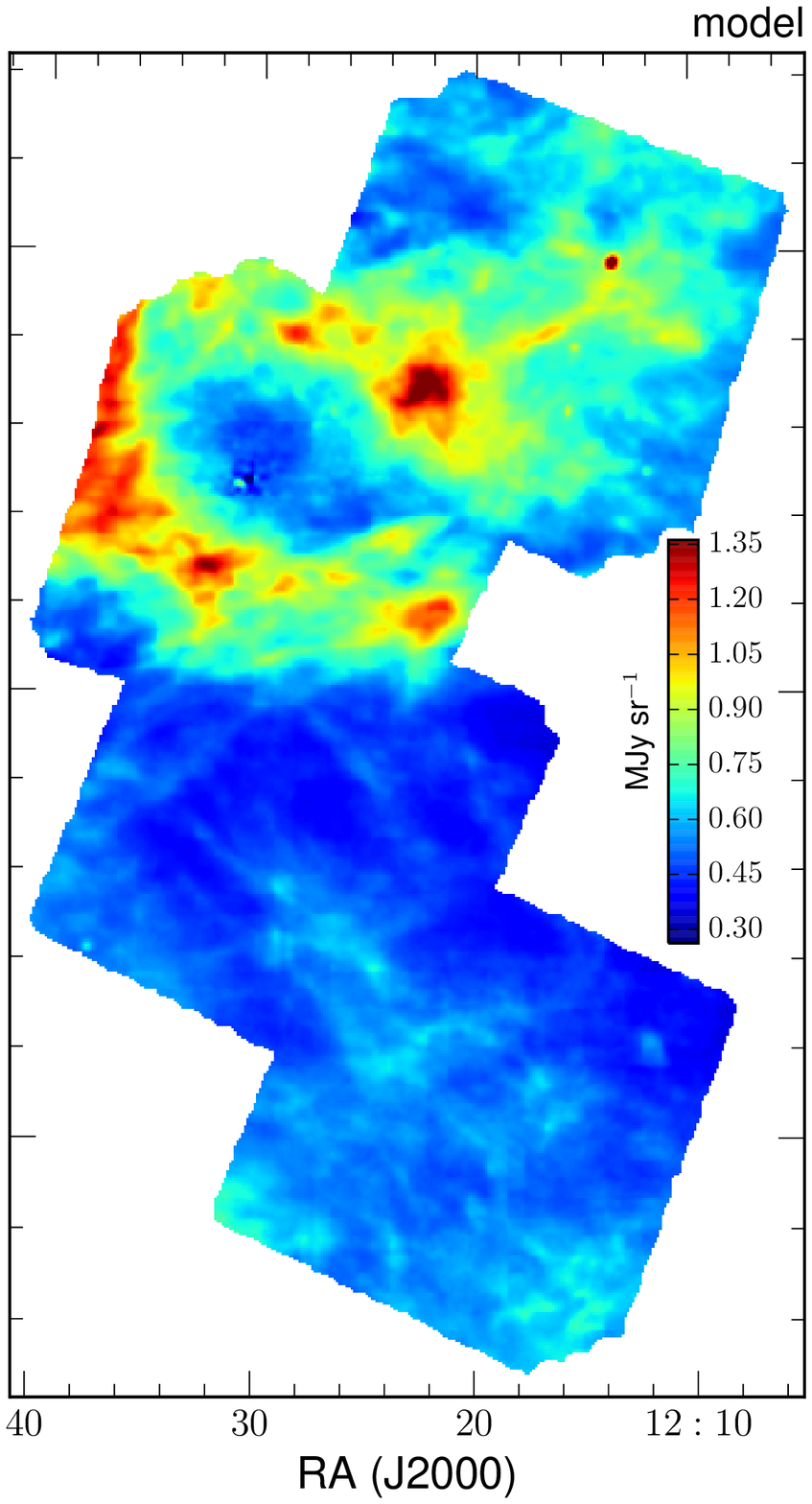}\includegraphics[scale=\scala,trim = 4bp 0 6.5bp 0,clip]{\figdir/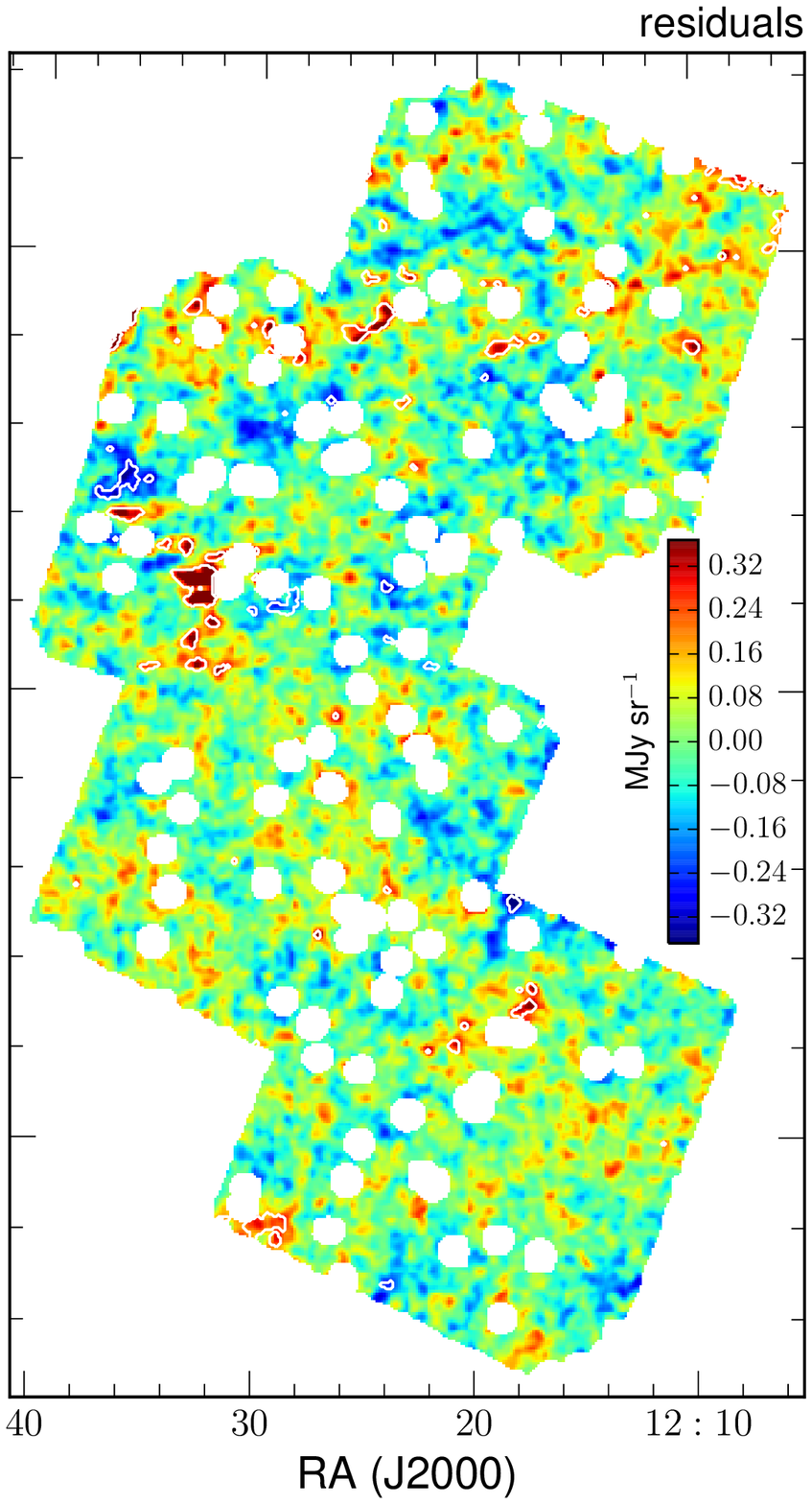}
\caption{Same as Fig.~\ref{fig:modres250}, but for the SPIRE $350\mu$m band}
\label{fig:modres350}
\end{figure*}

Fig.~\ref{fig:modres60} to \ref{fig:modres850} show the observed dust emission, the dust 
emission modelled from the $\ion{H}{i}$ observations, and the residuals in the IRAS
60 and $100\mu$m bands, the SPIRE and HFI $350\mu$m bands, the SPIRE $500\mu$m band and
the HFI $550\mu$m band, and the HFI $850\mu$m band, respectively. Images follow the same
criteria used to present the results for the SPIRE $250\mu$m band in Fig.~\ref{fig:modres250}.

\begin{figure*}
\includegraphics[scale=\scala,trim = 0 0 8.0bp 0,clip]{\figdir/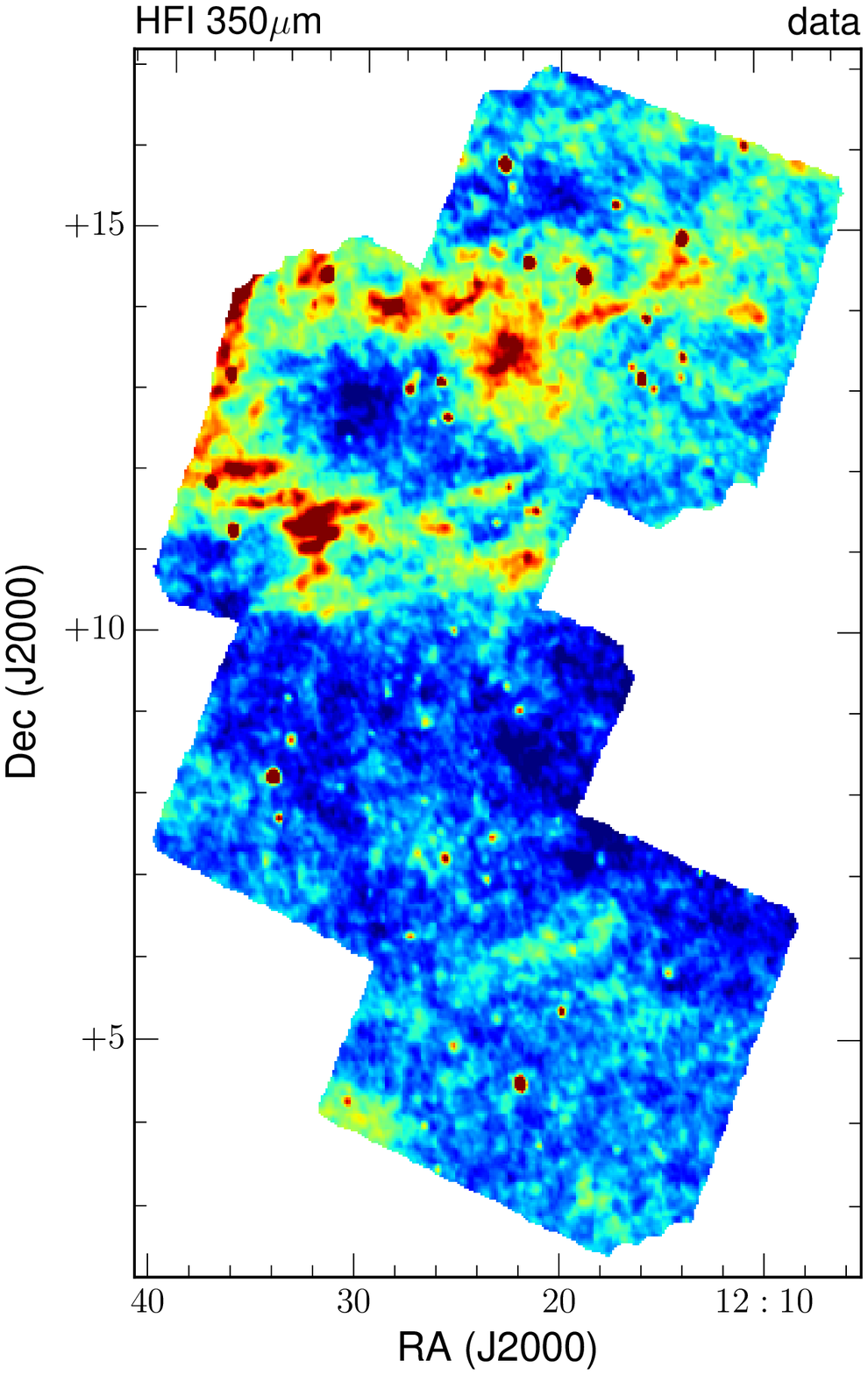}\includegraphics[scale=\scala,trim = 4bp 0 6.5bp 0,clip]{\figdir/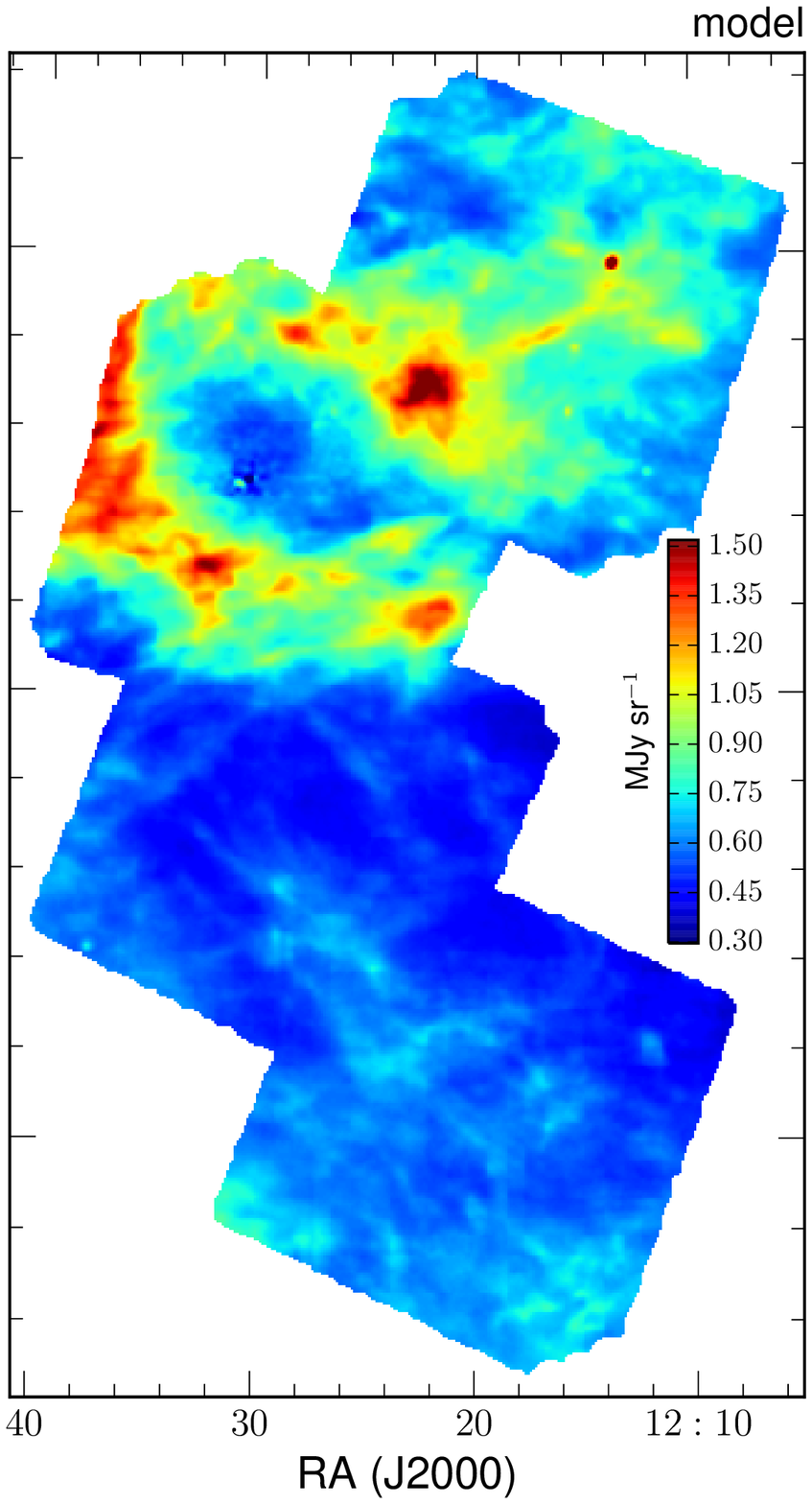}\includegraphics[scale=\scala,trim = 4bp 0 6.5bp 0,clip]{\figdir/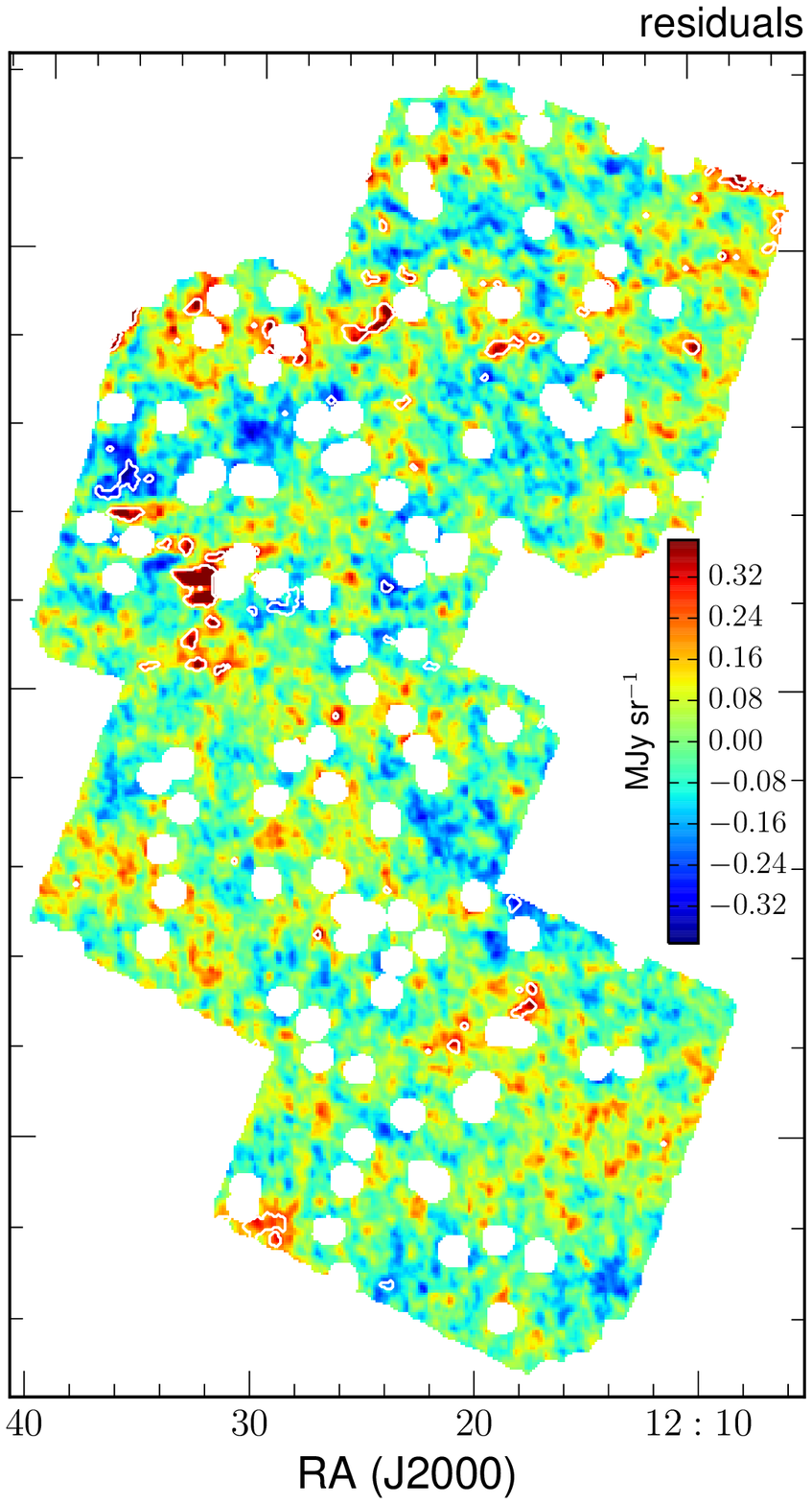}
\caption{Same as Fig.~\ref{fig:modres250}, but for the HFI $350\mu$m band}
\end{figure*} 

\begin{figure*}
\includegraphics[scale=\scala,trim = 0 0 8.0bp 0,clip]{\figdir/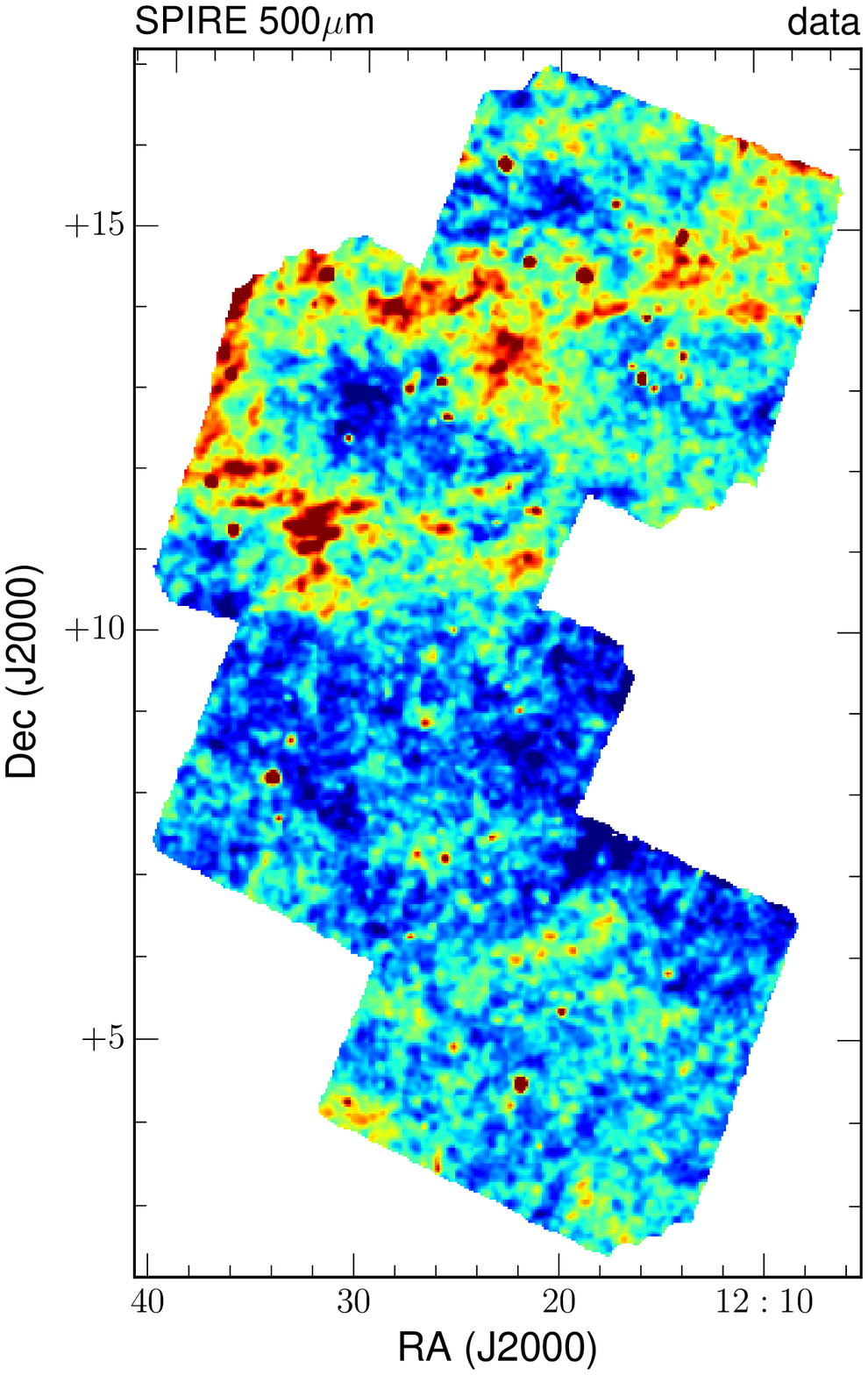}\includegraphics[scale=\scala,trim = 4bp 0 6.5bp 0,clip]{\figdir/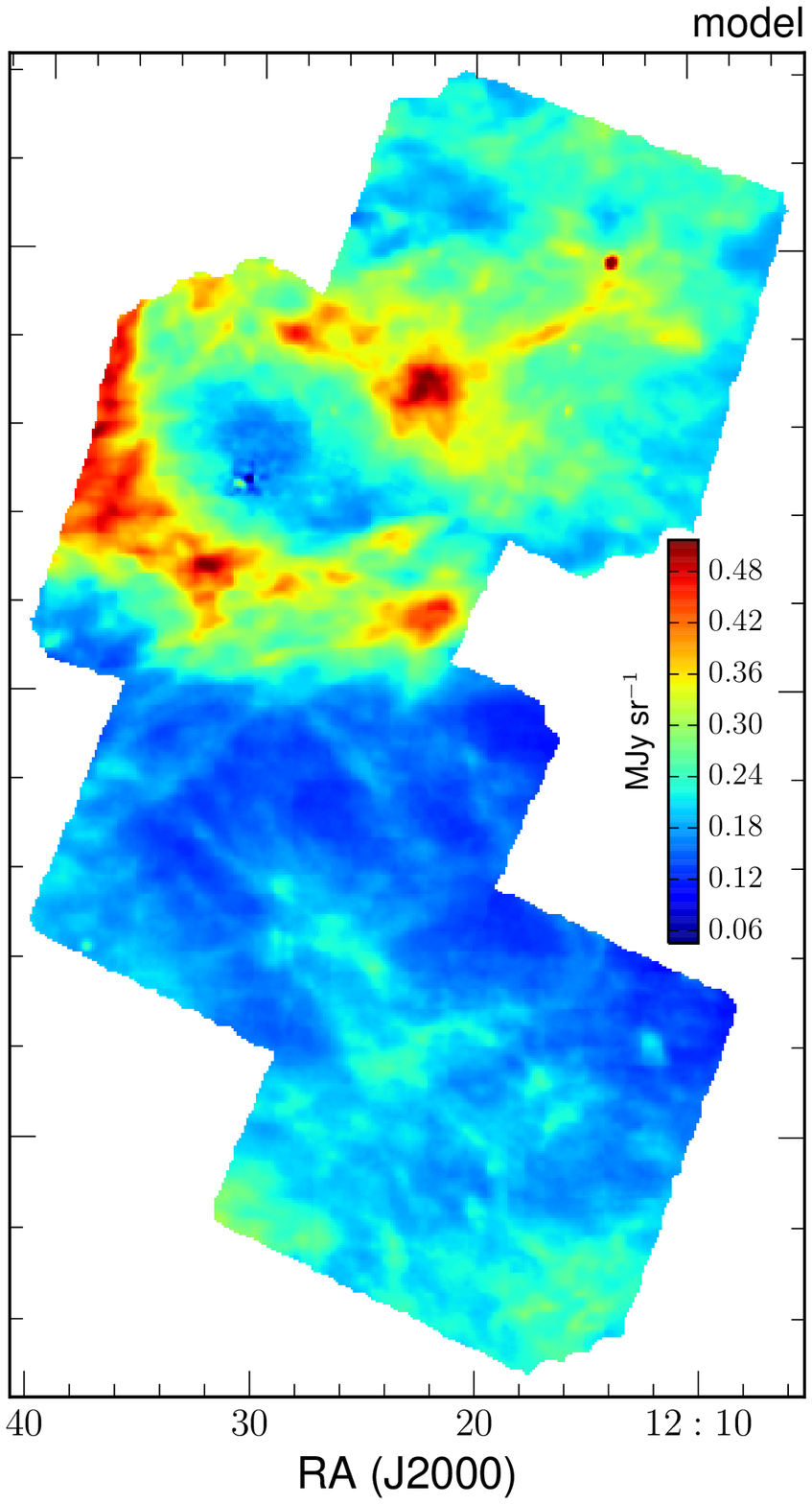}\includegraphics[scale=\scala,trim = 4bp 0 6.5bp 0,clip]{\figdir/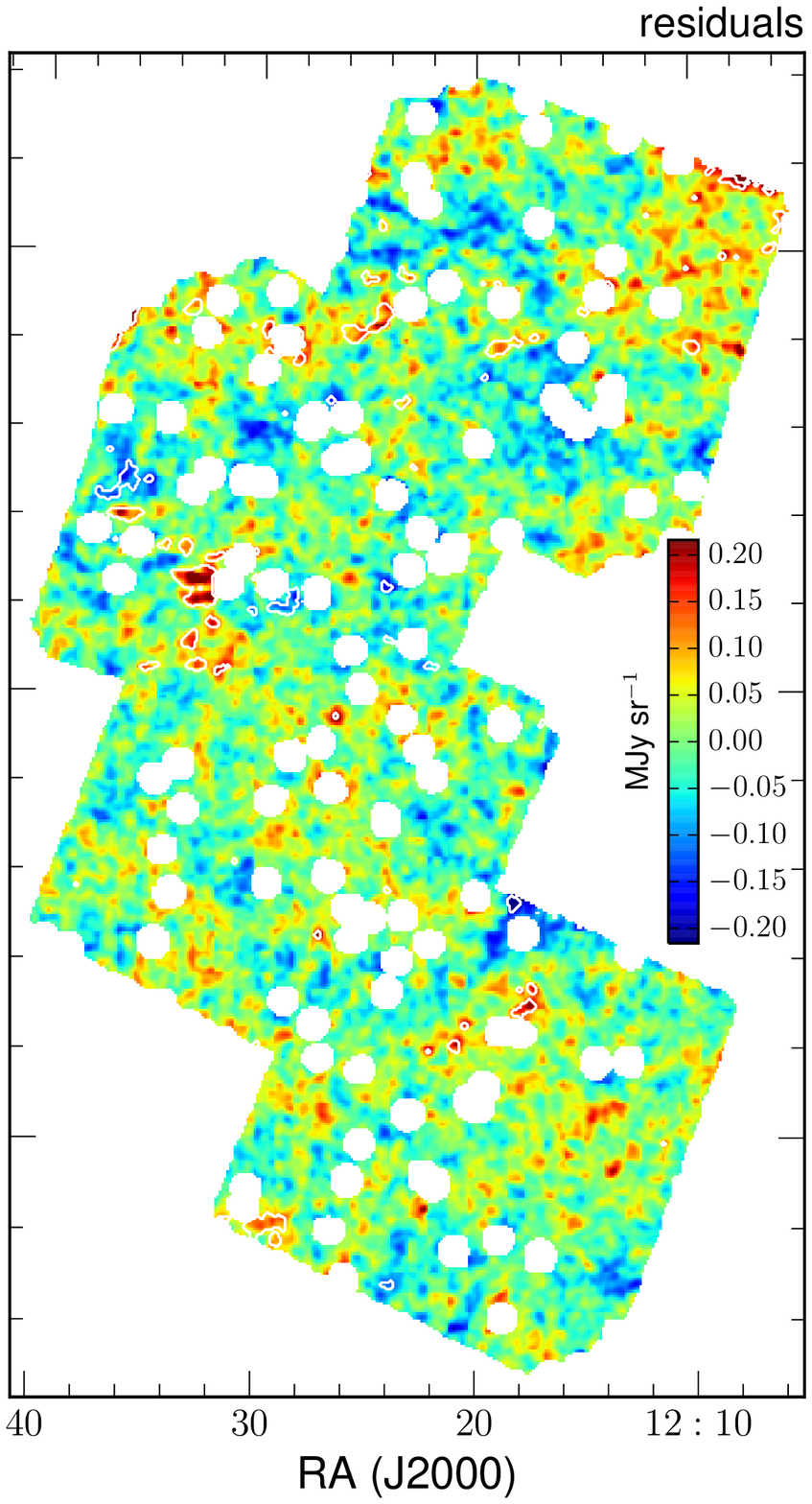}
\caption{Same as Fig.~\ref{fig:modres250}, but for the SPIRE $500\mu$m band}
\end{figure*} 

\begin{figure*}
\includegraphics[scale=\scala,trim = 0 0 8.0bp 0,clip]{\figdir/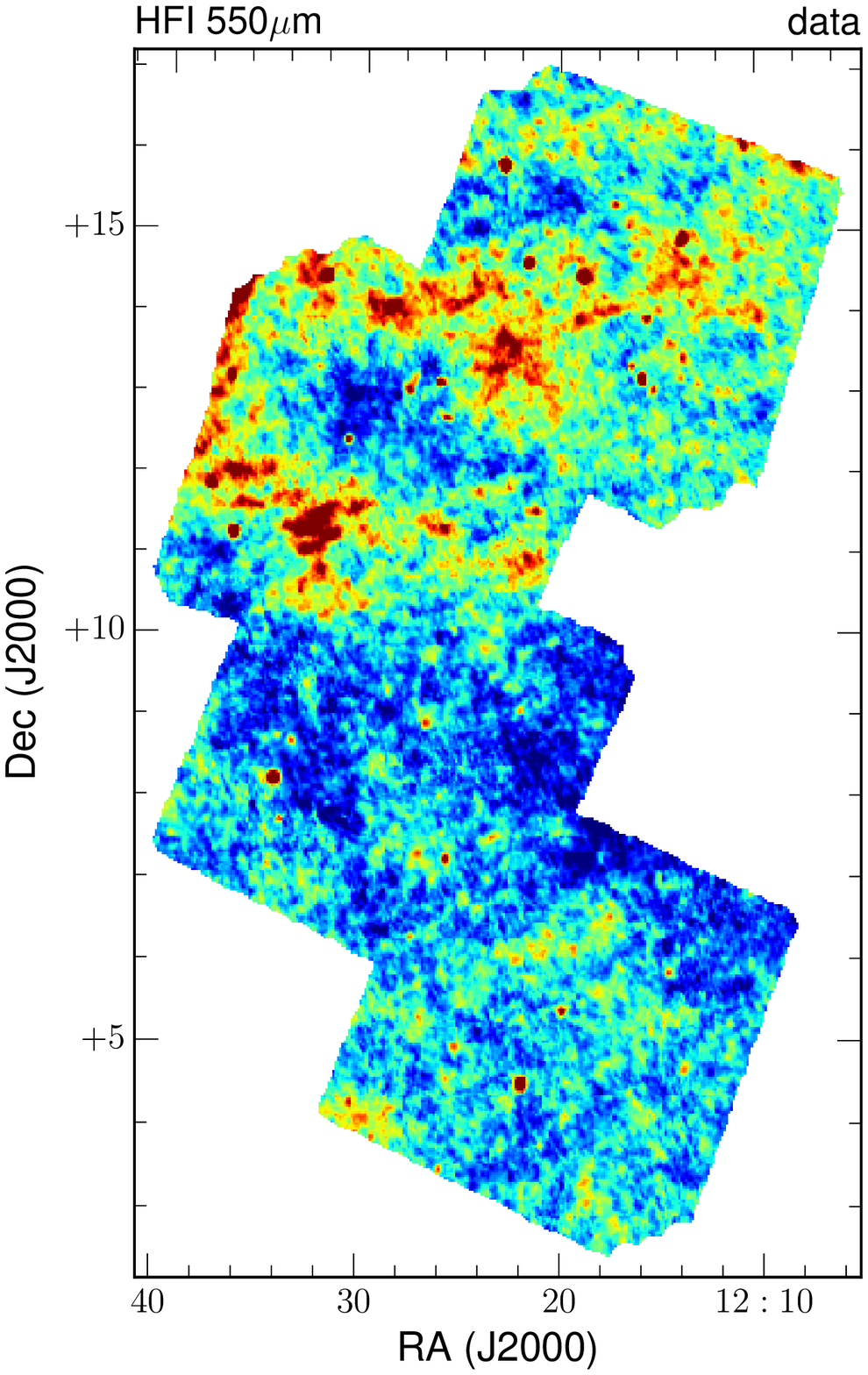}\includegraphics[scale=\scala,trim = 4bp 0 6.5bp 0,clip]{\figdir/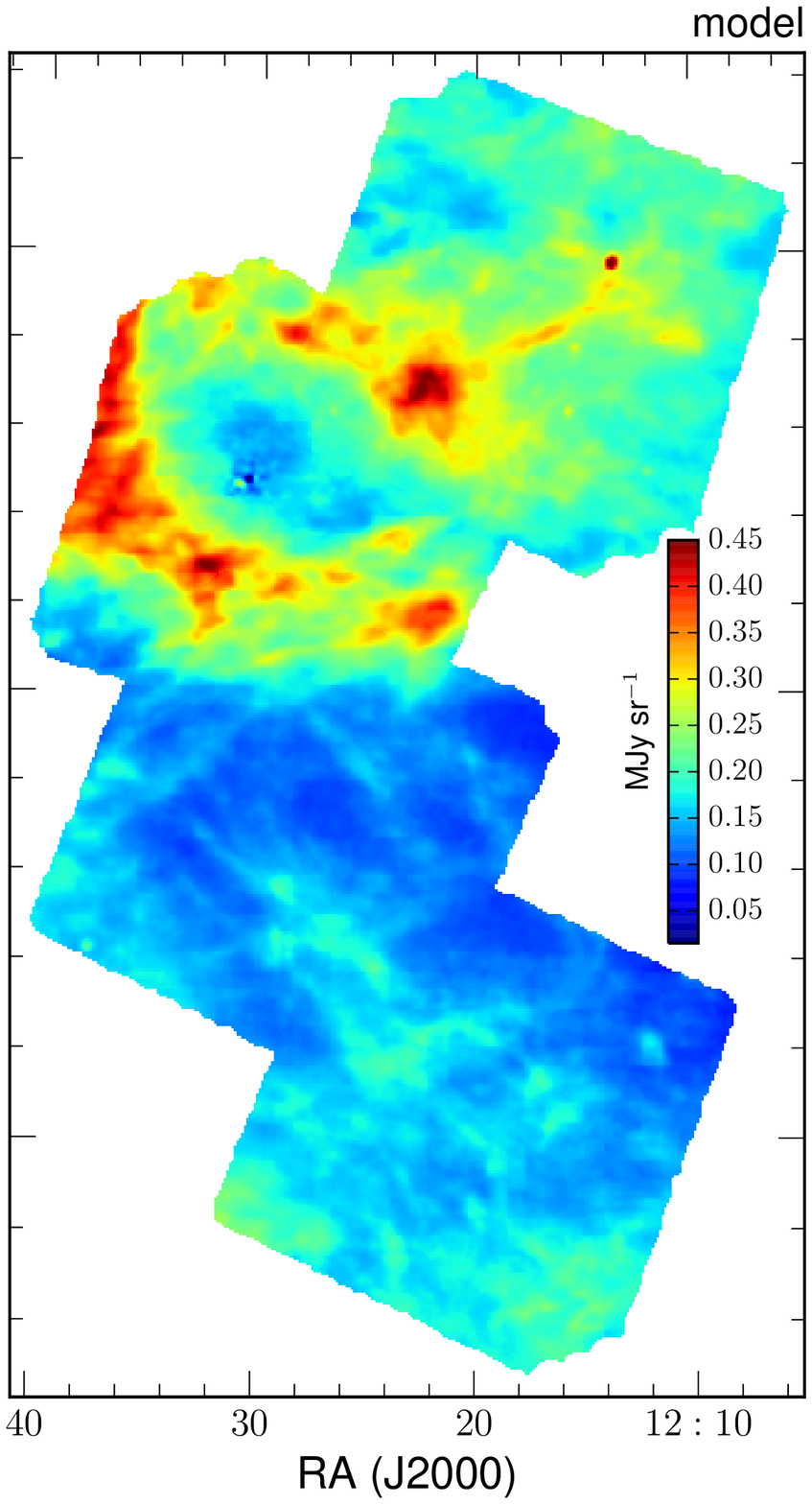}\includegraphics[scale=\scala,trim = 4bp 0 6.5bp 0,clip]{\figdir/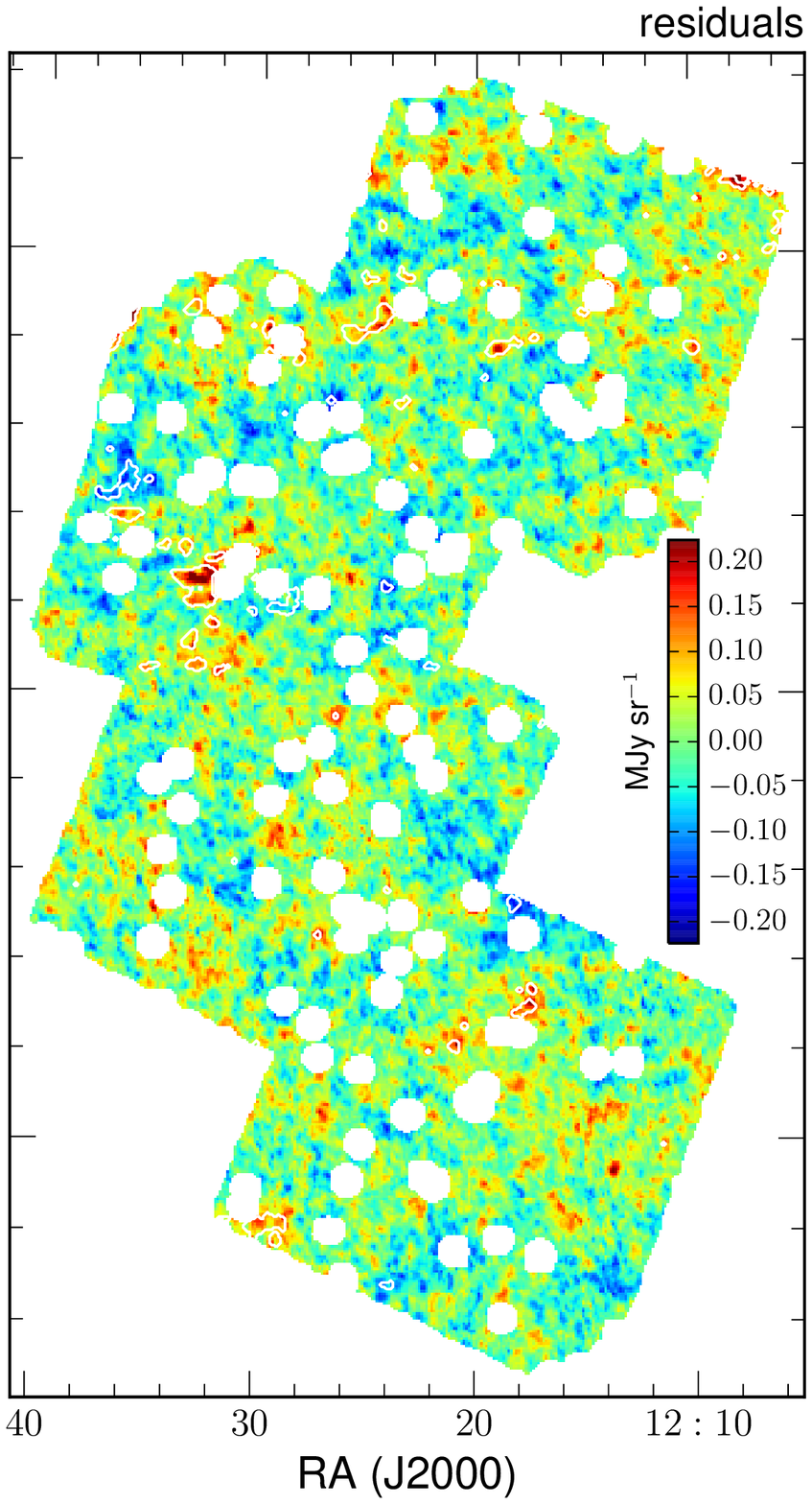}
\caption{Same as Fig.~\ref{fig:modres250}, but for the HFI $550\mu$m band}
\end{figure*} 

\begin{figure*}
\includegraphics[scale=\scala,trim = 0 0 8.0bp 0,clip]{\figdir/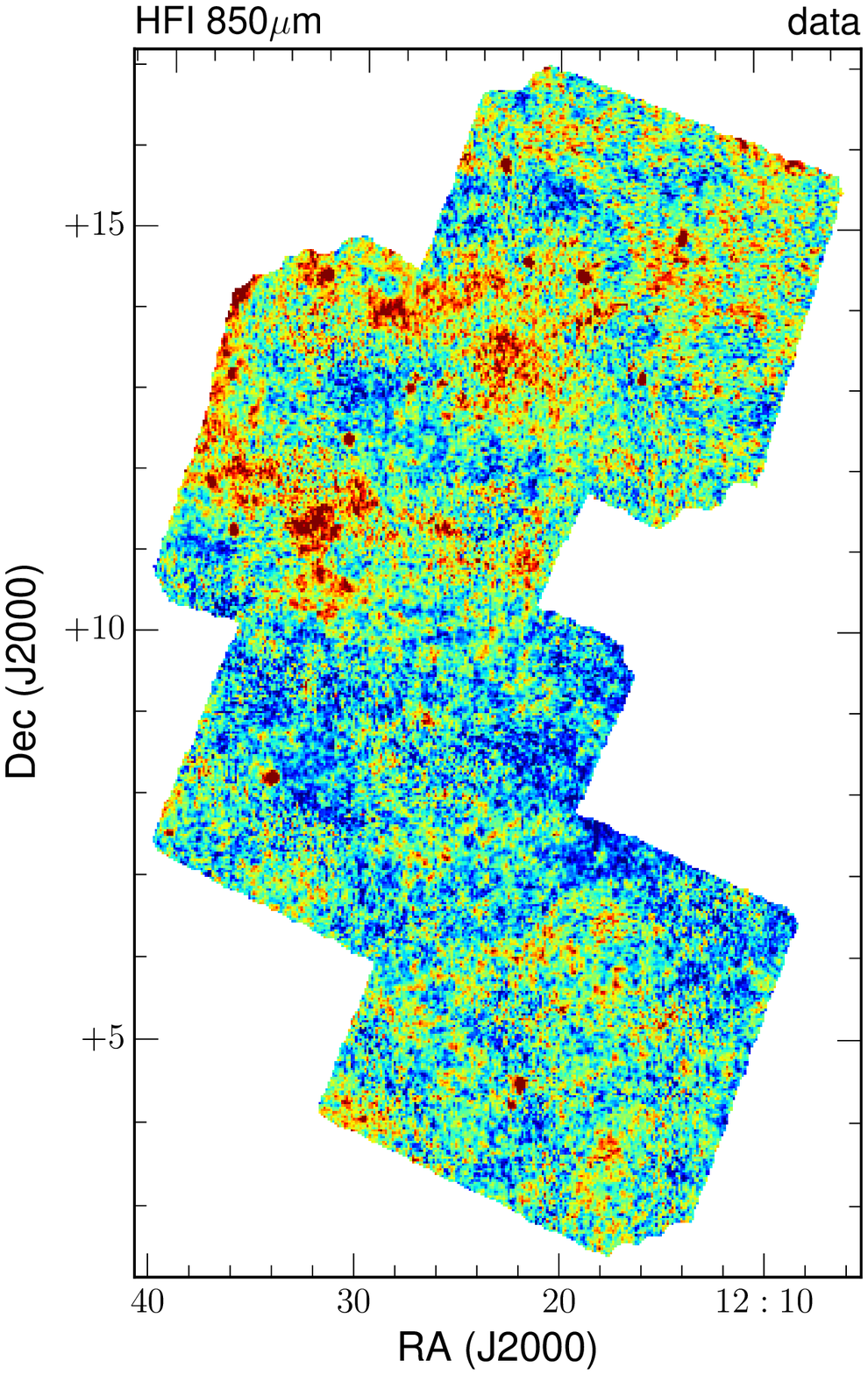}\includegraphics[scale=\scala,trim = 4bp 0 6.5bp 0,clip]{\figdir/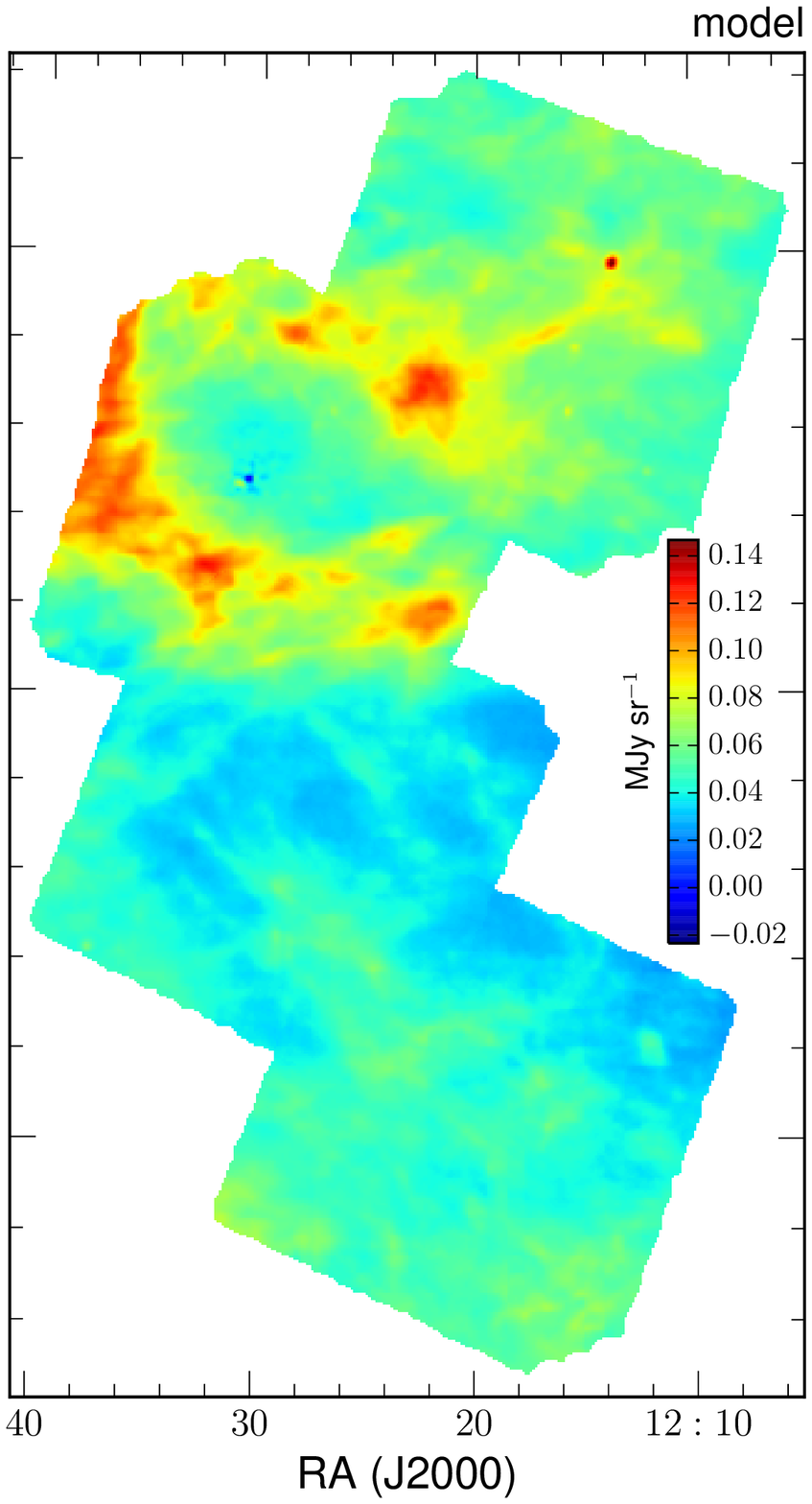}\includegraphics[scale=\scala,trim = 4bp 0 6.5bp 0,clip]{\figdir/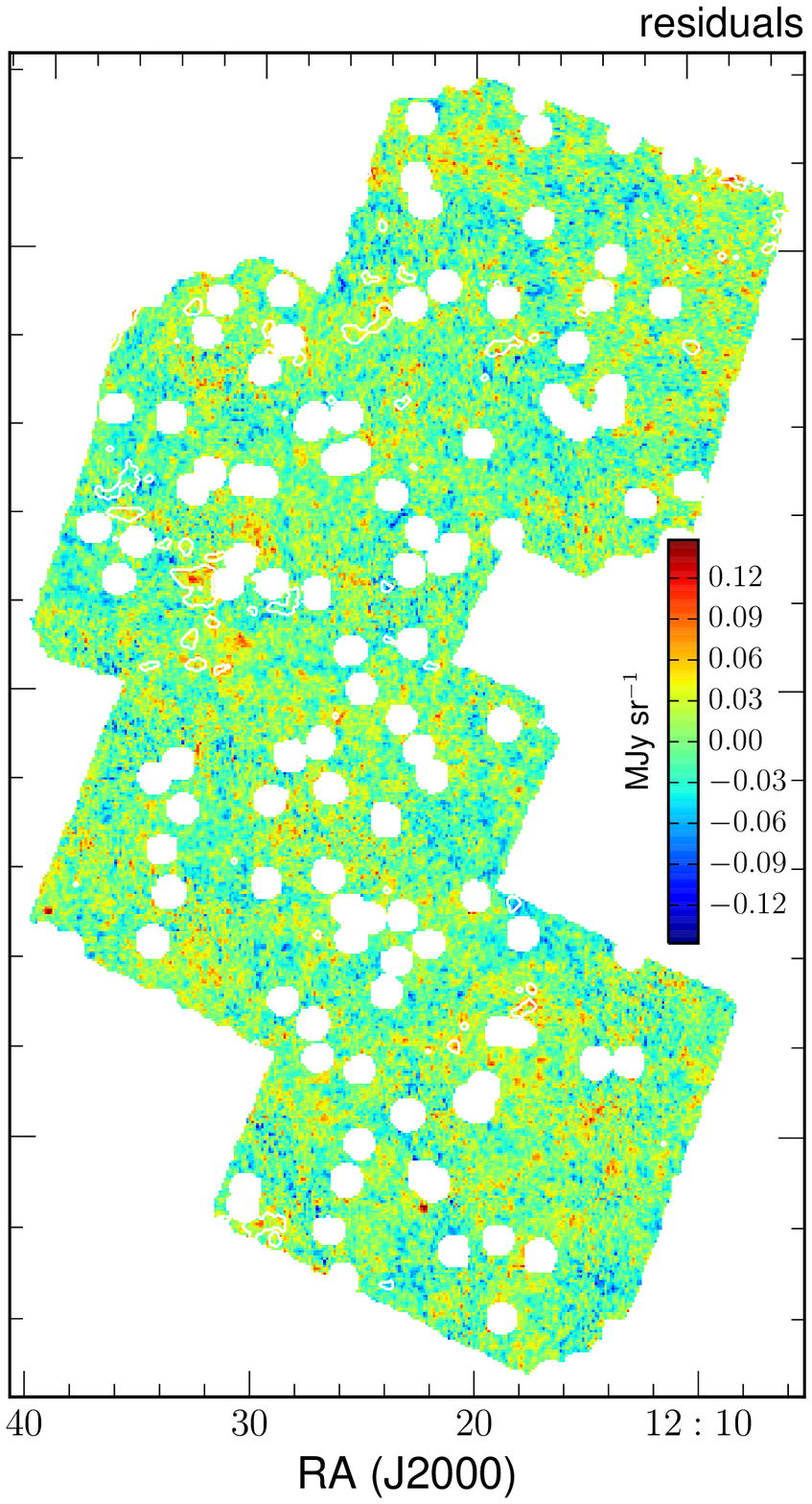}
\caption{Same as Fig.~\ref{fig:modres250}, but for the HFI $850\mu$m band}
\label{fig:modres850}
\end{figure*} 

\end{appendix}

\end{document}